\pdfoutput=1
\documentclass[a4paper,11pt]{article}

\usepackage{lineno}
\usepackage{jheppub}
\usepackage[utf8]{inputenc}
\usepackage{amsmath}
\usepackage{amsthm}
\usepackage{amssymb}
\usepackage{enumitem}   
\usepackage[english]{babel}
\usepackage{url}
\usepackage{mathtools}
\usepackage{bbold}
\usepackage{slashed}
\usepackage{multirow}
\usepackage{lipsum}
\usepackage{xcolor}
\usepackage{float}
\usepackage{revsymb}
\usepackage{braket}
\usepackage{siunitx}
\usepackage{xfrac}
\usepackage{physics}
\usepackage{caption}
\usepackage{subcaption}
\usepackage{comment}

\addtocontents{toc}{\protect\setcounter{tocdepth}{2}}

\allowdisplaybreaks

\newcommand{\bm}[1]{\boldsymbol{#1}}

\definecolor{dgreen}{rgb}{0,0.6,0.0}

\begin{document}


\preprint{
\begin{tabular}{l}
MS-TP-26-19\\
KEK-TH-2858
\end{tabular}
}

\title{Axion-driven spontaneous leptogenesis, precisely}

\author[a]{Konstantin Kuckenberg,}
\author[b]{Martin A.\ Mojahed,}
\author[a,c]{Kai~Schmitz,}
\author[d]{Hidenaga Watanabe}

\affiliation[a]{Institute for Theoretical Physics, University of M\"unster, 48149 M\"{u}nster, Germany}
\affiliation[b]{Istituto Nazionale di Fisica Nucleare (INFN) Sezione di Roma, 00185 Rome, Italy}
\affiliation[c]{Kavli IPMU (WPI), UTIAS, The University of Tokyo, Kashiwa, Chiba 277-8583, Japan}
\affiliation[d]{Institute of Particle and Nuclear Studies, KEK, Tsukuba, Ibaraki 305-0801, Japan}
\affiliation[d]{Graduate University for Advanced Studies (Sokendai),1-1 Oho, Tsukuba, Ibaraki 305-0801, Japan}

\emailAdd{kkuckenb@uni-muenster.de}
\emailAdd{martin.mojahed@roma1.infn.it}
\emailAdd{kai.schmitz@uni-muenster.de}
\emailAdd{hidenaga@post.kek.jp}


\abstract{We revisit a minimal scenario for spontaneous leptogenesis where the asymmetry generation is driven by a heavy axion-like field evolving via standard misalignment. We focus on the simplest framework in which lepton number violation originates from the dimension-five Weinberg operator and contributions from standard thermal leptogenesis become negligible. The asymmetry is computed by solving a set of transport equations, which incorporate fully flavor-dependent interaction rates for the Weinberg operator that are presented here for the first time. We also derive a simple algebraic master formula for the generated $B\!-\!L$ asymmetry, valid for arbitrary axion velocities and classically shift-symmetric couplings to the Standard Model, that reproduces the $B\!-\!L$-production dynamics of the full set of transport equations in controlled parameter regions. Along the way, we quantify the importance of low-scale neutrino masses and the two mass orderings on the final prediction for the generated asymmetry. Finally, we assess the cosmological viability of different axion couplings by accounting for entropy dilution from late-time axion decays and imposing constraints from baryonic isocurvature perturbations. We demonstrate that explaining the baryon asymmetry via standard Majoron misalignment is under severe tension due to dilution from late-time decays. In contrast, axions coupled to strong or weak sphalerons can successfully produce the observed asymmetry for decay constants near the energy scale of grand unification and axion oscillation temperatures $T_{\text{osc}} \gtrsim 10^{11\cdots12}$ GeV, provided a cosmological history featuring high-scale inflation and efficient reheating.
}


\maketitle


\section{Introduction}

Axions are pseudoscalar fields whose leading interactions with Standard Model (SM) gauge bosons and fermions are governed by (classically) shift-symmetric, dimension-five operators. Following the proposal of the QCD axion as a solution to the strong $CP$ problem nearly half a century ago~\cite{Peccei:1977hh,Peccei:1977ur,Weinberg:1977ma,Wilczek:1977pj}, the axion paradigm has broadened considerably, with axions appearing in many field and string theory frameworks to address open problems in particle physics and cosmology. Their approximate shift symmetry protects a naturally flat potential, making axions attractive ingredients for inflationary model building~\cite{Freese:1990rb}. An axion is also central in the relaxion mechanism~\cite{Graham:2015cka}, which addresses the electroweak hierarchy problem. In addition, axions constitute one of the most compelling and well-studied dark-matter candidates~\cite{Preskill:1982cy,Abbott:1982af,Dine:1982ah}. Finally, the existence of axions is well motivated from ultraviolet (UV) completions such as string theory, where axions arise generically from the dimensional reduction of higher-dimensional form fields in compactifications~\cite{Ibanez:1986xy,Banks:2003sx,Svrcek:2006yi}.

Axions may have also played an active role in generating the baryon asymmetry of the Universe (BAU), commonly expressed by the baryon-to-photon ratio $\eta_B^{\rm{obs}}=n_b/n_\gamma=(6.12\pm0.04)\times10^{-10}$~\cite{Planck:2018vyg,ParticleDataGroup:2024cfk}. A popular axion-based framework is spontaneous baryogenesis~\cite{Cohen:1987vi,Cohen:1988kt}, in which a coherently evolving axion background with non-vanishing velocity spontaneously violates $CPT$ symmetry, acting as an effective chemical potential that shifts the energies of particles relative to those of antiparticles. In the presence of particle-number-violating reactions, the axion-induced bias can lead to a net asymmetry between particle and antiparticles even if the thermal plasma remains close to thermal equilibrium. The spontaneous baryogenesis framework is particularly compatible with inflationary cosmology~\cite{Starobinsky:1980te,Guth:1980zm}, since inflation can homogenize the axion field over our observable patch of the Universe and set initial conditions that support coherent post-inflationary evolution. Existing model realizations differ in the origin of $B$ or $B\!-\!L$ violation, in the axion couplings to the Standard Model (SM), in the microphysics responsible for initiating axion motion, and in the resulting axion motion~\cite{Chiba:2003vp,Takahashi:2003db,Kusenko:2014uta,Ibe:2015nfa,Takahashi:2015waa,Bae:2018mlv,Co:2019wyp,Domcke:2020kcp,Co:2020xlh,Co:2020jtv,Foster:2022ajl,Chao:2023ojl,Chun:2023eqc,Barnes:2024jap,Datta:2024xhg,Berbig:2025hlc,Peng:2025sri,Asadi:2025cvm,Chun:2025abp,Takahashi:2026ngu,Chun:2026jgn}. In the most minimal\,---\,and arguably best motivated\,---\,scenario, the axion acquires a displaced field value during inflation and relaxes toward its minimum after inflation ends, reaching the bottom of the potential for the first time when the Hubble parameter $H$ drops to the scale of the axion mass $m_a$. To generate the observed BAU, this scenario requires the axion to start oscillating at temperatures far above the electroweak phase transition, and consequently also requires a source of $B\!-\!L$ violation to avoid complete washout by weak sphalerons~\cite{Kuzmin:1985mm}. To this end, it is common to include the Weinberg operator~\cite{Weinberg:1979sa}, which simultaneously provides $B\!-\!L$ violation and a natural explanation for the smallness of active neutrino masses.

In this paper, we present a comprehensive analysis of this minimal spontaneous leptogenesis (LG) scenario with the Weinberg operator as the only source of lepton-number violation (LNV). To this end, we present flavor-dependent LNV interaction rates mediated by the Weinberg operator, expressed directly in terms of low-energy neutrino parameters, which to the best of our knowledge, is the first time such a result appears in the literature. The LNV rates are incorporated into a set of Boltzmann equations describing the complete evolution of SM-particle asymmetries in the axion background, which we solve numerically. Our setup of Boltzmann equations enables a systematic study of spontaneous LG, including a quantitative assessment of the sensitivity of the generated BAU to the choice of lepton-flavor basis, neutrino mass ordering, and the absolute neutrino mass scale.

In our analysis, we consider axion couplings to weak sphalerons, strong sphalerons, and the $B\!-\!L$ current (as in standard Majoron realizations~\cite{Chikashige:1980ui,Gelmini:1980re}), and track the full time evolution of the axion velocity. In addition, we develop a simplified semi-analytic treatment of the evolution of the $B\!-\!L$ asymmetry for axions with generic couplings, based on an instantaneous-equilibration approximation of SM interaction rates. This approximation allows a reduction of the coupled set of Boltzmann equations to a single algebraic master equation that governs the evolution of $B\!-\!L$ asymmetry. We compare this result with numerical solutions to the full set of Boltzmann equations, finding good agreement. Although we focus on standard misalignment here, we emphasize that both the full set of Boltzmann equations and the algebraic master formula can be readily applied to scenarios with generic axion evolution.

The remainder of this paper is organized as follows. In Sec.~\ref{sec:2}, we review the fundamentals of spontaneous baryogenesis, connect the Weinberg operator to low-energy neutrino data, and present our results for the fully flavor-dependent interaction rates induced by the Weinberg operator. Sec.~\ref{sec:3} establishes the network of Boltzmann equations governing the evolution of SM-particle asymmetries. We solve these equations numerically across various temperature scales, varying the axion mass, couplings, and neutrino parameters. We also introduce an algebraic master formula for the $B\!-\!L$ asymmetry that accurately reproduces the fully numerical predictions for the $B\!-\!L$ asymmetry. Some technical details of these transport equations and the algebraic approximation are collected in App.~\ref{appendix:BEq} and App.~\ref{appendix:algebraic}, respectively. In Sec.~\ref{sec:4}, we evaluate the viability of embedding spontaneous LG into a consistent cosmological history and impose constraints from baryonic isocurvature perturbations and late-time entropy dilution from axion decays to map out the successful parameter space for the BAU. Supplementary numerical results are provided in App.~\ref{appendix:NumericResults}, and we present our conclusions in Sec.~\ref{sec:conclusion}.

\section{Evolution of the axion field}
\label{sec:2}
The purpose of the following section is to first outline the qualitative features of spontaneous baryogenesis and introduce our setup, including the axion velocity and its couplings to the SM. We then present the flavor-dependent interaction rates mediated by the Weinberg operator and express them in terms of low-scale neutrino parameters. These results will be used as inputs in subsequent sections about asymmetry generation. 

\subsection{Equation of motion}
\label{sec:EoM}

Models of spontaneous baryogenesis commonly assume a coupling between a current $j_A^\mu$ associated with some $U(1)_A$ symmetry, where $A$ is typically taken to be baryon number $B$, and a homogeneous axion field $a(t)$ with a decay constant $f$, 
\begin{align}
\label{eq:muAeff}
    \mathcal{L}_a\supset \frac{\partial_\mu a}{f}\,j_A^\mu=\frac{\dot{a}}{f}\,j^0_A=\mu_A^{\rm eff}j^0_A.
\end{align}
This $CPT$-violating coupling can be regarded as an external chemical potential, $\mu_A^{\rm eff}$, which induces non-vanishing chemical potentials $\mu_i$ for all particle species $i$ carrying $A$ charge. At least some of the particles charged under $A$ generally also carry additional global $U(1)$ charges. If reactions in the plasma violate one of these global charges\,---\,call it $C$\,---\,then these reactions, in the presence of the external chemical potential bias induced by the axion motion, will drive the thermal plasma into a state of nonzero equilibrium charge density $q_C^{\rm eq}$. The mapping between $q_C^{\rm eq}$ and the effective external chemical potential $\mu_A^{\rm eff}$ can be determined by identifying the processes that are involved in driving the plasma to its new equilibrium state; see Ref.~\cite{Domcke:2020kcp} for general and explicit formulae.

To ensure that the axion field is homogeneous on super-Hubble scales, we assume that the global symmetry yielding the axion as a pseudo-Nambu--Goldstone boson breaks well before the end of inflation and is never restored in the post-inflationary Universe. Thus
\begin{align}
    \label{eq:f>}
    f>H_I,\, T_{\text{max}},
\end{align}
where $H_I$ denotes the Hubble scale during inflation and $T_{\text{max}}$ is the maximal temperature attained in the post-inflationary Universe. Under these assumptions, the initial field value of the axion at the end of inflation is
\begin{align}
    a_i=f\,\theta_i,
\end{align}
with $\theta_i\in (-\pi, \,\pi]$ an \textit{a priori} random misalignment angle, which is treated as a free parameter in our observable Universe.\footnote{In contrast, if the symmetry is broken or restored after inflation ($f< H_I$), the initial field value varies randomly ($\theta_i \in (-\pi, \pi]$) across causally disconnected regions. This leads to spatial domains with at least $\mathcal{O}(1)$ contrasts in the local baryon yield, including potential regions of anti-matter dominance. We therefore restrict our analysis to the pre-inflationary regime, where the spatial background for $\dot{a}(t)$ remains manifestly homogeneous.} The subsequent evolution of the homogeneous mode is governed by the equation of motion (EoM) of the axion field, 
\begin{align}
    \label{EoM}
    \ddot{a}+3H(t)\,\dot{a}+\partial_a V(a)\approx 0 \,,
\end{align}
where we neglect the backreaction from the produced charge asymmetries, i.e., we assume that any plasma-induced dissipation term in the axion EoM is subdominant compared to Hubble friction over the parameter range of interest; see Sec.~\ref{sec:constraints} for an extended discussion on this point. In the following, we shall approximate the axion potential by its harmonic expansion about a minimum, $V(a)=\sfrac{1}{2}\,m_a^2a^2$. This is, in general, a sound approximation for a generic axion potential if the relevant field excursion during the epoch of interest remains sufficiently close to the minimum, or at least not near the hilltop, so that anharmonic corrections are negligible. Accordingly, we restrict our numerical analysis to initial misalignments in the range $\theta_i\in [0,1]$. Possible temperature dependence of the axion potential can be incorporated by promoting $m_a\rightarrow m_a(T)$, but we will consider a temperature-independent mass satisfying $m_a<H_I$. While we need not specify the origin of the axion mass, let us note that it may arise e.\,g.\ from an additional anomalous coupling of the axion to the gauge fields of a strongly coupled hidden sector or from a generic source of explicit breaking of the global symmetry yielding the axion as a pseudo-Nambu--Goldstone boson~\cite{Holdom:1982ex,Kamionkowski:1992mf,Barr:1992qq,Rubakov:1997vp}. 

In a radiation-dominated Universe with a fixed number of relativistic degrees of freedom, where $H(t) = 1/(2t)$, the EoM in Eq.~\eqref{EoM}, together with our choice of scalar potential, $V(a)=\sfrac{1}{2}\,m_a^2a^2$, admits a simple analytic solution for the misalignment angle,
\begin{align}
\label{eq:theta}
    \theta=\theta_i\left(\frac{2}{m_a t}\right)^{\frac{1}{4}}\Gamma\left(\frac{5}{4}\right)J_{1/4}(m_a t),
\end{align}
where $J_n(x)$ denotes the Bessel function of the first kind, $\Gamma(x)$ is the gamma function, and we have set $\dot{\theta}(0)=0$. It is convenient to parametrize the onset of oscillations in terms of a temperature scale $T_{\rm osc}$, defined via the relation
\begin{align}
    \label{eq:maTosc}
    3H(T_{\text{osc}})=m_a,
\end{align}
and to introduce a new temperature variable $z \equiv T_{\rm osc}/T$. Since $T\propto t^{-1/2}$ during radiation domination, one may write $m_a t\equiv c'z^2$ where $c'=\frac{3}{2}\sqrt{\frac{g_*(T_{\text{osc}})}{g_*(T)}}$. In terms of these variables, the temperature-normalized axion velocity becomes
\begin{align}
    \label{eta(T)}
    \eta(T)\equiv\frac{\dot{\theta}(T)}{T}= -\theta_i\,\left(\frac{m_a}{T_{\rm{osc}}}\right)\,\left(\frac{2z^2}{c'}\right)^{1/4}\Gamma\left(\frac{5}{4}\right)J_{5/4}\left(c'z^2\right).
\end{align}
We emphasize that, in this paper, we shall work with this expression for $\eta(T)$, which goes beyond the approximation employed by the authors of Ref.~\cite{Domcke:2020kcp}, $\eta(T) \propto z^{-1/2} \sin(z^2-1)$. 

\subsection{Coupling to the Standard Model}

We will focus on the following three representative axion--SM couplings in the Lagrangian,
\begin{align}
    \label{axionBL}
    &\delta_{B-L}\frac{\partial_\mu a}{f}J^\mu_{B-L}, \quad &&\text{Majoron-like coupling}, \\
    \label{axionGG}
    &-C_s\frac{a}{f}\frac{g_s^2}{32\pi^2}G^a_{\mu\nu}\widetilde{G}^{a\mu\nu}, \quad &&\text{axion coupled to strong sphalerons}, \\
    \label{axionWW}
    &-C_w\frac{a}{f}\frac{g_w^2}{32\pi^2}W^a_{\mu\nu}\widetilde{W}^{a\mu\nu}, \quad &&\text{axion coupled to weak sphalerons},
\end{align}
where $C_s$ and $C_w$ denote strong and weak anomaly coefficients, respectively, and $\delta_{B-L}$ is a coefficient that is e.\,g.\ $1/2$ for a standard Majoron.
Via the couplings in Eqs.~\eqref{axionBL}–\eqref{axionWW}, the axion velocity~\eqref{eta(T)} acts as an external bias in the Boltzmann equations for the SM chemical potentials, directly affecting $B\!-\!L$ violating, strong sphaleron, and weak sphaleron processes, respectively. More precisely, in the case of the Majoron-like coupling, nonzero $\dot{a}$ will induce effective chemical potentials for all species charged under $B\!-\!L$ following the logic in Eq.~\eqref{eq:muAeff}. Meanwhile, the coupling to either strong or weak sphalerons will introduce a bias in the chemical-equilibrium condition associated with the respective interaction. 

\subsection{Weinberg operator}
\label{subsec:WeinbergOperator}
A coherently evolving axion background with $\eta(T)\neq 0$ breaks $CPT$ spontaneously by selecting a preferred time direction, generating an effective chemical potential that biases particles against antiparticles. In the presence of baryon and/or lepton number violation, this bias drives the generation of a net baryon asymmetry. In the spontaneous LG setup considered here, the required $B+L$ violation is provided by weak sphalerons, while $B\!-\!L$ violation is induced by the dimension-five Weinberg operator~\cite{Weinberg:1979sa},
\begin{align}
    \label{Weinberg}
    \mathcal{L}_5
    &=\frac{c_{\alpha\beta}}{\Lambda}\left(\overline{L^C_\alpha}\Tilde{H}^*\right)\left(\Tilde{H}^\dagger L_\beta\right)+{\rm{h.c.}} \,.
\end{align}
Here, $L_\alpha$ and $H$ denote the SM lepton and Higgs doublets, $\tilde H \equiv i\sigma_2 H^\ast$ is the hypercharge-conjugate of $H$, and $C$ denotes charge conjugation. The symmetric matrix $c_{\alpha\beta}$ contains the Wilson coefficients in flavor space, while $\Lambda$ parametrizes the UV scale at which this operator is generated. For instance, assuming $\mathcal{O}(1)$ UV couplings, $\Lambda$ corresponds to the right-handed neutrino (RHN) mass scale in the type-I seesaw model~\cite{Minkowski:1977sc,Yanagida:1979as,Gell-Mann:1979vob,Mohapatra:1979ia,Yanagida:1980xy,Schechter:1980gr}. Throughout, we assume $T\ll \Lambda$, so that the effective description in Eq.~\eqref{Weinberg} remains valid. In a type-I seesaw interpretation, this corresponds to sufficiently heavy RHNs, e.\,g., near the energy scale of grand unification, and the assumption $T\ll \Lambda$ ensures that the contribution from standard thermal LG is negligible.\footnote{In other words, the thermal production of RHNs is negligible, such that a non-negligible lepton asymmetry from RHN decays would require a non-thermal source of RHN production (which we do not consider).}

After electroweak symmetry breaking, Eq.~\eqref{Weinberg} yields the Majorana neutrino mass matrix,
\begin{align}
    \label{cab1}
    \frac{c_{\alpha\beta}}{\Lambda}=\frac{m_{\alpha\beta}}{2v^2}, \qquad \expval{H}=v\approx 174\,\text{GeV},
\end{align}
with
\begin{align}
\label{eq:m_alphabeta}
m_{\alpha\beta}=\sum_{i=1}^3 U_{\alpha i}^\ast U_{\beta i}^\ast m_i,
\end{align}
and hence
\begin{align}
    \label{eq:m_alphabetasq}
    \abs{m_{\alpha\beta}}^2=\sum_{i,j}U_{\alpha i}^*U_{\alpha j}U^*_{\beta i}U_{\beta j}m_im_j.
\end{align}
We further distinguish between normal (NO) and inverted mass ordering (IO), 
\begin{align}
    \label{eq:m_NO}
    &\text{NO:}\quad m_1<m_2<m_3, \quad m_2=\sqrt{m_1^2+\Delta m_{21}^2}, \quad m_3=\sqrt{m_1^2+\Delta m_{31}^2},\\
    &\text{IO}:\quad m_3<m_1<m_2, \quad m_2=\sqrt{m_3^2+\abs{\Delta m_{32}}^2}, \quad m_1=\sqrt{m_3^2+\abs{\Delta m_{32}}^2-\Delta m_{21}^2}, \nonumber
\end{align}
and parameterize the Pontecorvo–Maki–Nakagawa–Sakata (PMNS)~\cite{Pontecorvo:1957qd,Maki:1962mu} matrix as
\begin{align}
    &U =
\begin{pmatrix}
1 & 0 & 0 \\
0 & c_{23} & s_{23} \\
0 & -s_{23} & c_{23}
\end{pmatrix}
\begin{pmatrix}
c_{13} & 0 & s_{13} e^{-i\delta_{\rm CP}} \\
0 & 1 & 0 \\
-s_{13} e^{i\delta_{\rm CP}} & 0 & c_{13}
\end{pmatrix}
\begin{pmatrix}
c_{12} & s_{12} & 0 \\
-s_{12} & c_{12} & 0 \\
0 & 0 & 1
\end{pmatrix} \, \mathrm{diag}\left(e^{i\alpha_1},\, e^{i\alpha_2},\, 1\right) \,.
\end{align}
Here $s_{ij}\equiv \sin\theta_{ij}$, $c_{ij}\equiv \cos\theta_{ij}$, $\delta_{\rm CP}$ denotes the Dirac phase, and $\alpha_i$ are the two Majorana phases. In our numerical analysis, we fix all oscillation parameters, except for the Majorana phases $\alpha_1$ and $\alpha_2$, to their best-fit values from Ref.~\cite{Esteban:2024eli}. Only a subset of our analysis has any dependence on the Majorana phases. For this part, we set $\alpha_1=\alpha_2=0$ in all relevant plots, namely Figs.~\ref{fig:BasisComparison} and~\ref{fig:ParameterSpace1_Tau_Basis}. We have checked that the resulting impact on our results from varying $\alpha_1, \alpha_2$ is at most an $\mathcal{O}(\lesssim 1\%)$ effect. Finally, we treat the lightest neutrino mass as a free parameter and scan over its allowed range.

Finally, the Weinberg operator induces $\Delta L=2$ LNV scatterings in the thermal plasma. Below, we present the corresponding equilibrium reaction densities for three physically relevant basis choices in the charged-lepton flavor space.
\begin{itemize}
    \item The charged-lepton flavor basis, hereafter denoted as the $(e, \mu, \tau)$ basis. This basis is suitable for $T\lesssim T_\mu$, where $T_\mu$ denotes the equilibration temperature of the $\mu$-Yukawa interaction.
    \item The hybrid $(1, 2, \tau)$ basis, which is a particularly suitable choice for $T_\tau\gtrsim T\gtrsim T_\mu$, , where $T_\tau$ denotes the equilibration temperature of the $\tau$-Yukawa interaction.
    \item The neutrino-mass basis, hereafter denoted as the $(1, 2, 3)$ basis. This basis is a particularly suitable choice for $T\gg T_\tau$.
\end{itemize}
We will outline the structural differences and implications of working in these respective bases at the beginning of Section~\ref{sec:3}.

In the $(e,\,\mu,\,\tau)$ basis, the equilibrium reaction densities can be written directly in terms of low-energy neutrino parameters as follows,\footnote{In the appropriate flavor-blind limit, one can compare our results with those obtained from the standard type-I seesaw calculation. Our result in Eq.~\eqref{rate1} reproduces the standard type-I seesaw expression given in Ref.~\cite{Pilaftsis:2003gt}. For Eq.~\eqref{rate2}, we find that our result is a factor of $3$ larger than that in Ref.~\cite{Pilaftsis:2003gt}, but in full agreement with an explicit first-principles calculation in Ref.~\cite{Frossard:2013swa}.  
}
\begin{align}
    \label{rate1}
    \gamma^{\text{eq}}_{\text{sub}}(H L_\alpha\rightarrow H^* L_\beta^\dagger)&=\frac{3T^6\,\abs{m_{\alpha\beta}}^2}{8\pi^5v^4}=\sum_{i,j}\left(U_{\alpha i}^*U_{\alpha j}U^*_{\beta i}U_{\beta j}m_im_j\right)\frac{3T^6}{8\pi^5v^4},  \\
    \label{rate2}
    \gamma^{\text{eq}}(L_\alpha L_\beta\rightarrow H^*H^*)&=\frac{3T^6\,\abs{m_{\alpha\beta}}^2}{8(1+\delta_{\alpha\beta})\pi^5v^4}=\sum_{i,j}\left(U_{\alpha i}^*U_{\alpha j}U^*_{\beta i}U_{\beta j}m_im_j\right)\frac{3T^6}{8(1+\delta_{\alpha\beta})\pi^5v^4}.
\end{align}
To the best of our knowledge, Eqs.~\eqref{rate1} and~\eqref{rate2} constitute the first explicit presentation of these fully flavor-resolved Weinberg-operator rates in the $(e,\,\mu,\,\tau)$ basis in the literature.  

Next, in the $(1,\,2,\,\tau)$-basis, where the $e-\mu$ sub-block of the symmetric Majorana mass matrix $m_{\alpha\beta}$ is diagonal,\footnote{Explicitly, this is achieved by rotating the mass matrix with a unitary matrix $V$, such that $V^T\begin{pmatrix}
    m_{ee} & m_{e\mu}\\
    m_{e\mu} & m_{\mu\mu}
\end{pmatrix}V=\begin{pmatrix}
    \tilde{m}_{11} & 0\\
    0 & \tilde{m}_{22}
\end{pmatrix}$.} the Weinberg rates can be written in a similar form as above, where the flavor indices now run over $\alpha, \beta \in \{1, 2, \tau\}$
\begin{align}
    \label{rate1-12tau}
    \gamma^{\text{eq}}_{\text{sub}}(H L_\alpha\rightarrow H^* L_\beta^\dagger)&=\frac{3T^6\,\abs{\tilde{m}_{\alpha\beta}}^2}{8\pi^5v^4},  \\
    \label{rate2-12tau}
    \gamma^{\text{eq}}(L_\alpha L_\beta\rightarrow H^*H^*)&=\frac{3T^6\,\abs{\tilde{m}_{\alpha\beta}}^2}{8(1+\delta_{\alpha\beta})\pi^5v^4},
\end{align}
where 
\begin{align}
     \label{eq:tildem_alphabetasq}
    \abs{\Tilde{m}_{11}}^2&=\frac{1}{2} \left( |m_{ee}|^2 + 2|m_{e\mu}|^2 + |m_{\mu\mu}|^2 - \Delta \right),\nonumber \\
    \abs{\Tilde{m}_{22}}^2&=\frac{1}{2} \left( |m_{ee}|^2 + 2|m_{e\mu}|^2 + |m_{\mu\mu}|^2 + \Delta \right), \nonumber \\
    \Tilde{m}_{12}&=\Tilde{m}_{21}=0, \nonumber \\
    \abs{\Tilde{m}_{\tau\tau}}^2&=\abs{m_{\tau\tau}}^2,\nonumber \\
    \abs{\tilde{m}_{1\tau}}^2 &= \abs{\tilde{m}_{\tau 1}}^2=\cos^2\theta |m_{e\tau}|^2 + \sin^2\theta |m_{\mu\tau}|^2 - \sin(2\theta) \text{Re}(e^{-i\delta} m_{e\tau}^* m_{\mu\tau}), \nonumber \\
    \abs{\tilde{m}_{2\tau}}^2 &=\abs{\tilde{m}_{\tau2}}^2= \sin^2\theta |m_{e\tau}|^2 + \cos^2\theta |m_{\mu\tau}|^2 + \sin(2\theta) \text{Re}(e^{-i\delta} m_{e\tau}^* m_{\mu\tau}).
\end{align}
Here the rotation angle $\theta$, the phase $\delta$, and the real numbers $A$ and $\Delta$ are given by
\begin{align}
    m_{ee}^*m_{e\mu}+m^*_{e\mu}m_{\mu\mu}&=Ae^{i\delta}, \quad \tan(2\theta)=\frac{2A}{\abs{m_{\mu\mu}}^2-\abs{m_{ee}}^2}, \nonumber \\
    \Delta &= \sqrt{(|m_{\mu\mu}|^2 - |m_{ee}|^2)^2 + 4A^2}.
\end{align}
Note that by using Eq.~\eqref{eq:m_alphabeta}, all interaction rates can be expressed purely in terms of the PMNS matrix and active neutrino masses. The fact that we analyze solutions to Boltzmann equations employing these flavor-resolved Weinberg-operator rates as part of our analysis represents another important step beyond the current state of the art in the literature.

Finally, the Weinberg rate in the $(1,\,2,\,3)$-basis is by construction diagonal, and takes the following form
\begin{align}
    \label{rate1-123}
    \gamma^{\text{eq}}_{\text{sub}}(H L_i\rightarrow H^* L_j^\dagger)&=\frac{3T^6m_im_j}{8\pi^5v^4}\delta_{ij},  \\
    \label{rate2-123}
    \gamma^{\text{eq}}(L_i L_j\rightarrow H^*H^*)&=\frac{3T^6m_im_j}{16\pi^5v^4}\delta_{ij}.
\end{align}
The standard flavor-independent result, which can be identified with the total LNV rate in the $(1,\,2,\,3)$-basis, can then be written as, 
\begin{align}
    \label{eq:gammaWB}
    \gamma_{\text{WB}}^{\Delta L=2}=\gamma^{\text{eq}}_{\text{sub}}(H L\rightarrow H^* L)+2\gamma^{\text{eq}}(L L\rightarrow H^*H^*)=\frac{3\overline{m}^2T^6}{4\pi^5v^4}\equiv \kappa\frac{\overline{m}^2T^6}{v^4},
\end{align}
where
\begin{align}
    \overline{m}^2=m_1^2+m_2^2+m_3^2, \qquad \kappa = \frac{3}{4\pi^5} \simeq 2.45\cdot 10^{-3}.
\end{align}
Since this paper focuses on spontaneous LG at $T \gtrsim 10^{10}$ GeV, our analysis relies on the interaction rates given in Eqs.~\eqref{rate1-12tau}–\eqref{rate2-123}.

\section{Generation of the baryon asymmetry}
\label{sec:3}
In this section, we compute the asymmetries generated through spontaneous LG. We first present sets of Boltzmann equations governing the temperature evolution of SM chemical potentials.  
Next, we derive a simple algebraic estimate for the resulting $B\!-\!L$ asymmetry, based on the limiting assumption that each SM interaction is either fully out of equilibrium or fully equilibrated. We then compare fully numerical solutions of the full set of Boltzmann equations with the simplified algebraic approach and extract the $B\!-\!L$ asymmetry produced by standard spontaneous LG across a wide range of axion-oscillation temperatures for the three axion couplings in Eqs.~\eqref{axionBL}–\eqref{axionWW}. As detailed below, no single set of semi-classical kinetic equations can capture the entire physical dynamics across all regimes. Consequently, we solve the Boltzmann equations for two physically motivated choices of lepton-flavor bases and compare their numerical results in Section~\ref{sec:baryonasymmetry}. This allows us to quantify the intrinsic uncertainty associated with performing calculations in a fixed flavor basis. Following these comparisons, all remaining calculations are performed in a single basis motivated below. For completeness, we also compare with results obtained using a simplified flavor-independent approximation for the LNV rates, and comment on existing results in the literature. Additional numerical results are provided in Appendix~\ref{appendix:asymprod}.\\

\noindent\textbf{Lepton-flavor basis:} The classical Boltzmann framework adopted in this work involves two distinct interactions that favor different lepton-flavor bases: the Weinberg operator and the $\tau$-Yukawa coupling. Specifically, the Weinberg operator selects the $(1,\,2,\,3)$ basis, whereas spectator processes sourced by charged-lepton Yukawas favor the $(e,\,\mu,\,\tau)$ basis. Within the temperature regimes of interest, only the $\tau$-Yukawa coupling is dynamically relevant, as the interaction rates mediated by $Y_{\mu,e}$ are highly suppressed relative to the Hubble rate. The system thus reduces to two competing bases: $(1,\,2,\,3)$ and $(1,\,2,\,\tau)$. Because the projection operators for these two bases do not commute, nontrivial off-diagonal correlations (quantum coherence effects) and their subsequent damping are physically expected to occur, particularly in the regime where the $\tau$-Yukawa interaction starts to equilibrate. Fully capturing these dynamics requires a quantum kinetic treatment based on density-matrix equations~\cite{Sigl:1993ctk}, which is beyond the scope of this paper and left for future work. 

Restricting our present description to classical Boltzmann equations, no single basis choice remains globally valid across the entire temperature range~\cite{Abada:2006fw,Nardi:2006fx}. To quantify the resulting systematic uncertainty, we first compare the numerical results obtained using both bases. In light of the numerical comparisons, we argue that the Weinberg-favored basis represents the best choice, and for the remainder of our analysis we perform all calculations in this basis.

\subsection{Boltzmann equations}
\label{sec:BEqs}
Generating a viable $B\!-\!L$ asymmetry requires the axion to develop a non-vanishing velocity before $\Delta L=2$ Weinberg interactions become inefficient. The axion velocity biases the thermodynamic equilibrium state of the SM plasma, enabling the Weinberg operator to generate the asymmetry. Parametrically, these LNV interactions start to freeze out below $T \sim 10^{13}$ GeV. Barring additional $B\!-\!L$-violating processes, this asymmetry remains frozen at all lower temperatures, where it provides the basis for a nonvanishing baryon asymmetry. 

Accordingly, we restrict our attention to temperatures well above $10^{10}\,\text{GeV}$. In this regime, SM interactions do not efficiently distinguish the first two generations of right-handed charged leptons, nor can they resolve the three lightest quark flavors. It is therefore sufficient to work with a reduced set of $12$ (rather than $16$) chemical potentials, $\{\mu_i\}$, where $i$ denotes
\begin{align}
    i=\tau,\,L_1,\,L_2,\,L_{3,\tau},\,t,\,c,\,b,\,q_{uds},\,Q_1,\,Q_2,\,Q_3,\,H.
\end{align}
Following the formalism laid out in Ref.~\cite{Domcke:2020kcp}, we have derived Boltzmann equations that describe how each $\mu_i$ changes with temperature. The results are listed below, first for the $(1,\,2,\,\tau)$ basis and then for the $(1,\,2,\,3)$ basis. \\

\noindent$\mathbf{(1,\,2,\,\tau)}$\textbf{-basis:} Let us first consider the basis expected to provide the best description of the system for $T_\tau\gtrsim T\gg T_\mu$. The chemical potential associated with right-handed quarks are only directly affected by strong sphalerons and Yukawa interactions,
\begin{small}
\begin{align}
\frac{d}{d\ln T}\left(\frac{\mu_t}{T}\right) &=-\frac{\gamma_{\text{SS}}(T)}{3H(T)}\left(\frac{-3\mu_{q_{uds}} - \mu_t-\mu_c - \mu_b+ 2\mu_{Q_1}+2\mu_{Q_2} + 2\mu_{Q_3}}{T}
- C_s\,\eta(T)\right)\nonumber\\
&
-\frac{\gamma_t(T)}{3H(T)}\left(\frac{-\mu_t + \mu_{Q_3}+ \mu_H}{T}\right),
\\[6pt]
\frac{d}{d\ln T}\left(\frac{\mu_c}{T}\right) &=-\frac{\gamma_{\text{SS}}(T)}{3H(T)}\left(\frac{-3\mu_{q_{uds}} - \mu_t - \mu_b - \mu_c+ 2\mu_{Q_{1}} +2\mu_{Q_{2}}+ 2\mu_{Q_3}}{T}
- C_s\,\eta(T)\right)\nonumber\\
&
-\frac{\gamma_c(T)}{3H(T)}\left(\frac{-\mu_c + \mu_{Q_2} + \mu_H}{T} \right),
\\[6pt]
\frac{d}{d\ln T}\left(\frac{\mu_b}{T}\right) &=-\frac{\gamma_{\text{SS}}(T)}{3H(T)}\left(\frac{-3\mu_{q_{uds}} - \mu_t-\mu_c - \mu_b+ 2\mu_{Q_1}+2\mu_{Q_2} + 2\mu_{Q_3}}{T}
- C_s\,\eta(T)\right)\nonumber\\
&-\frac{\gamma_b(T)}{3H(T)}\left(\frac{-\mu_b + \mu_{Q_3} - \mu_H}{T} \right),\\[6pt]
\frac{d}{d\ln T}\left(\frac{\mu_{q_{uds}}}{T}\right) &=-\frac{\gamma_{\text{SS}}(T)}{3H(T)}
\left(\frac{-3\mu_{q_{uds}} - \mu_t-\mu_c - \mu_b+ 2\mu_{Q_1}+2\mu_{Q_2} + 2\mu_{Q_3}}{T}- C_s\,\eta(T)\right).
\end{align}
\end{small}

The chemical potentials of quark doublets are affected by strong sphalerons, weak sphalerons, and Yukawa interactions,
\begin{small}
\begin{align}
    \frac{d}{d\ln T}\left(\frac{\mu_{Q_i}}{T}\right) &=\frac{\gamma_{\text{WS}}(T)}{2H(T)}\left(\frac{\mu_{L_1} +\mu_{L_2}+ \mu_{L_\tau} + 3\mu_{Q_{1}} +3\mu_{Q_{2}} + 3\mu_{Q_3}}{T}- C_w\,\eta(T)\right)\nonumber\\&\quad+ \frac{\gamma_{\text{SS}}(T)}{3H(T)}
\left(\frac{-3\mu_{q_{uds}} - \mu_t-\mu_c - \mu_b+ 2\mu_{Q_1}+2\mu_{Q_2} + 2\mu_{Q_3}}{T}- C_s\,\eta(T)\right)\nonumber \\
&\quad + \delta_{i2}\frac{\gamma_c(T)}{6H(T)}
\left(\frac{-\mu_c + \mu_{Q_2} + \mu_H}{T} \right) +\delta_{i3} \frac{\gamma_t(T)}{6H(T)}
\left(\frac{-\mu_t + \mu_{Q_3} + \mu_H}{T} \right)\nonumber \\
&\quad+\delta_{i3} \frac{\gamma_b(T)}{6H(T)}\left(\frac{-\mu_b + \mu_{Q_3} - \mu_H }{T}\right).
\end{align}
\end{small}

At the temperature scales under consideration, $\tau$ is the only right-handed lepton that may be (approximately) in equilibrium and thus potentially resolved from the remaining two lepton flavors. We thus only keep $\mu_\tau$ and set $\mu_e = \mu_\mu = 0$. Meanwhile, lepton doublets are affected by weak sphalerons and interactions induced by the Weinberg operator,  
\begin{small}
\begin{align}
\frac{d}{d\ln T}\left(\frac{\mu_\tau}{T}\right) &=-\frac{\gamma_\tau(T)}{H(T)}\left(\frac{-\mu_\tau + \mu_{L_\tau} - \mu_H}{T} \right),\\
\label{eq:muL_full}
\frac{d}{d\ln T}\left(\frac{\mu_{L_i}}{T}\right)&=\frac{\gamma_{\rm WS}(T)}{2H(T)}
\left(\frac{\mu_{L_{1}}+\mu_{L_{2}} + \mu_{L_{\tau}}
+3\mu_{Q_{1}} + 3\mu_{Q_{2}} + 3\mu_{Q_{3}}}{T}- C_w\,\eta(T)\right)
\nonumber\\
&+\sum_{\beta=i,\tau}\left(\frac{\gamma^{\text{eq}}_{\text{sub}}(HL_i\rightarrow H^*L_\beta^*)+(1+\delta_{i\beta})\gamma^{\text{eq}}(L_i L_\beta\rightarrow H^*H^*)}{{(T^3/6)H(T)}}\right)\nonumber \\
    &\times \left(\frac{2\mu_H+\mu_{L_i}+\mu_{L_\beta}}{T}+2\delta_{B-L}\eta(T)\right), \\
\frac{d}{d\ln T}\left(\frac{\mu_{L_\tau}}{T}\right)&=\frac{\gamma_{\rm WS}(T)}{2H(T)}
\left(\frac{\mu_{L_{1}}+\mu_{L_{2}} + \mu_{L_{\tau}}
+3\mu_{Q_{1}} + 3\mu_{Q_{2}} + 3\mu_{Q_{3}}}{T}- C_w\,\eta(T)\right)
\nonumber\\
&+ \frac{\gamma_{\tau}(T)}{2H(T)}
\left(\frac{-\mu_{\tau} + \mu_{L_{\tau}} - \mu_H}{T}\right)\nonumber \\
&+\sum_{\beta=1,2,\tau}\left(\frac{\gamma^{\text{eq}}_{\text{sub}}(HL_\tau\rightarrow H^*L_\beta^*)+(1+\delta_{\tau\beta})\gamma^{\text{eq}}(L_\tau L_\beta\rightarrow H^*H^*)}{{(T^3/6)H(T)}}\right)\nonumber \\
    &\times \left(\frac{2\mu_H+\mu_{L_\tau}+\mu_{L_\beta}}{T}+2\delta_{B-L}\eta(T)\right).
\end{align}
\end{small}

Finally, the SM Higgs doublet is affected by all Yukawa interactions and interactions induced by the Weinberg operator,
\begin{small}
\begin{align}
\label{eq:muH_full}
\frac{d}{d\ln T}\left(\frac{\mu_H}{T}\right) &=-\frac{\gamma_\tau(T)}{4H(T)}\left(\frac{-\mu_\tau + \mu_{L_\tau} - \mu_H}{T} \right)+ \frac{\gamma_t(T)}{4H(T)}
\left(\frac{-\mu_t + \mu_{Q_3} + \mu_H}{T} \right)\nonumber\\&\quad- \frac{\gamma_b(T)}{4H(T)}\left(\frac{-\mu_b + \mu_{Q_3} - \mu_H}{T} \right) + \frac{\gamma_c(T)}{4H(T)}
\left(\frac{-\mu_c + \mu_{Q_2} + \mu_H}{T} \right)
\nonumber \\
&+\frac{1}{2}\sum_{i=1,2}\sum_{\beta=i,\tau}\left(\frac{\gamma^{\text{eq}}_{\text{sub}}(HL_i\rightarrow H^*L_\beta^*)+(1+\delta_{i\beta})\gamma^{\text{eq}}(L_i L_\beta\rightarrow H^*H^*)}{{(T^3/6)H(T)}}\right)\nonumber \\
    &\times \left(\frac{2\mu_H+\mu_{L_i}+\mu_{L_\beta}}{T}+2\delta_{B-L}\eta(T)\right)\nonumber \\
&+\frac{1}{2}\sum_{\beta=1,2,\tau}\left(\frac{\gamma^{\text{eq}}_{\text{sub}}(HL_\tau\rightarrow H^*L_\beta^*)+(1+\delta_{\tau\beta})\gamma^{\text{eq}}(L_\tau L_\beta\rightarrow H^*H^*)}{{(T^3/6)H(T)}}\right)\nonumber \\
    &\times \left(\frac{2\mu_H+\mu_{L_\tau}+\mu_{L_\beta}}{T}+2\delta_{B-L}\eta(T)\right).
\end{align}
\end{small}
Details about our numerical implementation of the various interaction rates can be found in App.~\ref{appendix:BEq}.\\

\noindent$\mathbf{(1,\,2,\,3)}$\textbf{-basis:} Next, we consider the basis expected to provide the best description of the system for $T\gg T_\tau$. Since its regime of validity is determined by the $\tau$-Yukawa interaction rates being far from equilibrium, we will ignore the impact of the $\tau$-Yukawa interaction in the following. In this basis, the Boltzmann equations for the quarks are the same as those in the $(1,\,2,\,\tau)$ basis, modulo a replacement $L_{1,\,2,\,\tau}\rightarrow L_{1,2,3}$. The evolution equations for the lepton and Higgs doublets read, 
\begin{small}
\begin{align}
\frac{d}{d\ln T}\left(\frac{\mu_{L_i}}{T}\right)&=\frac{\gamma_{\rm WS}(T)}{2H(T)}
\left(\frac{\mu_{L_{1}}+\mu_{L_{2}} + \mu_{L_{3}}
+3\mu_{Q_{1}} + 3\mu_{Q_{2}} + 3\mu_{Q_{3}}}{T}- C_w\,\eta(T)\right)
\nonumber\\
&+\left(\frac{\gamma^{\text{eq}}_{\text{sub}}(HL_i\rightarrow H^*L_i^*)+2\gamma^{\text{eq}}(L_i L_i\rightarrow H^*H^*)}{{(T^3/6)H(T)}}\right)\left(\frac{2\mu_H+2\mu_{L_i}}{T}+2\delta_{B-L}\eta(T)\right), \\
\frac{d}{d\ln T}\left(\frac{\mu_H}{T}\right) &=\frac{\gamma_t(T)}{4H(T)}
\left(\frac{-\mu_t + \mu_{Q_3} + \mu_H}{T} \right)- \frac{\gamma_b(T)}{4H(T)}\left(\frac{-\mu_b + \mu_{Q_3} - \mu_H}{T} \right)\nonumber \\
&+ \frac{\gamma_c(T)}{4H(T)}
\left(\frac{-\mu_c + \mu_{Q_2} + \mu_H}{T} \right)
\nonumber \\
&+\frac{1}{2}\sum_{i}\left(\frac{\gamma^{\text{eq}}_{\text{sub}}(HL_i\rightarrow H^*L_i^*)+2\gamma^{\text{eq}}(L_i L_i\rightarrow H^*H^*)}{{(T^3/6)H(T)}}\right)\nonumber \\
&\times\left(\frac{2\mu_H+2\mu_{L_i}}{T}+2\delta_{B-L}\eta(T)\right).
\end{align}
\end{small}
\noindent Meanwhile, we set $\frac{d}{d\ln T}\left(\frac{\mu_\tau}{T}\right)=0$, i.e. $\mu_\tau$ is treated as a conserved quantity on the same level as $\mu_{e}$ and $\mu_{\mu}$.\\

\noindent\textbf{Flavor-blind approximation:} Finally, let us consider a flavor-blind limit—as adopted, for instance, in Ref.~\cite{Domcke:2020kcp}—where the Weinberg operator acts equally on all lepton flavors. Unlike our setup, it assumes a Weinberg rate proportional to the identity matrix in flavor space. In the flavor-blind limit of the $(1,\,2,\,3)$ or $(1,\,2,\,\tau)$-bases, the Boltzmann equations for quarks and right-handed charged leptons are unaffected by this limit. Moreover, in the temperature regimes of interest, one can replace $L_1,\,L_2\rightarrow L_{12}$, and the corresponding evolution equations for lepton and Higgs doublets read (where the flavor index now run over $\alpha \in \{12, 3(\tau)\}$)
\begin{small}
\begin{align}
\label{eq:muL_notfull}
    \frac{d}{d\ln T}\left(\frac{\mu_{L_\alpha}}{T}\right)&=
\frac{(1/3)\gamma_{\text{WB}}^{\Delta L=2}(T)}{H(T)\,(T^3/6)}\left(\frac{2\mu_H+2\mu_{L_{3(\tau)}}\delta_{\alpha3(\tau)}+2\mu_{L_{12}}\delta_{\alpha12}}{T}+2\delta_{B-L}\eta(T)\right)+..., \\
\label{eq:muH_notfull}
\frac{d}{d\ln T}\left(\frac{\mu_H}{T}\right) &=
\frac{{(1/3)}\gamma_{\text{WB}}^{\Delta L=2}(T)}{H(T)\,(T^3/6)}\left(\frac{2\mu_H+2\mu_{L_{12}}}{T}+2\delta_{B-L}\eta(T)\right)\nonumber \\
&+\frac{(1/6)\gamma_{\text{WB}}^{\Delta L=2}(T)}{H(T)\,(T^3/6)}\left(\frac{2\mu_H+2\mu_{L_{3(\tau)}}}{T}+2\delta_{B-L}\eta(T)\right)+...,
\end{align}
\end{small}\par\noindent
where the ellipses encode sphaleron and Yukawa interactions, which are unmodified compared to the equations above apart from the replacement $\gamma_{\text{WS}}(\mu_{L_1}+\mu_{L_2})\rightarrow 2\gamma_{\text{WS}}\mu_{L_{12}}$.
\\

\noindent\textbf{Solution scalings:} By inspecting Eq.~\eqref{eta(T)} and the Boltzmann equations listed above, it can be seen that $\theta_i$ acts purely as an overall multiplicative factor in the source terms. Consequently, by defining rescaled chemical potentials $\tilde{\mu}_i \equiv \mu_i / (T\theta_i)$, the equations governing the evolution of $\tilde{\mu}_i$ become independent of $\theta_i$. This observation offers a major practical advantage: we only need to solve the system of evolution equations once to find $\mu_{B-L} / (T\theta_i)$, from which the baryon asymmetry for any value of $\theta_i$ can be immediately obtained by simple rescaling. 

Furthermore, if the axion possesses only one coupling from Eqs.~\eqref{axionBL}–\eqref{axionWW}, say $C_s$, the definition $\tilde{\mu}_i \equiv \mu_i / (T\theta_iC_s)$ removes all $\theta_i$ and $C_s$ dependence from the evolution equations. Consequently, we only need to solve the system once for each individual coupling, after which we can scale the results to find the $B\!-\!L$ asymmetry for any choice of $\theta_i$ and $C_s$. This approach remains valid in realistic multi-coupling scenarios, provided one coupling provides the dominant contribution to the asymmetry.

\subsection{Algebraic approximation}
\label{subsec:algebraic}

In regimes where an interaction in the Boltzmann equations is highly efficient, the process reaches chemical equilibrium and imposes a constraint on the chemical potentials of the participating species. Conversely, when an interaction is highly inefficient, it effectively decouples, giving rise to a conservation law among a subset of the chemical potentials. Consequently, within a given temperature range, one can adopt a simplified treatment where each SM interaction is approximated as either fully efficient or completely decoupled, thereby yielding a system of linear constraints or conservation laws for the chemical potentials.

This algebraic approach was systematized in Ref.~\cite{Domcke:2020quw} within the context of wash-in LG (see also Refs.~\cite{Antaramian:1993nt,Fong:2015vna,Domcke:2022kfs,Schmitz:2023pfy,Mojahed:2024mvb,Mojahed:2025vgf,Mojahed:2025exj,Mojahed:2025odq} for related work), enabling the full set of Boltzmann equations for all SM chemical potentials to be reduced to a smaller system governing $\mu_{\Delta_\alpha}$, where $\Delta_\alpha\equiv B/3-L_\alpha$. The purpose of the following section is to generalize this algebraic framework to accommodate a generic realization of spontaneous LG in an arbitrary chemical background, which may or may not feature an additional contribution from wash-in LG.

To make contact with previous literature~\cite{Domcke:2020quw,Domcke:2022kfs,Mojahed:2025vgf}, let us first note that the Boltzmann equation for $\mu_{\Delta_\alpha}$ takes the following form, 
\begin{align}
    \label{wilg}
    (\partial_t+3H)q_{\Delta_\alpha}=\sum_\beta \gamma_{\alpha\beta}^w\frac{\mu_{L_\beta}+\mu_H}{T},
\end{align}
where $q_C=\bar{\mu}_CT^2/6$ for a conserved charge $C$, with 
\begin{align}
    \bar{\mu}_C=\sum_in_i^Cg_i\mu_i,
\end{align}
where $n_i^C$ are charge vectors and $g_i$ multiplicities, see Refs.~\cite{Domcke:2020kcp,Domcke:2020quw} for detailed definitions. At any given temperature, the $n$ chemical potentials of the SM species are subject to $m < n$ constraints from linearly independent equilibrium processes and $n - m$ conservation laws, yielding a total of $n$ linear equations. Solving this linear system for the individual chemical potentials, the combination $\mu_{L_\alpha}+\mu_H$ can be expressed as
\begin{align}
    \label{eq:L+H}
    \mu_{L_\alpha}+\mu_H=\bar{\mu}^0_\alpha-\sum_\beta C_{\alpha\beta}\bar{\mu}_{\Delta_\beta},
\end{align}
where $C_{\alpha\beta}$ is the standard flavor coupling matrix~\cite{Barbieri:1999ma,Abada:2006fw,Nardi:2006fx,Abada:2006ea,Blanchet:2006be,Antusch:2006cw} induced by the
spectator processes in the thermal plasma~\cite{Buchmuller:2001sr,Garbrecht:2014kda,Garbrecht:2019zaa}. The first term on the RHS of Eq.~\eqref{eq:L+H} can be written in the following form, 
\begin{align}
    \label{eq:mu0}
    \bar{\mu}^0_\alpha=\sum_{C\neq\Delta_\alpha}S_{\alpha C}\bar{\mu}_C+\left(\dot{\theta}\delta_{B-L}+\sum_{I\neq \text{LNV}}S'_{\alpha I}n_S^I\dot{\theta}\right),
\end{align}
where the first term on the RHS is a source term relevant for wash-in LG, while the terms in the parenthesis are relevant for spontaneous LG. Here, $n_S^I$ is a source vector that specifies the charge of the axion involved in the process $I$~\cite{Domcke:2020kcp}. Specifically, for an axion coupling to sphalerons of type $J$, the charge vector is $n_S^I=C_J\delta_{IJ}$, where $C_J$ here denotes the relevant anomaly coefficient. For an axion coupled to a current $J_Q=\sum n_i^QJ_i$ we have $n_S^I=\sum_i n_i^Qn_i^I$ where $n_i^I$ is the component of the vector $n^I$ that specifies how the SM species $i$ participates in the relevant equilibrated interaction $I$~\cite{Domcke:2020quw}. Upon substituting Eq.~\eqref{eq:L+H} into Eq.~\eqref{wilg}, we obtain
\begin{align}
    -\left(\partial_t + 3 H\right) q_{\Delta_\alpha} & =  - \sum_\beta \Gamma_{\alpha\beta}^{\rm w} \left(q_\beta^0 - \sum_\sigma C_{\beta\sigma}\, q_{\Delta_\sigma}\right) \,,
\label{eq:wilg2}
\end{align}
where $q_\alpha^0 \equiv \sfrac{1}{6}\:\bar{\mu}_\alpha^0 T^2$, $q_{\Delta_\alpha} \equiv \sfrac{1}{6}\:\bar{\mu}_{\Delta_\alpha}T^2$, and $\Gamma_{\alpha\beta} = 6/T^3\,\gamma_{\alpha\beta}$~\cite{Mojahed:2025vgf}.
Equation~\eqref{eq:wilg2} takes the exact same form as the wash-in contribution in standard LG, with the sole difference that $q_\beta^0$ now encodes axion-induced contributions in addition to any non-vanishing (approximately) conserved charges.

The total washout rate is often dominated by a single interaction channel, at least in a limited temperature regime. When that is the case, it can be expressed in factorized form as $\Gamma_{\alpha\beta}^w=P_{\alpha\beta}\Gamma^w$.
The overall temperature dependence is captured by the flavor-blind rate $\Gamma^w$, whereas the matrix $\bm{P}$ specifies how the washout is distributed in flavor space. When this factorization applies, Eq.~\eqref{eq:wilg2} can be solved exactly~\cite{Domcke:2020quw}. In particular, the washout rate induced by the Weinberg operator reads, 
\begin{align}
    \gamma_{\alpha\beta}^w&=\sum_\sigma 2(\delta_{\alpha\beta}+\delta_{\sigma\beta})\left[\gamma^{\text{eq}}_{\text{sub}}(H L_\alpha\rightarrow H^* L_\sigma^\dagger)+(1+\delta_{\alpha\sigma})\gamma^{\text{eq}}(L_\alpha L_\sigma\rightarrow H^*H^*)\right].
\end{align}
For the three flavor-bases introduced in Sec.~\ref{subsec:WeinbergOperator}, we have 
\begin{align}
    \Gamma^w(T)&=\frac{9T^3}{\pi^5v^2},\\
    P_{\alpha\beta}^{(e,\,\mu,\,\tau)}&=\frac{1}{v^2}\left[\abs{m_{\alpha\beta}}^2+\sum_\sigma \delta_{\alpha\beta}\abs{m_{\alpha\sigma}}^2\right],\\
    P_{\alpha\beta}^{(1,\,2,\,\tau)}&=\frac{1}{v^2}\left[\abs{\tilde{m}_{\alpha\beta}}^2+\sum_\sigma \delta_{\alpha\beta}\abs{\tilde{m}_{\alpha\sigma}}^2\right],\\
    P_{\alpha\beta}^{(1,\,2,\,3)}&=2\delta_{\alpha\beta}\frac{m_\alpha^2}{v^2},
\end{align}
where $\abs{m_{\alpha\beta}}^2$, $\abs{\tilde{m}_{\alpha\beta}}^2$, and $m_\alpha$ are given in Eqs.~\eqref{eq:m_alphabetasq},~\eqref{eq:tildem_alphabetasq}, and \eqref{eq:m_NO}, respectively, and the flavor-blind rate is the same independent of choice of basis.

Since the Boltzmann equations in Sec.~\ref{sec:BEqs} are expressed via chemical potentials, we retain this framework rather than adopting charge densities as in Ref.~\cite{Domcke:2020quw}, noting that switching between the two is trivial.\footnote{Indeed, in our notation, quantities like $\bar{\mu}_{\Delta_\alpha}$ (or more generally $\bar{\mu} _C$) are nothing but charge asymmetries expressed in units of a chemical potential.} For arbitrary initial conditions at reheating after inflation, $\mu^{0}_{\Delta\beta}(T_0)$, we obtain
\begin{align}
    \label{eq:Master}
    \frac{\mu_{\Delta_\alpha}(T)}{T}&=E_{\alpha\beta}(T_0,T)\,\frac{\mu_{\Delta_\beta}^0(T_0)}{T_0}+\int_{T_0}^TdT'\left[\frac{\partial}{\partial T'}\,E_{\alpha\beta}(T',T)\right]\frac{\mu^{\rm{eq}}_{\Delta_\beta}(T')}{T'} \,,
\end{align}
where $\mu_{\Delta_\alpha}^{\rm{eq}}$ is the equilibrium attractor in the presence of LNV interactions. This attractor can be split into two contributions, 
\begin{align}
    \label{qdeltaalphaeq}
    \mu_{\Delta_\alpha}^{\rm{eq}}&=\mu_{\Delta_\alpha}^{{\rm{eq}},C}+\mu_{\Delta_\alpha}^{{\rm{eq}},\theta}, \vphantom{\bigg]}\\
    \label{qWILG}
    \mu_{\Delta_\alpha}^{{\rm{eq}},C}(T)&=\sum_\beta\sum_{C\neq \Delta\alpha}C_{\alpha\beta}^{-1}(T)\,S_{\beta C}\,\bar{\mu}_C, \vphantom{\bigg]}\\
    \label{qTheta}\mu_{\Delta_\alpha}^{{\rm{eq}},\theta}(T)&=\dot{\theta}(T)\sum_\beta C_{\alpha\beta}^{-1}(T)\bigg[\delta_{B-L}+\sum_{I\neq \text{LNV}} S'_{\beta I}(T)\,n_S^I\bigg] \,.
\end{align}
where $\mu_{\Delta_\alpha}^{{\rm{eq}},C}$ is nonzero in the presence of nonzero conserved charge densities in the thermal plasma, relevant for wash-in LG, and $\mu_{\Delta_\alpha}^{{\rm{eq}},\theta}$ is nonzero in the presence of a moving axion with non-vanishing couplings to the SM. The exponential matrix $E_{\alpha\beta}$ encodes how LNV reactions drive the thermal plasma towards the equilibrium attractor $\mu_{\Delta_\alpha}^{\rm{eq}}$, and is given by
\begin{align}   
    \label{eq:Edef}
    \bm{E}(T_0,T)&=\exp{\int_{T_0}^TdT'\:\frac{\Gamma_W(T')}{H(T')T'}\,\bm{P}\bm{C}(T')} \,.
\end{align}
Eq.~\eqref{eq:Master}, together with the definitions in Eqs.~\eqref{qdeltaalphaeq}--\eqref{eq:Edef}, constitutes our algebraic master formula for the $B\!-\!L$ asymmetry from spontaneous and wash-in LG. 

A few comments on our result are in order. First, in the limit $\dot{\theta}=0$, one recovers the standard wash-in LG result, cf.\ Eq.~(7) in Ref.~\cite{Domcke:2020quw}.\footnote{To see this, integrate by parts in Eq.~\eqref{eq:Master} and impose $\partial_{T'}\,(\mu^{\rm{eq}}_{\Delta_\beta}(T')/T')=0$, which is a direct consequence of charge conservation, $\partial_{T'}\,(q^{\rm{eq}}_{\Delta_\beta}(T')/s(T'))=0$, where $s$ denotes entropy density.} Second, in the absence of non-vanishing conserved charges, one can define rescaled chemical potentials $\tilde{\mu}_{\Delta_\alpha} \equiv \mu_{\Delta_\alpha} / (T\theta_i)$, such that Eq.~\eqref{qdeltaalphaeq} becomes independent of $\theta_i$. Third, the matrices $\bm{C}$, $\bm{S}$, and $\bm{S}'$ can be computed in any given temperature regime, as shown in the Appendix of Ref.~\cite{Domcke:2020quw}, where explicit computations of $\bm{C}$ and $\bm{S}$ were carried out. In App.~\ref{appendix:algebraic}, we present fully general expressions for all three matrices, which are readily applicable to any scenario of spontaneous LG with arbitrary axion couplings in a thermal plasma of any chemical composition, spanning temperature regimes from $T=10^{15}$ GeV down to the electroweak phase transition (EWPT). For convenience, we reproduce below the specific components of these general results required for our current analysis, where we narrow the discussion to the axion couplings in Eqs.~\eqref{axionBL}--\eqref{axionWW}, and a thermal plasma where all (quasi) conserved charges $\mu_C$, $C\neq \Delta_\alpha$, are assumed to be negligible.\footnote{In our numerical analysis, we employ the following numerical values for the various equilibration temperatures under consideration~\cite{Domcke:2020quw}: $T_{\rm SS}=2.8\cdot 10^{13}$ GeV, $T_{\rm WS}=2.5\cdot 10^{12}$ GeV, $T_{b\tau}=1.4\cdot 10^{12}$ GeV, $T_c=1.2\cdot 10^{11}$ GeV, $T_\mu=4.7\cdot 10^9$ GeV.} The expressions below are sufficient to reconstruct the equilibrium attractor, $\mu_{\Delta_\alpha}^{{\rm{eq}},\theta}(T)$, for temperatures $T\gtrsim 10^{10}$ GeV. The algebraic results coincide for the $(1,\,2,\,3)$ and $(1,\,2,\,\tau)$ bases for $T>T_\tau$, while for $T_\tau>T>T_\mu$ we present separate formulae for the two bases. 

\bigskip
\noindent\textbf{\boldmath{$T\gtrsim T_{\rm SS}:$}}
\begin{footnotesize}
\begin{align}
    \bm{C}=\begin{pmatrix}
\frac{2}{3} & \frac{1}{6} & \frac{1}{6} \\
\frac{1}{6} & \frac{2}{3} & \frac{1}{6} \\
\frac{1}{6} & \frac{1}{6} & \frac{2}{3}
\end{pmatrix} \,, \quad \bm{S}' = \bordermatrix{ 
    & n_S^{\rm SS} & n_S^{\rm WS} \cr
     & 0&0 \cr
     & 0&0 \cr
   & 0&0 \cr 
} \,, \quad \Tr{\mu_{\Delta_\alpha}^{{\rm{eq}},\theta}}=3\,\delta_{B-L}\,\dot{\theta} \,.
\end{align}
\end{footnotesize}

\noindent\textbf{\boldmath{$T_{\rm SS}\gtrsim T\gtrsim T_{\rm WS}:$}}
\begin{footnotesize}
\begin{align}
    \bm{C}&=\begin{pmatrix}
\frac{15}{23} & \frac{7}{46} & \frac{7}{46} \\
\frac{7}{46} & \frac{15}{23} & \frac{7}{46} \\
\frac{7}{46} & \frac{7}{46} & \frac{15}{23}
\end{pmatrix} \,, \quad \bm{S}' = \bordermatrix{ 
    & n_S^{\rm SS} & n_S^{\rm WS} \cr
     & \frac{-3}{46}&0 \cr
     & \frac{-3}{46}&0 \cr
   & \frac{-3}{46}&0 \cr 
} \,, \quad 
\Tr{\mu_{\Delta_\alpha}^{{\rm{eq}},\theta}} =\left(\frac{69\delta_{B-L}}{22}-\frac{9n_S^{\rm SS}}{44}\right)\dot{\theta} \,.
\end{align}
\end{footnotesize}

\noindent\textbf{\boldmath{$T_{\rm WS}\gtrsim T\gtrsim T_{b\tau}:$}}
\begin{footnotesize}
\begin{align}
    \bm{C}&=\begin{pmatrix}
\frac{202}{345} & \frac{59}{690} & \frac{59}{690} \\
\frac{59}{690} & \frac{202}{345} & \frac{59}{690} \\
\frac{59}{690} & \frac{59}{690} & \frac{202}{345}
\end{pmatrix} \,, \quad \bm{S}' = \bordermatrix{ 
    & n_S^{\rm SS} & n_S^{\rm WS} \cr
     & -\frac{19}{115} & \frac{2}{15} \cr
     & -\frac{19}{115} & \frac{2}{15} \cr
   & -\frac{19}{115} & \frac{2}{15} \cr 
} \,, \quad  \Tr{\mu_{\Delta_\alpha}^{{\rm{eq}},\theta}}=\left(\frac{115\delta_{B-L}}{29}-\frac{19n_S^{\rm SS}}{29}+\frac{46n_S^{\rm WS}}{87}\right)\dot{\theta} \,.
\end{align}
\end{footnotesize}

\noindent\textbf{\boldmath{$T_{b\tau}\gtrsim T\gtrsim T_c:$}}
$(1,\,2,\,\tau)$ basis:
\begin{footnotesize}
\begin{align}
    \bm{C}&=\begin{pmatrix}
\frac{237}{460} & \frac{7}{460} & \frac{8}{115} \\
\frac{7}{460} & \frac{237}{460} & \frac{8}{115} \\
\frac{8}{115} & \frac{8}{115} & \frac{53}{115}
\end{pmatrix} \,, \quad \bm{S}'= \bordermatrix{ 
    & n_S^{\rm SS} & n_S^{\rm WS} \cr
     & -\frac{51}{460} & \frac{17}{115} \cr
     & -\frac{51}{460} & \frac{17}{115} \cr
   & -\frac{9}{115} & \frac{12}{115} \cr 
} \,, \quad \Tr{\mu_{\Delta_\alpha}^{{\rm{eq}},\theta}}=\left(5\delta_{B-L}-\frac{n_S^{\rm SS}}{2}+\frac{2n_S^{\rm WS}}{3}\right)\dot{\theta}.
\end{align}
\end{footnotesize}
$(1,\,2,\,3)$ basis: 
\begin{footnotesize}
\begin{align}
    \bm{C}&=\begin{pmatrix}
\frac{8}{15} & \frac{1}{30} & \frac{1}{30} \\
\frac{1}{30} & \frac{8}{15} & \frac{1}{30} \\
\frac{1}{30} & \frac{1}{30} & \frac{8}{15}
\end{pmatrix} \,, \quad \bm{S}'= \bordermatrix{ 
    & n_S^{\rm SS} & n_S^{\rm WS} \cr
     & -\frac{1}{10} & \frac{2}{15} \cr
     & -\frac{1}{10} & \frac{2}{15} \cr
   & -\frac{1}{10} & \frac{2}{15} \cr 
} \,,\quad \Tr{\mu_{\Delta_\alpha}^{{\rm{eq}},\theta}}=\left(5\delta_{B-L}-\frac{n_S^{\rm SS}}{2}+\frac{2n_S^{\rm WS}}{3}\right)\dot{\theta}.
\end{align}
\end{footnotesize}

\noindent\textbf{\boldmath{$T_{c}\gtrsim T\gtrsim T_\mu:$}}
$(1,\,2,\,\tau)$ basis:
\begin{footnotesize}
\begin{align}
\bm{C}&=\begin{pmatrix}
\frac{585}{1178} & -\frac{2}{589} & \frac{26}{589} \\
-\frac{2}{589} & \frac{585}{1178} & \frac{26}{589} \\
\frac{26}{589} & \frac{26}{589} & \frac{251}{589}
\end{pmatrix} \,, \quad \bm{S}'= \bordermatrix{ 
    & n_S^{\rm SS} & n_S^{\rm WS} \cr
     & -\frac{381}{2356} & \frac{86}{589} \cr
     & -\frac{381}{2356} & \frac{86}{589} \cr
   & -\frac{87}{589} & \frac{60}{589} \cr 
} \,, \quad \Tr{\mu_{\Delta_\alpha}^{{\rm{eq}},\theta}}=\left(\frac{17\delta_{B-L}}{3}-\frac{8n_S^{\rm SS}}{9}+\frac{20n_S^{\rm WS}}{27}\right)\dot{\theta} \,.
\end{align}
\end{footnotesize}
$(1,\,2,\,3)$ basis:
\begin{footnotesize}
\begin{align}
\bm{C}&=\begin{pmatrix}
\frac{199}{390} & \frac{2}{195} & \frac{2}{195} \\
\frac{2}{195} & \frac{199}{390} & \frac{2}{195} \\
\frac{2}{195} & \frac{2}{195} & \frac{199}{390}
\end{pmatrix} \,, \quad \bm{S}'= \bordermatrix{ 
    & n_S^{\rm SS} & n_S^{\rm WS} \cr
     & -\frac{41}{260} & \frac{2}{15} \cr
     & -\frac{41}{260} & \frac{2}{15} \cr
   & -\frac{41}{260} & \frac{2}{15} \cr 
} \,, \quad \Tr{\mu_{\Delta_\alpha}^{{\rm{eq}},\theta}}=\left(\frac{130\delta_{B-L}}{23}-\frac{41n_S^{\rm SS}}{46}+\frac{52n_S^{\rm WS}}{69}\right)\dot{\theta} \,.
\end{align}
\end{footnotesize}

\begin{figure}
    \centering
    {{\includegraphics[width=0.99\textwidth]{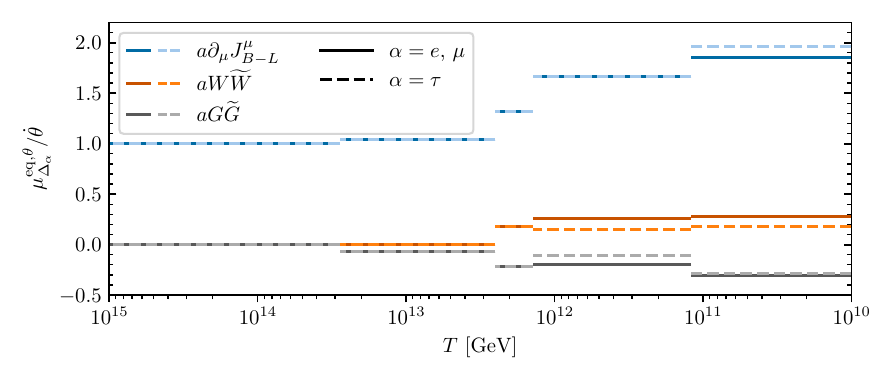}}}
    \caption{Evolution of $\mu_{\Delta_\alpha}^{{\rm{eq}},\theta}(T)/\dot{\theta}$ from $T=10^{15}$ GeV to $T=10^{10}$ GeV for the axion coupling to the $B\!-\!L$ current (blue), to strong sphalerons (gray), and to weak sphalerons (red/orange), all in the algebraic approximation in the $(1,\,2,\,\tau)$ basis. The blue, gray, and red curves are normalized by $\delta_{B-L}$, $C_s$, and $C_w$, respectively. See text for details. }
    \label{fig:Equilibrium}
\end{figure}

\paragraph{Equilibrium attractor:} Fig.~\ref{fig:Equilibrium} presents the results for the equilibrium attractor, $\mu_{\Delta_\alpha}^{{\rm{eq}},\theta}(T)$, normalized to $\dot{\theta}$ for the three axion couplings defined in Eqs.~\eqref{axionBL}–\eqref{axionWW} in the algebraic approximation in the $(1,\,2,\,\tau)$-basis. Solid lines represent the $\alpha=1,2$ components, which take on identical values across the temperature range considered here, while dashed lines stand for $\alpha=\tau$. The colors correspond to the following coupling scenarios: blue for Majoron-like, gray for strong sphalerons, and red/orange for weak sphalerons. The results are normalized by $\delta_{B-L}$, $C_s$, and $C_w$, respectively. Notably, in the algebraic approximation, the effect of strong (weak) sphalerons on $\mu_{\Delta_\alpha}^{{\rm{eq}},\theta}(T)$ vanishes for $T>T_{\rm SS}\,(T_{\rm WS})$, which explains why the red and blue curves deviate from zero only below these respective temperatures.

Finally, our algebraic result can be used to highlight a fundamental limitation of axion-driven spontaneous LG via standard misalignment. As shown, rapid LNV interactions drive the plasma toward an equilibrium state with
\begin{align}
    \label{eq:analytic1}
    \frac{\mu_{\Delta_\alpha}^{{\rm{eq}},\theta}(T)}{T}\approx \mathcal{O}\left(0.1\cdots1\right)\frac{\dot{\theta}}{T},
\end{align}
where the numerical coefficient depends on the specific axion--SM coupling and temperature. The chemical potentials $\mu_{\Delta_\alpha}$ relate to the BAU as follows, 
    \begin{align}
    \label{eq:analytic2}
    \frac{\mu_{B-L}(T)}{T}\equiv\Tr{\frac{\mu_{\Delta_\alpha}}{T}}=\frac{4\pi^2g_{*s}}{15}\,Y_{B-L}=c_{\text{sph}}^{-1}\,\frac{4\pi^2g_{*s}}{15}\,Y_B \,,
\end{align}
where $c_{\text{sph}}$ is the sphaleron-conversion factor~\cite{Harvey:1990qw,Laine:1999wv}. Comparing Eqs.~\eqref{eq:analytic1} and~\eqref{eq:analytic2}, and noting that the value of $\dot{\theta}/T$ around the time of the first oscillation is proportional to $T_{\text{osc}}$ (see Eqs.~\eqref{eq:maTosc} and \eqref{eta(T)}), reveals that for low $T_{\text{osc}}$, the axion velocity due to standard misalignment will no longer suffice to produce the observed BAU, irrespective of the efficiency of the LNV interactions.

\subsection{Numerical results}
\label{sec:baryonasymmetry}
Having established the formalism for spontaneous LG mediated by the Weinberg operator, we turn to a systematic numerical analysis. In this section, we numerically solve the Boltzmann equations presented in Sec.~\ref{sec:BEqs} across a wide range of neutrino and axion parameters. First, we examine several benchmarks to discuss the qualitative features of the evolution and assess the reliability of the algebraic approximation. We then illustrate how the generated $B\!-\!L$ asymmetry changes with $T_{\text{osc}}$ for various axion couplings and neutrino parameters, comparing the full numerical solutions against their algebraic counterparts. Finally, we comment on the impact of flavor-resolved effects in the Weinberg operator.

\begin{figure}
    \centering
    \includegraphics[width=\textwidth]{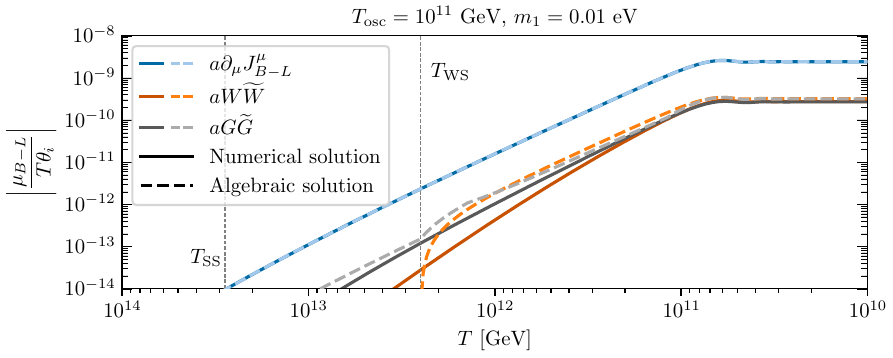}\\
    \includegraphics[width=\textwidth]{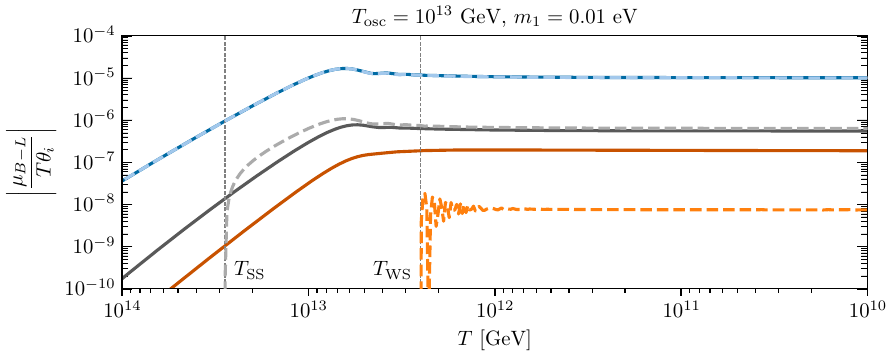}
    \caption{Example solutions of $\abs{\mu_{B-L}/(T\theta_i)}$ for NO with $m_1=0.01$ eV, and $T_{\text{osc}}=10^{11}$ GeV (upper panel) $T_{\text{osc}}=10^{13}$ GeV (lower panel). Solid lines show solutions to the full set of Boltzmann equations presented in Sec.~\ref{sec:BEqs}, in the $(1,\,2,\,3)$ basis, and dashed lines show the corresponding algebraic solution presented in Sec.~\ref{subsec:algebraic}. Different colors indicate axion coupling to the $B\!-\!L$ current (blue), weak sphalerons (red/orange), and strong sphalerons (gray). The blue, gray, and red results are normalized by $\delta_{B-L}=1$, $C_s=1$, and $C_w=1$, respectively. The gray vertical dashed lines represent the equilibration temperatures for strong and weak sphalerons, respectively: $T_{\rm SS}=2.8\cdot 10^{13}$ GeV and $T_{\rm WS}=2.5\cdot 10^{12}$ GeV.}
    \label{fig:BPEvolution}
\end{figure}

\noindent\paragraph{Benchmark results:} In Fig.~\ref{fig:BPEvolution}, we show benchmark solutions of $\abs{\mu_{B-L}/(T\theta_i)}$ for normal neutrino mass hierarchy in the $(1,\,2,\,3)$ basis with $m_1=0.01$ eV, with different colors indicating results for the axion coupling to the $B\!-\!L$ current (blue), to strong sphalerons (gray), and to weak sphalerons (red/orange). Solutions to the full set of evolution equations are indicated by solid lines, dashed lines show the algebraic solution.

The upper panel shows our results for $T_{\text{osc}}=10^{11}$ GeV. In this case, the asymptotic values of $\abs{\mu_{B-L}/(T\theta_i)}$, obtained using the algebraic approximation, are in very good agreement with those obtained by solving the full set of Boltzmann equations for all three axion couplings. This occurs because the oscillation temperature is significantly lower than the equilibration temperatures for both strong sphalerons (gray curves) and weak sphalerons (red/orange curves). In particular, the orange algebraic solution is zero (by construction) above the weak sphaleron temperature $T_{\rm WS}=2.5\cdot 10^{12}$ GeV, and then significantly overshoots the full numerical result shortly after $T_{\rm WS}$ is reached. This discrepancy arises because the algebraic approach by construction assumes instantaneous chemical equilibrium, whereas the full treatment correctly captures the gradual approach to equilibration, and the two solutions converge at lower temperatures as weak sphalerons fully equilibrate. Conversely, the agreement between the two methods for axion coupling to $J_{B-L}$ is excellent at all temperatures. This consistency arises because, in this Majoron-like scenario, the axion velocity couples directly to LNV interactions, which are treated on an equal footing in the full Boltzmann approach and the algebraic approach. 

The lower panel shows the case for $T_{\text{osc}}=10^{13}$ GeV. Here, the agreement between the algebraic approximation and the full Boltzmann solution is good for the $J_{B-L}$ coupling, decent for the strong sphalerons, and poor for the weak sphalerons. The reason for the poor performance of the algebraic approach for the weak-sphaleron coupling in this case is that $T_{\text{osc}}>T_{\rm WS}$. In the full treatment, the bulk of the asymmetry is generated while the axion rolls and undergoes its initial oscillations. Conversely, the algebraic approach misses these early contributions because weak sphalerons are treated as completely decoupled at these high temperatures. Consequently, the algebraic method only begins to accumulate asymmetry below $T \approx T_{\rm WS}$, a regime where the axion velocity and LNV interaction rates are already significantly suppressed, leading to a substantial underestimation of the final asymmetry. In contrast, strong sphalerons are active at $T > T_{\text{osc}}$, leading to efficient asymmetry generation. In this case, the algebraic approach overestimates the final asymmetry compared to the full Boltzmann solution, as it assumes strong sphalerons are instantaneously and fully equilibrated at $T_{\rm SS} = 2.8 \cdot 10^{13}$ GeV.

\noindent\paragraph{Generalities:} As a general rule, the algebraic framework tracks the full Boltzmann evolution for the Majoron-like coupling very well. For axion--sphaleron interactions, however, the validity of the approximation hinges on the relationship between $T_{\text{osc}}$ and the sphaleron equilibration temperature ($T^{\text{sph}}_{\text{eq}}$). In scenarios where $T_{\text{osc}} < T^{\text{sph}}_{\text{eq}}$, the method successfully captures the asymptotic asymmetry, although it slightly overshoots the yield immediately following equilibration. On the other hand, if $T_{\text{osc}} > T^{\text{sph}}_{\text{eq}}$, the algebraic treatment entirely misses the early dynamical phase\,---\,the first few oscillations where the bulk of the asymmetry is generated\,---\,resulting in a severe under-prediction of the final $B\!-\!L$ yield.

Finally, our results are obtained by integrating the system of Boltzmann equations in a radiation-dominated background, starting from a high reheating temperature (for definiteness, we set $T_{\rm{rh}}=10^{15}$ GeV) with vanishing initial chemical potentials for all SM species. As long as $T_{\text{rh}}\gg T_{\rm{osc}}$, the final asymmetry remains insensitive to the precise choice of $T_{\text{rh}}$.\footnote{The condition $T_{\text{rh}}\gg T_{\rm{osc}}$ can also be recast as a requirement on the inflaton decay rate $\Gamma_\phi$ relative to the axion mass: $H(T_{\rm rh})\simeq \Gamma_\phi \gg m_a$. Ref.~\cite{Kusenko:2014uta} discusses spontaneous LG during reheating.} If this hierarchy in temperatures is violated, the axion dynamics must be explicitly tracked throughout reheating after inflation by solving the axion equation of motion alongside the coupled evolution of the energy densities of radiation and the inflaton field. While technically straightforward, such an extension is inherently model-dependent\,---\,requiring the specification of a concrete inflaton and reheating sector\,---\,and therefore falls outside the scope of our present model-independent analysis.

\begin{figure}
    \centering
    {{\includegraphics[width=0.99\textwidth]{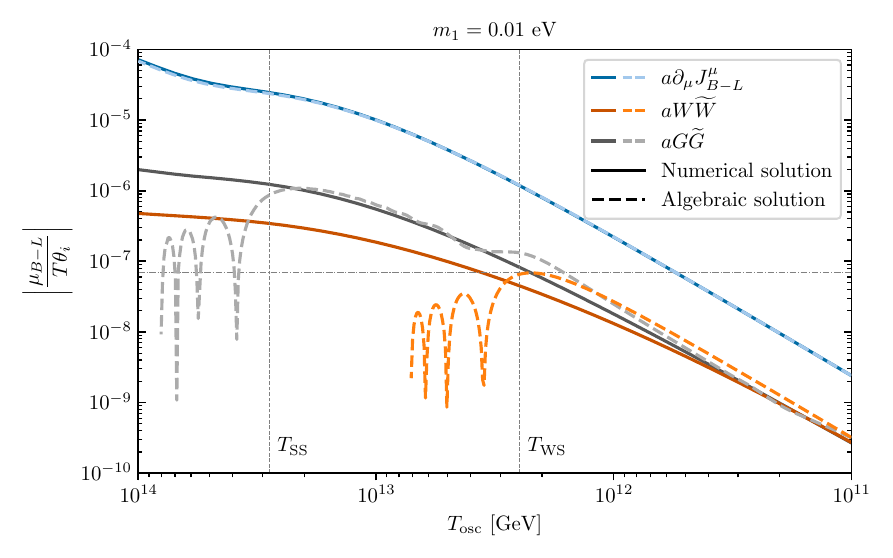}}}
    \caption{Value of $|\mu_{B-L}/(T\theta_i)|$, evaluated at $T=10^{10}\text{ GeV}$ for NO with $m_1=0.01$~eV, as a function of $T_{\rm{osc}}$ and axion coupling to the $B\!-\!L$ current (blue), strong sphalerons (gray), and weak sphalerons (red/orange). Solid lines show the results obtained by solving the full set of Boltzmann equations, while dashed curves show the algebraic approximation. The horizontal dash-dotted curve corresponds to the observed asymmetry setting $\theta_i = 1$, and the two vertical dashed curves mark $T_{\rm SS}$ (left) and $T_{\rm WS}$ (right), respectively. The orange and gray dashed curves have been cut, because the approximation for the algebraic solution breaks down and no longer yields physically meaningful results for $T_{\rm{osc}} > T_{\rm{eq}}^{\rm{sph}}$.}
    \label{fig:ParameterSpace1}
\end{figure}

\noindent\paragraph{Full versus algebraic solutions:} Fig.~\ref{fig:ParameterSpace1} shows $\abs{\mu_{B-L}/(\theta_iT)}$ evaluated at $T=10^{10}\text{ GeV}$ as a function of $T_{\rm{osc}}$ for axion coupling to $J_{B-L}$ (blue), strong sphalerons (gray), and weak sphalerons (red/orange). The computations were carried out in the $(1,\,2,\,3)$-basis with $m_1=0.01$ eV. The solid lines show the results obtained by solving the full set of Boltzmann equations, while the dashed curves show the algebraic approximations. The horizontal dashed curve shows the value corresponding to the observed BAU, normalized by $\theta_i$, for the case where the asymmetry is not further diluted later in the cosmological history, and the left (right) vertical line corresponds to $T_{\rm SS}$ $(T_{\rm WS})$. The figure shows that the algebraic approximation provides a reliable estimate of the asymmetry, up to $\mathcal{O}(10\%)$, for the Majoron-like coupling. Meanwhile, for axion coupling to sphalerons, the algebraic approximation is reliable up to a factor $\sim 2$ when $T_{\text{osc}}$ is at most as high as the sphaleron equilibration temperature. For higher oscillation temperatures, the approximation quickly breaks down, as discussed earlier in this section. Finally, the figure shows that for $\theta_i\approx 1$, an asymmetry compatible with the observed BAU can be generated for $T_{\text{osc}}\gtrsim 5\cdot 10^{11}$ GeV, $T_{\text{osc}}\gtrsim 2\cdot 10^{12}$ GeV, $T_{\text{osc}}\gtrsim 4\cdot 10^{12}$ GeV, for an axion coupled to $J_{B-L}$ with $\delta_{B-L}=1$, strong sphalerons with $C_s=1$, and weak sphalerons with $C_w=1$, respectively. Since the resulting asymmetry scales linearly with the anomaly coefficient, cf.\ the discussion between Eqs.~\eqref{eq:muH_full} and~\eqref{eq:muL_notfull}, we see that an axion coupled to $G\widetilde{G}$ with $C_s\gtrsim 10$, can generate a $B\!-\!L$ asymmetry that is comparable to or even larger than that generated by a Majoron.

\begin{figure}
    \centering
    {{\includegraphics[width=0.99\textwidth]{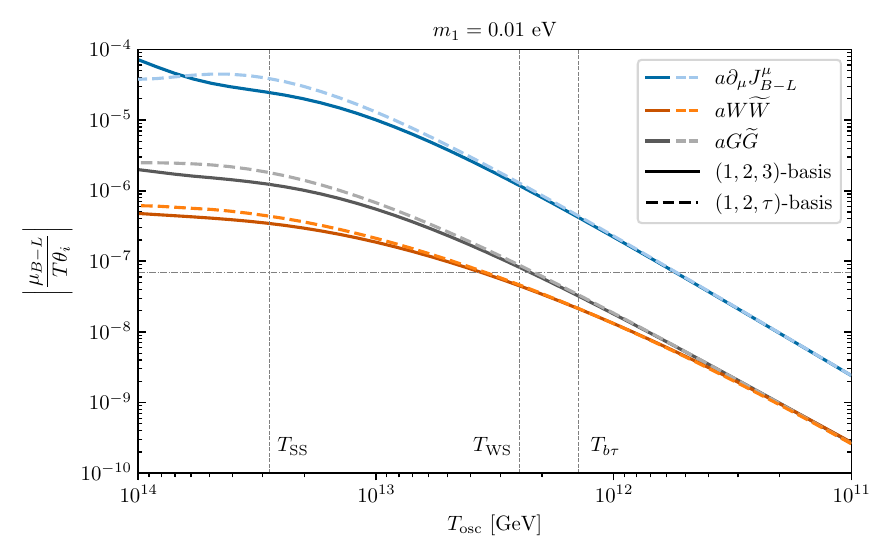}}}
    \caption{Value of $|\mu_{B-L}/(T\theta_i)|$, evaluated at $T=10^{10}\text{ GeV}$ for NO with $m_1=0.01$~eV, as a function of $T_{\rm{osc}}$ and axion coupling to the $B\!-\!L$ current (blue), strong sphalerons (gray), and weak sphalerons (red/orange). Solid lines show the results obtained in the $(1,\ 2,\ 3)$-basis, while dashed curves show the results in the $(1,\ 2,\ \tau)$-basis for $\alpha_1=\alpha_2=0$. 
     The horizontal dash-dotted line corresponds to the observed asymmetry setting $\theta_i = 1$, and the three vertical dashed lines mark $T_{\rm SS}=2.8\cdot 10^{13}$ GeV (left), $T_{\rm WS}=2.5\cdot 10^{12}$ GeV (middle) and $T_{b\tau}=1.4\cdot 10^{12}$ GeV (right), respectively.
    }
    \label{fig:BasisComparison}
\end{figure}

\noindent\paragraph{Choice of flavor basis:} Fig.~\ref{fig:BasisComparison} shows the results obtained by solving the full set of Boltzmann equations for the same benchmarks as in Fig.~\ref{fig:ParameterSpace1}, both in the $(1,\,2,\,3)$-basis (solid curves) and in the $(1,\,2,\,\tau)$-basis (dashed 
curves). The two bases deviate most strongly at $T_{\text{osc}}\sim 10^{13}\text{--}10^{14}\,\text{GeV}$, which is precisely the regime in which the $(1,\,2,\,3)$-basis is expected to provide a reliable description of the system. Conversely, for $T_{\text{osc}}\lesssim 10^{12}\,\text{GeV}$, where the $(1,\,2,\,\tau)$-basis becomes the appropriate description, the solid and dashed curves nearly coincide. The $(1,\,2,\,3)$-basis therefore yields accurate results over the entire range of $T_{\text{osc}}$: at high temperatures it is the correct description, while at low temperatures its predictions are numerically similar to those of the $(1,\,2,\,\tau)$-basis. This stems from the fact that in the freeze-in regime, where $T_{\text{osc}}$ is well below the decoupling temperature of the Weinberg operator, the $B-L$ asymmetry is proportional to the flavor-basis-independent quantity $\sum_i m_i^2$. Meanwhile, at large values for $T_{\text{osc}}$ the two bases yield significantly different results, as the $(1,\,2,\,\tau)$-basis introduces sizable spurious in that regime. This provides a robust a posteriori justification for employing the $(1,2,3)$ basis throughout the entire temperature range relevant to our analysis.\footnote{Moreover, a full density-matrix treatment is expected to produce curves in Fig.~4 that smoothly interpolate between the $(1,2,3)$ results at large $T_{\text{osc}}$ and the $(1,2,\tau)$ results at low $T_{\text{osc}}$. Whenever a diagonal flavor basis is a good description, the results of the appropriate Boltzmann-equation treatment and the results of a density-matrix treatment should agree.} Motivated by this, we present all remaining numerical results in the main text, in particular the parameter scans, in the $(1,\,2,\,3)$-basis, and collect some complementary results in App.~\ref{appendix:NumericResults}. Finally, it is worth emphasizing that in the $(1,\,2,\,3)$-basis, the LNV rates depend only on the neutrino masses, leaving the lightest neutrino mass (and the neutrino-mass ordering) as the only free parameter in that sector. Consequently, for a given mass ordering, the final baryon asymmetry is determined solely by the value of the lightest neutrino mass and is completely independent of the PMNS mixing parameters, including both the Dirac and Majorana CP phases. Remarkably enough, this means that the Weinberg interaction rates in the $(1,\,2,\,3)$-basis constitute a physical implication of the SM neutrino mass basis at extremely high temperatures, long before electroweak symmetry breaking, i.e., long before the SM neutrinos actually become massive\,---\,in other words, the Weinberg interaction rates in the $(1,\,2,\,3)$-basis foreshadow the light neutrino mass eigenstates long before these states are actually realized as dynamical states in the thermal plasma.

\begin{figure}
    \centering
    {{\includegraphics[width=0.99\textwidth]{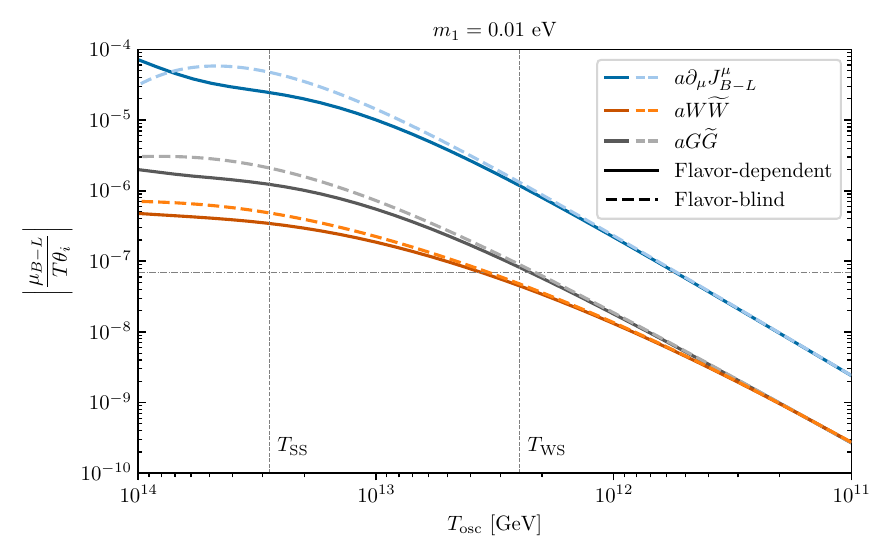}}}
    \caption{Value of $|\mu_{B-L}/(T\theta_i)|$, evaluated at $T=10^{10}\text{ GeV}$ for NO with $m_1=0.01$~eV, as a function of $T_{\rm{osc}}$ and axion coupling to the $B\!-\!L$ current (blue), strong sphalerons (gray), and weak sphalerons (red/orange). Solid lines show the flavor-dependent results, while dashed curves show the flavor-blind approximation. The horizontal dash-dotted curve corresponds to the observed asymmetry setting $\theta_i = 1$, and the two vertical dashed curves mark $T_{\rm SS}$ (left) and $T_{\rm WS}$ (right), respectively. }
    \label{fig:FlavorComparison}
\end{figure}

\paragraph{Comparison with the flavor-blind approximation:}
Fig.~\ref{fig:FlavorComparison} compares the results in the $(1,\,2,\,3)$-basis (solid curves, identical to those in Fig.~\ref{fig:ParameterSpace1} and~\ref{fig:BasisComparison}) with those obtained in the flavor-blind approximation 
of Eq.~\eqref{eq:muL_notfull} (dashed curves). The two treatments differ most markedly for $T_{\text{osc}} \gtrsim 10^{13}\,\text{GeV}$. There, the flavor-blind asymmetry initially grows with $T_{\text{osc}}$, exceeding the flavored result, but then flattens and even starts decreasing for the Majoron-like coupling. The flavored calculation, by contrast, yields an asymmetry that grows monotonically with $T_{\text{osc}}$ across the entire range considered.

This behavior can be understood as follows. At low $T_{\text{osc}}$, all LNV rates decouple well before the onset of the oscillations, the charges are produced in the freeze-in regime, and each $B/3-L_i$ yield is proportional to its rate. The total asymmetry then depends only on $\sum_i m_i^2$, and the two treatments agree by construction. The difference emerges once equilibration becomes relevant. In the flavor-blind case, all charges are sourced by a single universal rate set by the average $m^2_{\text{av}} \equiv \tfrac{1}{3}\sum_i m_i^2$, which for normal ordering with $m_1 = 0.01\,\text{eV}$ exceeds the physical rates of the two lighter flavors by roughly an order of magnitude. As $T_{\text{osc}}$ increases, the light flavors are thus driven toward their equilibrium asymmetries at artificially low $T_{\text{osc}}$, and the flavor-blind treatment overshoots the flavored result. For yet larger $T_{\text{osc}}$, the universal rate remains in equilibrium into the oscillatory regime, and all three 
charges then suffer from washout. In the flavored treatment the rates are instead hierarchical, $\gamma_i^{\text{eq}} \propto m_i^2$, with those of $L_1$ and $L_2$ smaller than that of $L_3$ by factors of ${\sim}25$ and 
${\sim}15$. Within the scanned range, only the third flavor equilibrates into the oscillatory regime, so only the $B/3-L_3$ component is suppressed at $T_{\text{osc}}\sim 10^{14}$ GeV, while the two lighter flavors still decouple before the onset of the oscillations and their contributions continue to grow with $T_{\text{osc}}$ --- yielding the monotonic 
behavior of the solid curves.

\begin{figure}
     \centering
     \begin{subfigure}[b]{\textwidth}
         \centering
         \includegraphics[width=\textwidth]{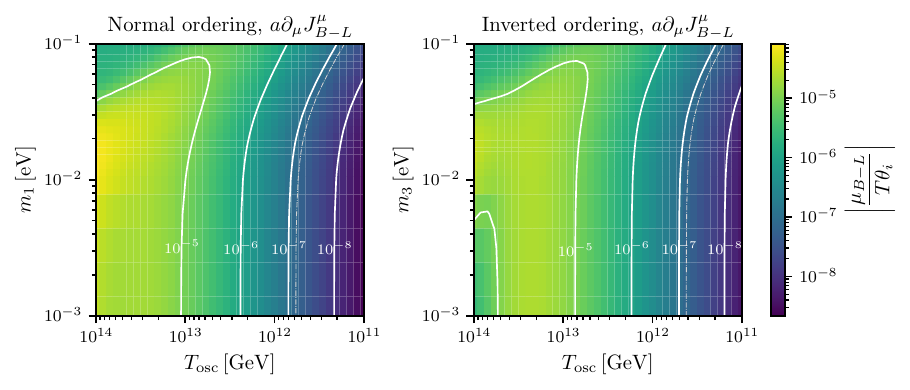}
     \end{subfigure}
     \hfill
     \begin{subfigure}[b]{\textwidth}
         \centering
         \includegraphics[width=\textwidth]{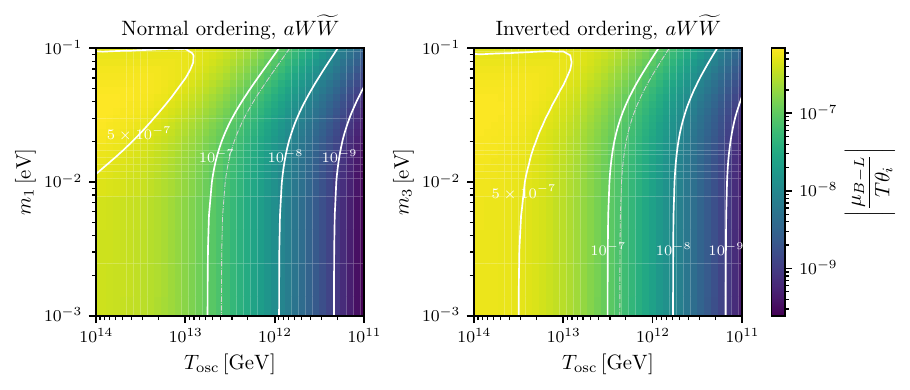}
     \end{subfigure}
     \hfill
     \begin{subfigure}[b]{\textwidth}
         \centering
         \includegraphics[width=\textwidth]{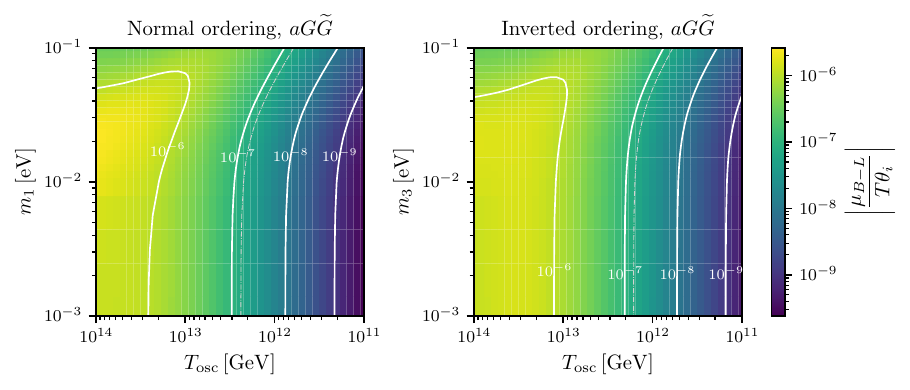}
     \end{subfigure}
        \caption{Parameter scan for $|\mu_{B-L}/(T\theta_{i})|$ evaluated at $T=10^{10}$ GeV over the smallest neutrino mass $m_1$ (NO) or $m_3$ (IO) and the oscillation temperature $T_{\rm osc}$ for the case of the axion being coupled to the $B\!-\!L$ current (top), the weak sphalerons (middle) and the strong sphalerons (bottom). The light gray dash-dotted line corresponds to the observed asymmetry setting $\theta_i = 1$.}
        \label{fig:Big Parameter Scan}
\end{figure}

\noindent\paragraph{Comparison with Ref.~\cite{Domcke:2020kcp}:} It is worth noting that the results presented in Fig.~\ref{fig:ParameterSpace1},~\ref{fig:BasisComparison} and~\ref{fig:FlavorComparison} differ qualitatively from those in Fig.~6 of Ref.~\cite{Domcke:2020kcp}. This discrepancy arises from three primary factors. First, the authors of Ref.~\cite{Domcke:2020kcp} impose an ad hoc axion decay temperature, $T_{\text{dec}}$, which alters the behavior near $T_{\text{osc}}\sim 10^{11}$ GeV; our analysis avoids this assumption. Second, they rely on a simplified approximation for the axion velocity, whereas we utilize the full expression in Eq.~\eqref{eta(T)}. This methodological difference leads to an order-of-magnitude offset in the calculated asymmetry across all considered values of $T_{\text{osc}}$ in the two figures.\footnote{Specifically, in our setup the axion-velocity scale remains significantly larger across all relevant temperatures, resulting in an order-of-magnitude enhancement in the produced asymmetry. Furthermore, Eq.~\eqref{eta(T)} incorporates the complete axion dynamics, whereas the approximation adopted in Ref.~\cite{Domcke:2020kcp} assumes a vanishing axion velocity for temperatures above $T_{\text{osc}}$.}
Finally, Ref.~\cite{Domcke:2020kcp} implements an LNV rate that is significantly larger than ours. Specifically, their LNV rate enters the Boltzmann equations in Eqs.~\eqref{eq:muL_notfull}--\eqref{eq:muH_notfull} with a prefactor three times larger than in our case, $\gamma_{\text{WB}}^{\Delta L=2}\rightarrow 3\,\gamma_{\text{WB}}^{\Delta L=2}$, while the rate itself is evaluated at a fixed neutrino mass scale, corresponding to $\overline{m}^2\approx \frac{3.00}{2.45}(0.05\,\text{eV})^2$ in our convention for the rate in Eq.~\eqref{eq:gammaWB}.\footnote{Numerically, their treatment of LNV rates corresponds in our formalism to adopting the flavor-blind approximation with $m_1 \simeq 0.05\text{ eV}$.} This enhancement of the LNV rates results in the asymmetry-generation tracking the axion oscillations more tightly. As a consequence, for $T_{\text{osc}}\sim 10^{13\cdots14}$ GeV, the $B-L$ asymmetry in Ref.~\cite{Domcke:2020kcp} is strongly suppressed by the axion velocity at high oscillation temperatures. In contrast, our rates, which were carefully derived in Section~\ref{subsec:WeinbergOperator}, induce a weaker coupling to the axion velocity, which leads to a smaller suppression of the $B-L$ asymmetry at $T_{\text{osc}}\sim 10^{13\cdots14}$ GeV.

\begin{figure}
    \centering
    \includegraphics[width=0.99\textwidth]{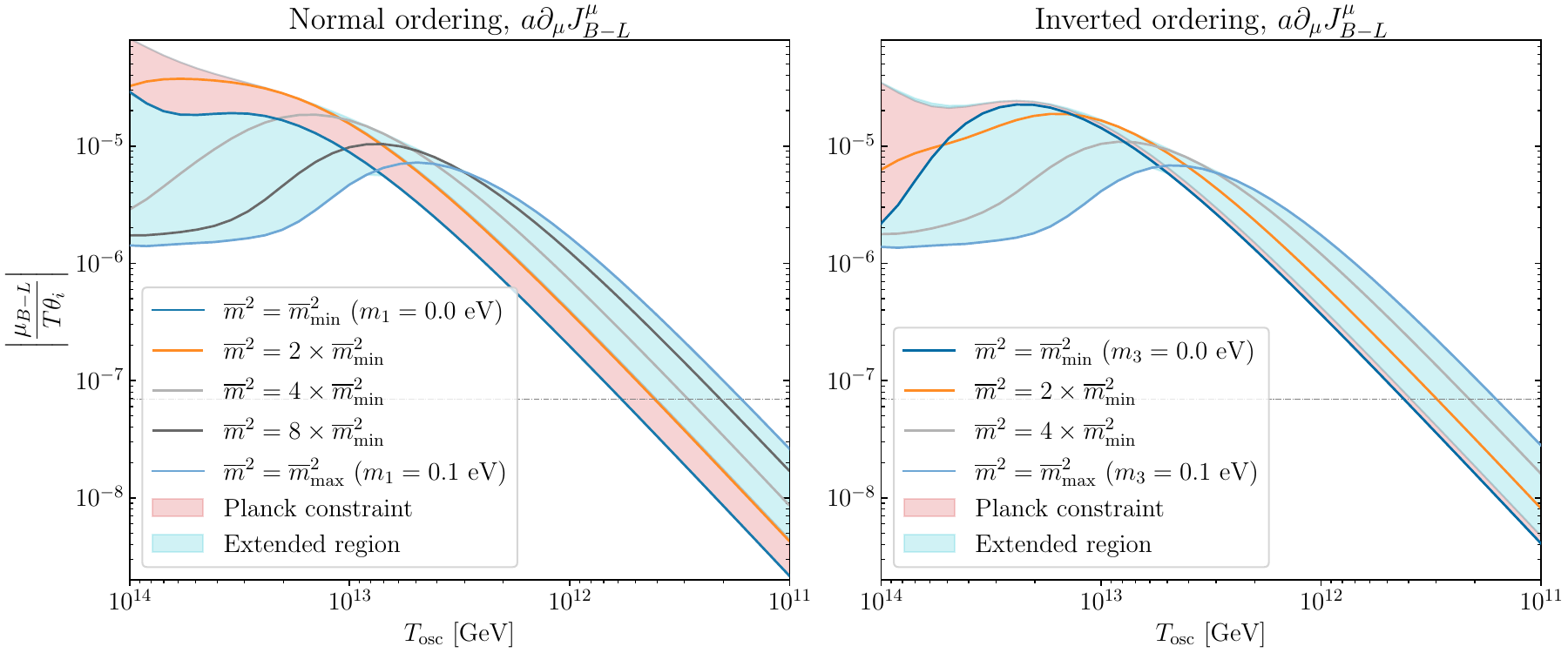}\\
    \includegraphics[width=0.99\textwidth]{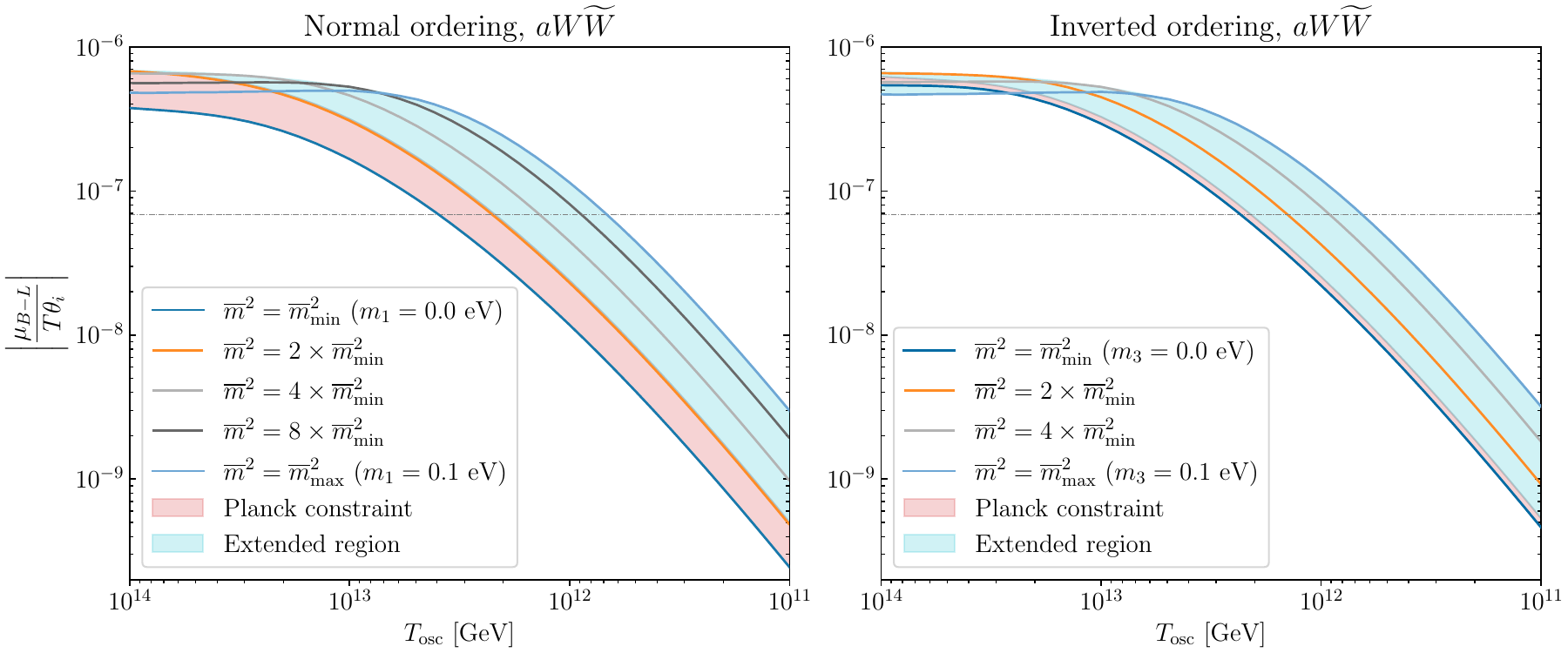}\\
    \includegraphics[width=0.99\textwidth]{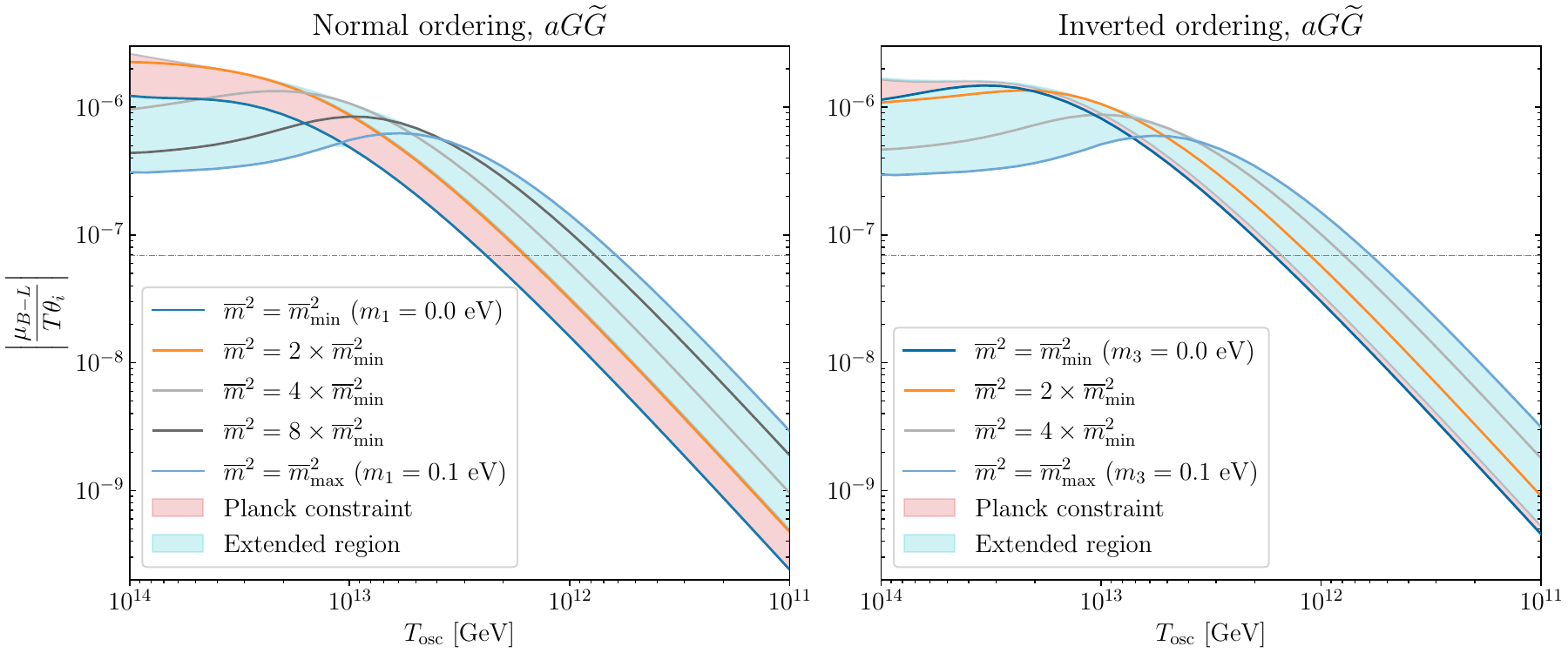}   
    \caption{Value of $|\mu_{B-L}/(T\theta_i)|$, evaluated at $T=10^{10}\text{ GeV}$ for NO (left) and IO (right), for different neutrino masses, as a function of $T_{\rm{osc}}$, and axion coupling to the $B\!-\!L$ current (top), weak sphalerons (middle), and strong sphalerons (bottom). The different curves correspond to different values of $\overline{m}^2 = m_1^2+m_2^2+m_3^2$. The two colored bands display different neutrino mass intervals. The red band belongs to the PLANCK 2018 upper limit, $\sum m_i \leq 0.12$ eV~\cite{Planck:2018vyg}, while the blue band shows a broader interval, $m_1$, $m_3\leq 0.1$ eV. The horizontal dash-dotted line corresponds to the observed asymmetry setting $\theta_i = 1$.}
    \label{fig:Band plot}
\end{figure}

\paragraph{Full parameter space:} Fig.~\ref{fig:Big Parameter Scan} and~\ref{fig:Band plot} show an extended analysis of the parameter space, where the smallest neutrino mass is no longer fixed to $m_1=0.01$ eV, but instead is treated as a free parameter to study how the $B\!-\!L$ asymmetry depends on it. Fig.~\ref{fig:Big Parameter Scan} shows $|\mu_{B-L}/(T\theta_i)|$ in the $T_{\rm osc}$\,--\,$m_1$-plane. The three rows correspond to the coupling to the $B\!-\!L$ current (top), the weak sphalerons (middle), and the strong sphalerons (bottom), while the columns are for NO (left) and IO (right). All the different cases exhibit some similarities in their behavior, which are the following: 
\begin{enumerate}
    \item For small neutrino mass scales, the $B\!-\!L$ asymmetry is not sensitive to a change in the lightest neutrino mass. This can be seen for the contour lines in Fig.~\ref{fig:Big Parameter Scan}, which are almost vertical lines for neutrino mass scales $m_1$, $m_3 < 0.01$ eV. 
    \item The dependence on the neutrino mass scale is non-trivial. For small $T_{\rm osc}$, the $B\!-\!L$ asymmetry grows with increasing $m_1$ or $m_3$, while for large $T_{\rm osc}$, the situation is more complicated, and for the upper left corner in the parameter space, the behavior is inverted, and the $B\!-\!L$ asymmetry can decrease with increasing neutrino mass scale. 
\end{enumerate}
The numerical dependence on the neutrino mass is further studied in Fig.~\ref{fig:Band plot}, where the values for $|\mu_{B-L}/(T\theta_i)|$ are plotted as a function of $T_{\rm osc}$, similar to Fig.~\ref{fig:ParameterSpace1}, but this time for different neutrino masses. The total rate of $B\!-\!L$ violation in the flavor-blind limit scales like $\overline m^2$ [see Eq.~\eqref{eq:gammaWB}], which suggests that the produced $B\!-\!L$ asymmetry should scale in proportion to $\overline m^2$.\footnote{See Eqs.~(9),~(20), and the subsequent section in Ref.~\cite{Kusenko:2014uta} for an extended discussion of this point.} Our analysis confirms that this relation yields a good description for low oscillation temperatures. For $T_{\rm osc} \sim 10^{11}$ GeV, an increase by a factor of two in $\overline m^2$ leads to an increase by a factor of two in $|\mu_{B-L}/(T\theta_i)|$ to good approximation. Meanwhile, the simple proportionality to $\overline m^2$ breaks down at larger $T_{\rm osc}$, and the dependence on the neutrino mass scale becomes more intricate. In addition, Fig.~\ref{fig:Band plot} contains bands that show the allowed region of values that $|\mu_{B-L}/(T\theta_i)|$ can take for different constraints on the neutrino masses. The red band shows the possible values for the PLANCK 2018 constraint, $\sum m_i \leq 0.12$ eV~\cite{Planck:2018vyg}, which corresponds to $m_1=0.030$ eV and $m_3=0.016$ eV for NO and IO, respectively. The blue band shows the possible values for a less conservative interval with $m_1\in [0, 0.1]$ eV and $m_3\in [0, 0.1]$ for NO and IO, respectively.

\paragraph{Impact of neutrino parameters:} As shown in Fig.~\ref{fig:Band plot}, varying the absolute neutrino mass scale or the neutrino-mass ordering can lead to a significant change in the final $B\!-\!L$ asymmetry. The impact of the different neutrino-mass parameters depends on the exact position in parameter space. In the case of NO, varying the neutrino mass scale within the range allowed by PLANCK 2018, leads to changes up to $100 \%$. Meanwhile, for IO, the bands tend to be narrower, with the exception of the coupling to the $B-L$ current, where varying the neutrino mass within PLANCK 2018 constraints can lead to changes of one order of magnitude for temperatures around $10^{14}$ GeV. Furthermore, as can also be seen from Figs.~\ref{fig:Big Parameter Scan} and~\ref{fig:Band plot}, the impact of NO vs.\ IO (for fixed neutrino-mass scale) is of the order of $0.1 \cdots 1000 \%$, depending on the value of $T_{\text{osc}}$ and the considered SM--axion coupling, where the biggest deviations are seen at high $T_{\text{osc}}$ for the Majoron-like coupling.

\section{Embedding into a consistent cosmology}
\label{sec:4}

Thus far, we have considered a framework for calculating the asymmetries generated through spontaneous LG, along with a simple algebraic approximation for the $B\!-\!L$ asymmetry. Up to this point, all axion parameters have been treated as free, without incorporating cosmological or consistency constraints. In this section, we shall embed spontaneous LG in a standard post-inflationary cosmological history where the energy budget of the Universe is determined by the radiation energy density and the energy density stored in the axion field. We will incorporate constraints from isocurvature perturbations, quantify entropy dilution of the $B\!-\!L$ asymmetry from axion decays, and discuss the relevance of the axion--SM coupling in the axion EoM. Finally, we reassess the viable parameter space for successful spontaneous LG driven by axion oscillations in light of these effects.

\subsection{Expansion history and cosmological constraints} 
\label{sec:constraints}

Let $Y_{B-L}^{\text{sLG}}$ denote the asymptotic value of the $B\!-\!L$ yield obtained by solving the Boltzmann equations in Sec.~\ref{sec:BEqs} while neglecting entropy injection from axion decays. Then, accounting for entropy production during axion decays, the final yield can be written as
\begin{align}
    \label{eq:YBLDil}
    Y_{B-L}=\Delta_a Y_{B-L}^{\text{sLG}},
\end{align}
where $\Delta_a$ is the entropy dilution factor~\cite{Kusenko:2014uta}.
Unless the axion dominates the energy budget of the Universe around the time of its decay, the associated entropy injection is small, and one may approximate $\Delta_a\simeq 1$.\footnote{For completeness, let us estimate $\Delta_a$ for such a case: Define the characteristic decay time $t_*$ by $H(t_*)\equiv c\Gamma_a$, where $c$ is an order-one number, and let us for simplicity consider the axion decay as instantaneous. Further, define $r_*\equiv \rho_a(t_*)/\rho_{\text{tot}}(t_*)$, where $\rho_{\text{tot}}(t_*)=\rho_R(t_*)+\rho_a(t_*)=3M_P^2\,(c\Gamma_a)^2$. Then, the temperature of the radiation bath right before and after axion decay can be written as $T_*^-=(1-r_*)^{1/4}\sqrt{c\Gamma_aM_*}$ and $T_*^+=\sqrt{c\Gamma_aM_*}$, where $M_* = (90/(\pi^2 g_*))^{1/2}M_P$, which implies $\Delta_a^{-1}\simeq (T_*^+/T_*^-)^3=(1-r_*)^{-3/4}$. Then, assuming no stage of intermediate axion domination, $r_*\lesssim 1/2$, one obtains $1\lesssim\Delta_a^{-1}\lesssim 2^{3/4}$.} The first objective of this section is to estimate $\Delta_a$ in the opposite regime, where the axion provides an early period of matter domination. We will henceforth denote 
\begin{align}
\label{eq:Delta_a}
\Delta_a \;\simeq\;
\begin{cases}
1\,, & \text{no intermediate axion domination}\,,\\[4pt]
\Delta_a'\,, & \text{intermediate axion domination}\,,
\end{cases}
\end{align}
to separate the two cases~\cite{Kusenko:2014uta}. Throughout this section, we will work within the instantaneous axion-decay approximation. 

\paragraph{Energy density, early matter domination, and entropy dilution:} 
In the analysis throughout this paper, we have assumed that reheating after inflation completes well before axion oscillations, and thereby treated the Universe as radiation-dominated at the onset of axion oscillations. The ratio between $\rho_a$ and $\rho_R$ is
\begin{align}
    \frac{\rho_a}{\rho_R}\bigg|_{T_{\text{osc}}}\simeq\frac{\sfrac{1}{2}\,m_a^2f_a^2\theta_i^2}{3M_P^2H^2_{\text{osc}}}=\frac{3}{2}\theta_i^2\frac{f_a^2}{M_P^2} \,,
\end{align}
whereupon it further increases like $\rho_a/\rho_R \propto 1/(R^3g_*T^4) \propto (g_{*s}/g_*)/T$, with $g_*$ and $g_{*s}$ denoting the energy and entropy degrees of freedom of the radiation bath, respectively. In the following, we will only consider times before SM neutrino decoupling, $T \gtrsim 1\,\textrm{MeV}$, such that we can use $g_*$ and $g_{*s}$ interchangeably to good approximation. Setting $g_{*s} = g_*$, we may then express the temperature at a hypothetical axion--radiation equality as
\begin{align}
    T_{\text{eq}}=\frac{3}{2}\theta_i^2\frac{f_a^2}{M_P^2}T_{\text{osc}}=\frac{3}{2}\theta_i^2\frac{f_a^2}{M_P^2}\sqrt{\frac{M_Pm_a}{3}}\left(\frac{90}{\pi^2g_*(T_{\text{osc}})}\right)^{1/4}.
\end{align}
Hence, to arrive at a period of axion domination, which requires $\Gamma_a<\ H_{\rm{eq}}$, the axion-decay rate must satisfy
\begin{align}
    \Gamma_a<\frac{3}{2\sqrt{2}}\theta_i^4\left(\frac{g_*(T_{\text{eq}})}{g_*(T_{\text{osc}})}\right)^{1/2}\left(\frac{f_a}{M_P}\right)^4m_a.
\end{align}
If this condition is satisfied, entropy dilution from axion decays becomes an important effect, and the entropy-dilution factor can be written as
\begin{align}
    \Delta_a'=\frac{g_{*s}(T_{\text{eq}})}{g_{*s}(T_{\text{dec}})}\left(\frac{T_{\text{eq}}}{T_{\text{dec}}}\right)^3\left(\frac{R_{\text{eq}}}{R_{\text{dec}}}\right)^3=\frac{T_{\text{dec}}}{T_{\text{eq}}},
\end{align}
where $R$ and $T_{\text{dec}}$ denote the Friedmann--Lema\^itre--Robertson--Walker scale factor and the reheating temperature after axion decays, respectively. 
In the instantaneous axion-decay limit in a matter-dominated Universe, we can use $H(t_{\text{dec}})=\sfrac{2}{3}\,\Gamma_a$, which gives 
\begin{align}
    T_{\text{dec}}=\sqrt{\frac{2M_P\Gamma_a}{3}}\left(\frac{90}{\pi^2g_*(T_{\text{dec}})}\right)^{1/4},
\end{align}
where $T_{\text{dec}}\gtrsim 10$ MeV is required to not spoil big-bang nucleosynthesis (BBN) predictions. Note that this is an implicit relation for $T_{\text{dec}}$, as both $g_*$ and the couplings entering $\Gamma_a$ evolve with temperature. In our numerical results, we evolve the SM gauge couplings using one-loop RGEs and fix the renormalization scale as $\mu=2\pi T$. 

By combining the results above, the entropy-dilution factor can be written in the form
\begin{align}
    \Delta_a'\simeq \frac{2\sqrt{2}}{3\theta_i^2}\left(\frac{M_P}{f_a}\right)^2\sqrt{\frac{\Gamma_a}{m_a}}\left(\frac{g_*(T_{\text{osc}})}{g_*(T_{\text{dec}})}\right)^{1/4}.
\end{align}
The tree-level results for the axion-decay rates associated with the three couplings considered in this work read, 
\begin{align}
        \Gamma_a^{WW}&=C_w^2\left(\frac{g_w^2}{32\pi^2}\right)^2\frac{3}{4\pi}\frac{m_a^3}{f_a^2}, \qquad \Gamma_a^{gg}=C_s^2\left(\frac{g_s^2}{32\pi^2}\right)^2\frac{2}{\pi}\frac{m_a^3}{f_a^2}, \nonumber \\ \Gamma_a^{\text{Majoron}}&=\frac{\delta_{B-L}^2}{12288\pi^5}\frac{m_a^5\overline{m}^2}{f_a^2v^4}.
\end{align}
It is worth noting that $\Gamma_a^{\text{Majoron}}$ is smaller than $\Gamma_a^{WW}$ and $\Gamma_a^{gg}$ by orders of magnitude for values of $m_a$ that result in non-negligible production of baryon asymmetry from oscillation-driven spontaneous LG. In fact, the requirement that the Majoron decays before BBN implies the following bound on the oscillation-temperature, or equivalently, the mass of the Majoron, 
\begin{align}
    T^{\text{Majoron-BBN}}_{\text{osc}}&\gtrsim 1.9\cdot 10^{13}\,\text{GeV}\left(\frac{f_a}{10^{15}\text{ GeV}}\right)^{1/5}\left(\frac{0.1\text{ eV}}{\overline{m}}\right)^{1/5} \,,\\
    m^{\text{Majoron-BBN}}_a&\gtrsim 4.5\cdot 10^{8}\,\text{GeV}\left(\frac{f_a}{10^{15}\text{ GeV}}\right)^{2/5}\left(\frac{0.1\text{ eV}}{\overline{m}}\right)^{2/5} \,.
    \label{eq:MajoronBBN}
\end{align}
Below, we list conditions for when entropy-dilution becomes relevant, the associated entropy-dilution factors, and the decoupling temperatures for the three different axion couplings:
\begin{align}
    \label{eq:WWDelta}
    \frac{a}{f}\frac{C_wg_w^2}{32\pi^2}W\widetilde{W}:&  \begin{cases}
m_a<3.7\cdot10^{11}\frac{\theta_i^2}{C_w}\left(\frac{0.55}{g_w}\right)^2\left(\frac{f_a}{10^{15}\,\text{GeV}}\right)^3\left(\frac{g_*(T_{\text{eq}})}{g_*(T_{{\text{osc}}})}\right)^{1/4}\,\text{GeV}, \\
\Delta_a'\simeq C_w\frac{2.6\cdot 10^{-3}}{\theta_i^2}\left(\frac{g_w}{0.55}\right)^2\left(\frac{10^{15}\,\text{GeV}}{f_a}\right)^3\left(\frac{m_a}{10^9\,\text{GeV}}\right)\left(\frac{g_*(T_{\text{osc}})}{g_*(T_{{\text{dec}}})}\right)^{1/4},\\
T_{\text{dec}}\simeq10.2C_w\left(\frac{106.75}{g_*(T_{\text{dec}})}\right)^{1/4}\left(\frac{g_w}{0.55}\right)^2\left(\frac{m_a}{10^9\,\text{GeV}}\right)^{3/2}\left(\frac{10^{15}\,\text{GeV}}{f_a}\right)\,\text{TeV},\end{cases} \\
\label{eq:GGDelta}
    \frac{a}{f}\frac{C_sg_s^2}{32\pi^2}G\widetilde{G}:&  \begin{cases}
        m_a< 1.9\cdot10^{11}\frac{\theta_i^2}{C_s}\left(\frac{0.6}{g_s}\right)^2\left(\frac{f_a}{10^{15}\,\text{GeV}}\right)^3\left(\frac{g_*(T_{\text{eq}})}{g_*(T_{{\text{osc}}})}\right)^{1/4}\,\text{GeV},\\
        \Delta_a'\simeq C_s\frac{5.1\cdot 10^{-3}}{\theta_i^2}\left(\frac{g_s}{0.6}\right)^2\left(\frac{10^{15}\,\text{GeV}}{f_a}\right)^3\left(\frac{m_a}{10^9\,\text{GeV}}\right)\left(\frac{g_*(T_{\text{osc}})}{g_*(T_{{\text{dec}}})}\right)^{1/4},\\
        T_{\text{dec}}\simeq 19.8C_s\left(\frac{106.75}{g_*(T_{\text{dec}})}\right)^{1/4}\left(\frac{g_s}{0.6}\right)^2\left(\frac{m_a}{10^9\,\text{GeV}}\right)^{3/2}\left(\frac{10^{15}\,\text{GeV}}{f_a}\right)\,\text{TeV},\end{cases} \\
        \label{eq:BLDelta}
    \frac{\partial_\mu a}{f}J^\mu_{B-L}:& \begin{cases} m_a<
    10^{13}\theta_i\left(\frac{f_a}{10^{15}\,\text{GeV}}\right)^{3/2}\sqrt{\frac{0.1\,\text{eV}}{\overline{m}}}\left(\frac{g_*(T_{\text{eq}})}{g_*(T_{{\text{osc}}})}\right)^{1/8}\,\text{GeV}, \\\Delta_a'
    \simeq \frac{9.5\cdot 10^{-9}}{\theta_i^2}\left(\frac{\overline{m}}{0.1\,\text{eV}}\right)\left(\frac{m_a}{10^9\,\text{GeV}}\right)^2\left(\frac{10^{15}\,\text{GeV}}{f_a}\right)^3\left(\frac{g_*(T_{\text{osc}})}{g_*(T_{{\text{dec}}})}\right)^{1/4},\\
    T_{\text{dec}}\simeq 70\left(\frac{10}{g_*}\right)^{1/4}\left(\frac{m_a}{10^9\,\text{GeV}}\right)^{5/2}\left(\frac{10^{15}\,\text{GeV}}{f_a}\right)\left(\frac{\overline{m}}{0.1\,\text{eV}}\right)\,\text{MeV},\end{cases}
\end{align}
where the gauge couplings entering $\Delta_a'$ are evaluated at $\mu=2\pi\,T_{\text{dec}}$.
Here, we clearly see that entropy dilution in the Majoron-like scenario is much more severe than the scenarios where the axion is coupled to strong or weak sphalerons.\footnote{For concreteness, recall that $m_a\propto T_{\text{osc}}^2$, and $T_{\text{osc}}=10^{12}$ GeV corresponds to $m_a\approx 4\cdot 10^6$ GeV.} In fact, by inspecting the results from Sec.~\ref{sec:baryonasymmetry}, it becomes clear that the dilution in the pure Majoron scenario drives the baryon asymmetry below its observed value if the following hierarchy of scales is realized $f_a\gtrsim T_{\rm{rh}} \gtrsim T_{\rm{osc}}\simeq \sqrt{M_P\,m_a}/g_*^{1/4}$, which is the regime of validity for the results presented throughout this paper. Hence, unless the Majoron has additional couplings that significantly reduce its lifetime, the ``vanilla'' case of Majoron-driven spontaneous LG via the Weinberg operator is ruled out. This observation is an important result of our paper.\footnote{Here, we are ignoring any additional lepton-number production from Majoron decay, for the following two reasons: First, in an effective framework with a single Weinberg operator---as considered in this paper---it is not possible to generate a non-zero $\epsilon_{CP}$ because CP violation requires interference between tree-level and loop-level diagrams with different physical phases. To calculate a non-zero $\epsilon_{CP}$, the Weinberg operator must be resolved into its UV completion (such as integrating in heavy right-handed neutrinos in a Type-I Seesaw mechanism) to evaluate the loop corrections properly. Second, lepton-number production from Majoron decay is only directly relevant for the BAU if it occurs before EW sphaleron-interactions freeze out. As can be seen from Eqs.~\eqref{eq:maTosc} and ~\eqref{eq:BLDelta}, this may only happen in a tiny region of the parameter space considered by us where $T_{\text{osc}}\sim 10^{14}$~GeV and $f_a\sim 10^{15}$ GeV.}

Let us compare our conclusion regarding the viability of ``vanilla'' Majoron-driven spontaneous LG via the Weinberg operator to related work in the literature. The authors of Ref.~\cite{Ibe:2015nfa}, e.g., the seminal paper on Majoron-driven spontaneous LG via the Weinberg operator, introduce higher-dimensional Planck-suppressed operators in order to achieve a shorter Majoron lifetime. It would be straightforward to combine our analysis with such an extension of the model. Meanwhile, the authors of Ref.~\cite{Chun:2023eqc} consider Majoron-driven spontaneous LG in the presence of on-shell RHNs in the thermal plasma, i.e., the extension of our scenario to lower RHN masses, $M_N \sim 10^{11\cdots13}\,\textrm{GeV}$. In this case, successful baryogenesis requires a Majoron mass in violation of the lower bound in Eq.~\eqref{eq:MajoronBBN}, which again rules out the simplest version of the model. The authors propose two solutions to this issue: first, one may introduce additional couplings between the Majoron and other fields, in order to make it more short-lived; second, one may break the parametric relation between the axion mass $m_a$ and its velocity $\dot\theta$ implied by Eq.~\eqref{eq:theta} by resorting to the kinetic misalignment mechanism~\cite{Co:2019jts}. This second solution is also adopted by the authors of Ref.~\cite{Takahashi:2026ngu}, who combine spontaneous LG based on the kinetic misalignment mechanism with the production of asymmetric dark matter (which in turn is related to the idea of asymgenesis~\cite{Mojahed:2025exj}, i.e., the co-genesis of asymmetric dark matter and the BAU in the context of wash-in LG). Meanwhile, for a scenario of Majoron-driven spontaneous LG in which the Majoron itself assumes the role of dark matter, see Ref.~\cite{Wada:2024cbe}.

The authors of Ref.~\cite{Chun:2025abp} also build on the work in Ref.~\cite{Chun:2023eqc} and complement the baryon asymmetry produced via spontaneous LG with a standard thermal LG contribution. Finally, it is worth mentioning that the authors of Ref.~\cite{Mojahed:2026eos} consider a scenario where the axion is the Goldstone boson of a spontaneously broken horizontal flavor symmetry. This scenario retains the appealing feature of the Majoron, namely, that it generates a direct coupling between the axion velocity and the Weinberg operator. However, it also generates extra couplings, which allow the axion to decay much earlier than the Majoron.

\paragraph{Baryonic isocurvature perturbations:} If the global symmetry giving rise to the axion is already spontaneously broken during inflation, i.e., in a so-called pre-inflationary axion scenario with an axion that is light during inflation, the initial misalignment angle $\theta_i$ acquires quantum fluctuations during inflation with $\delta\theta_i\simeq H_I/(2\pi f_a)$. Since $Y_B\propto \theta_i$ in the mechanism under study,\footnote{We assume an initial misalignment such that the generated asymmetry is linear in $\theta_i$, i.e., an initial phase away from a hilltop in the periodic potential, where anharmonic effects can become important~\cite{Kobayashi:2013nva}.} this induces baryonic isocurvature perturbations with power spectrum $\mathcal{P}_{B\gamma}\simeq(H_I/(2\pi f_a\theta_i))^2$. This baryonic isocurvature maps onto an effective cold-dark-matter (CDM) isocurvature suppressed by $(\Omega_b/\Omega_{\text{CDM}})^2\approx 0.035$. The PLANCK 2018~\cite{Planck:2018vyg} bound on uncorrelated CDM isocurvature 
then implies, 
\begin{align}
    \mathcal{P}_{B\gamma}\lesssim 2.4\cdot 10^{-9} \,,
\end{align}
which constrains the ratio of the axion decay constant and the inflationary Hubble scale, 
\begin{align}
    \label{eq:isocurvbound}
    \frac{H_I}{f_a}\lesssim 3\cdot 10^{-4}\,\theta_i.
\end{align}

\paragraph{Backreaction in the axion EoM:} As discussed in Sec.~\ref{sec:EoM}, our analysis is based on the classical EoM of the homogeneous axion background field. This approximation neglects the axion--SM coupling, which can potentially induce a relevant backreaction term in Eq.~\eqref{EoM}. In the case of the coupling to strong sphalerons, for instance, the $a\,G\widetilde{G}$ operator in the Lagrangian induces an additional friction term in the axion EoM, which then reads~\cite{McLerran:1990de}
\begin{align}
\ddot{a}+3H\,\dot{a}+\partial_a V  + \frac{C_s}{f}\frac{g_s^2}{32\pi^2} \left<G^a_{\mu\nu}\widetilde{G}^{a\mu\nu}\right> = 0\,.
\end{align}
In the thermal background and in the presence of nonvanishing chemical potentials for the SM quarks, the expectation value of the $G\widetilde{G}$ operator can be expressed as follows~\cite{Domcke:2020kcp},
\begin{small}
\begin{align}
\label{eq:EoMfriction}
\ddot{a}+3H\,\dot{a}+\partial_a V + \frac{C_s}{f}\,\Gamma_{\textrm{SS}}
\left(C_s \eta - \frac{-3\mu_{q_{uds}} - \mu_t-\mu_c - \mu_b+ 2\mu_{Q_1}+2\mu_{Q_2} + 2\mu_{Q_3}}{T}\right) = 0 \,,
\end{align}
\end{small}\par\noindent
where $\Gamma_{\rm SS}$ is the interaction density of the strong sphaleron processes~\cite{Moore:2010jd} (see App.~\ref{appendix:BEq}),
\begin{align}
\Gamma_{\rm SS} = \frac{\kappa_{\rm SS}\,\alpha_s^5}{2}\,T^4 \,, \qquad \kappa_{\rm SS} \simeq 270 \,, \qquad \alpha_s = \frac{g_3^2}{4\pi}  \,, \qquad g_3 \simeq 0.60 \,.
\end{align}

The Boltzmann equations listed above depend on $\Gamma_{\rm SS}$ through the transport coefficient $\gamma_{\rm SS} = 6\,\Gamma_{\rm SS}/T^3$. In particular, in order to assess the relevance of strong sphalerons in the Boltzmann equations, we need to compare $\gamma_{\rm SS}$ to the Hubble rate $H$. In the axion EoM, however, the situation is slightly different. Now, in order to compare the sphaleron friction term to the Hubble friction term, we need to identify the respective coefficients of $\dot{a}$, 
\begin{align}
\ddot{a}+3H\,\dot{a}+\partial_a V + \Upsilon_{\rm SS}\,\dot{a} + \cdots = 0 \,, \qquad \Upsilon_{\rm SS} = \frac{C_s^2}{f^2T}\,\Gamma_{\textrm{SS}} \,.
\end{align}
where the ellipsis stands for the contributions to the axion EOM from the chemical-potential terms in Eq.~\eqref{eq:EoMfriction}. In the presence of nonvanishing chemical potentials, these terms can reduce the impact of sphaleron friction; see Refs.~\cite{Drewes:2023khq,Berghaus:2025dqi} for recent work on this point. The effect of nonzero chemical potentials can notably be accounted for by replacing the dissipation rate $\Upsilon_{\rm SS}$ by an effective (and potentially heavily suppressed) rate $\Upsilon_{\rm SS}^{\rm eff}$. 

For our purposes, it is, however, not necessary to enter the discussion about the distinction between $\Upsilon_{\rm SS}$ and $\Upsilon_{\rm SS}^{\rm eff}$. Our intention in this subsection is merely to confirm that the omission of the sphaleron friction term in the axion EoM in Eq.~\eqref{EoM} was justified. In particular, we are interested in strong lower limits on the axion decay constant $f$ ensuring inefficient sphaleron friction under all circumstances, no matter if the axion has already begun oscillating, chemical potentials have already begun to build, or chemical equilibrium has been reached. To this end, it is sufficient to compare $\Upsilon_{\rm SS}$ rather than $\Upsilon_{\rm SS}^{\rm eff}$ to $H$,
\begin{align}
\frac{\Upsilon_{\rm SS}}{H} = \left(\frac{90}{\pi^2 g_*}\right)^{1/2}\frac{\kappa_{\rm SS}\,\alpha_s^5}{2} \frac{M_PT}{f^2/C_s^2} \,.
\end{align}
This ratio turns out to be suppressed, $\Upsilon_{\rm SS} / H \ll1 $, across the entire parameter region relevant for spontaneous LG. In particular, setting the temperature $T$ to its maximally allowed value, $T \sim f$, we find the following lower limit on the axion decay constant
\begin{align}
\frac{f}{C_s^2} \gtrsim \left(\frac{90}{\pi^2 g_*}\right)^{1/2}\frac{\kappa_{\rm SS}\,\alpha_s^5}{2} M_P\simeq 1.8\times 10^{12}\,\textrm{GeV} \,.
\end{align}
For smaller temperatures, $T \ll f$, the lower bound is obviously weaker. In view of these results, we conclude that sphaleron heating, even in its maximal form and without any suppression by the presence of nonzero chemical potentials, has no chance of playing a role in the axion EoM for the regions of parameter space considered in this work. 

Finally, it is trivial to carry our results for strong sphalerons over to the case of weak sphalerons. In this case, we find an even weaker lower limit on the axion decay constant,
\begin{align}
\frac{f}{C_w^2} \gtrsim \left(\frac{90}{\pi^2 g_*}\right)^{1/2}\frac{\kappa_{\rm WS}\,\alpha_w^5}{2} M_P\simeq 6.8\times 10^{10}\,\textrm{GeV} \,.
\end{align}
In the case of our third model, the coupling to the $B\!-\!L$ current, no nonperturbative dynamics are involved. The expectation value of $\partial_\mu J_{B-L}^\mu$ in the axion EoM can be simply expressed in terms of the expectation value of the source of $B\!-\!L$ violation in our model, i.e., the expectation value of the Weinberg operator. Above the electroweak phase transition, this expectation value, however, vanishes, which means that the Majoron--Weinberg coupling in the Lagrangian introduces no additional friction term in the axion EoM. 

\subsection{Final baryon asymmetry}
\label{sec:FinalParameter}
In this section, we present the final baryon asymmetry after entropy dilution from axion decays and show the resulting isocurvature constraints. As noted above, the Majoron's suppressed decay rate makes it decay either after the time of BBN, or before BBN but after dominating the Universe's energy density long enough that its eventual decay causes severe entropy dilution. This dilution suppresses the baryon asymmetry below the observed value for any axion parameters satisfying $f_a\gg T_{\text{osc}}$. However, should the Majoron acquire additional couplings that significantly shorten its lifetime, its efficient production of a lepton asymmetry, as demonstrated in Sec.~\ref{sec:baryonasymmetry}, would allow it to explain the BAU for oscillation temperatures $T_{\text{osc}}\gtrsim 10^{11}$ GeV. 

Due to the highly suppressed final baryon asymmetry in the Majoron scenario, we will focus exclusively on the axion couplings to strong and weak sphalerons in the rest of this section. Here, we will determine the regions in the $T_{\text{osc}}$--$f_a$ planes that are compatible with spontaneous LG. Below, we show results for normal mass ordering, while additional results for inverted mass ordering are included in App.~\ref{appendix:NumericResults}.\\ 

\begin{figure}[]%
    \centering
    {\includegraphics[width=0.499\textwidth]{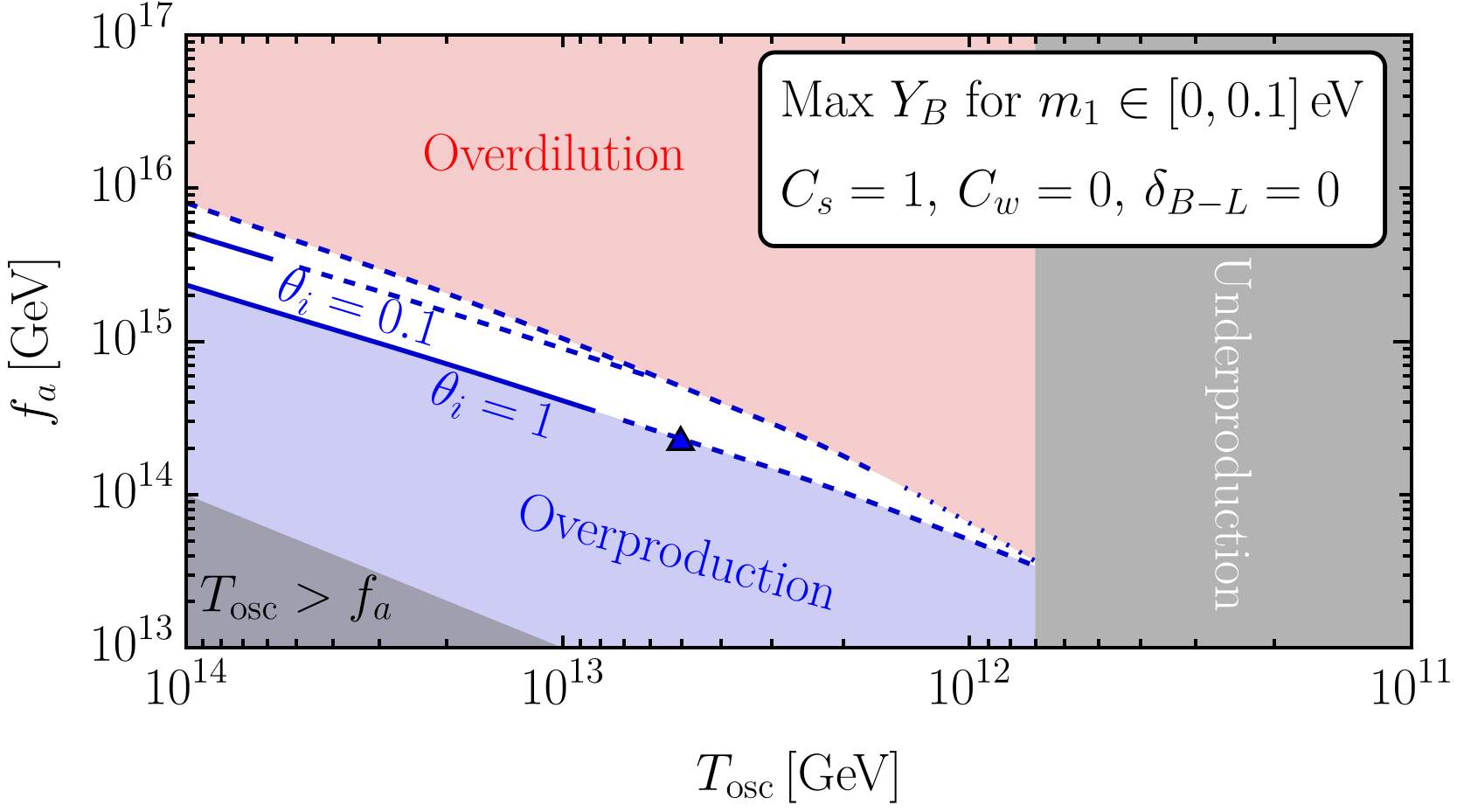}}\hfill
    {\includegraphics[width=0.499\textwidth]{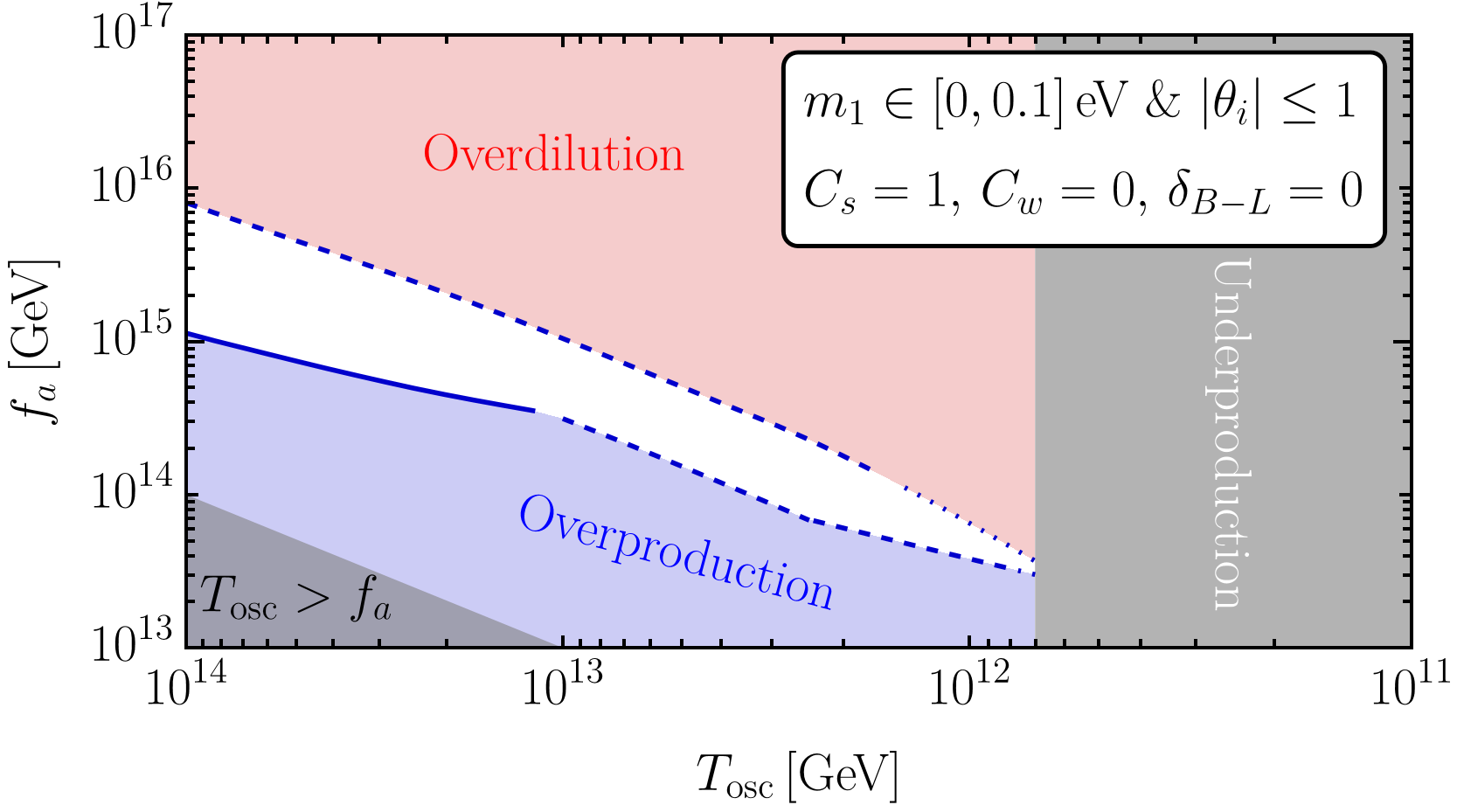}}\hfill
    {\includegraphics[width=0.499\textwidth]{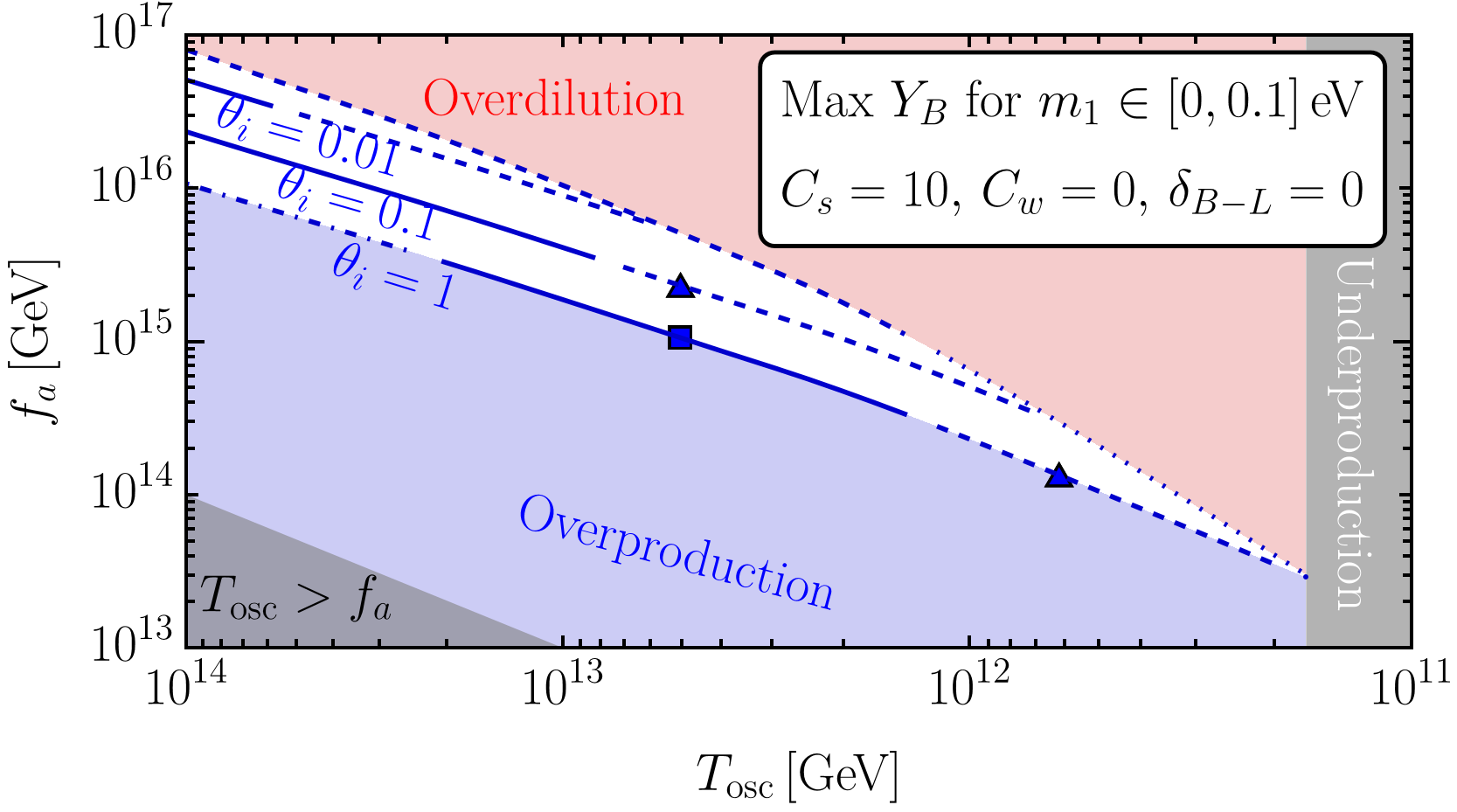}}\hfill
    {\includegraphics[width=0.499\textwidth]{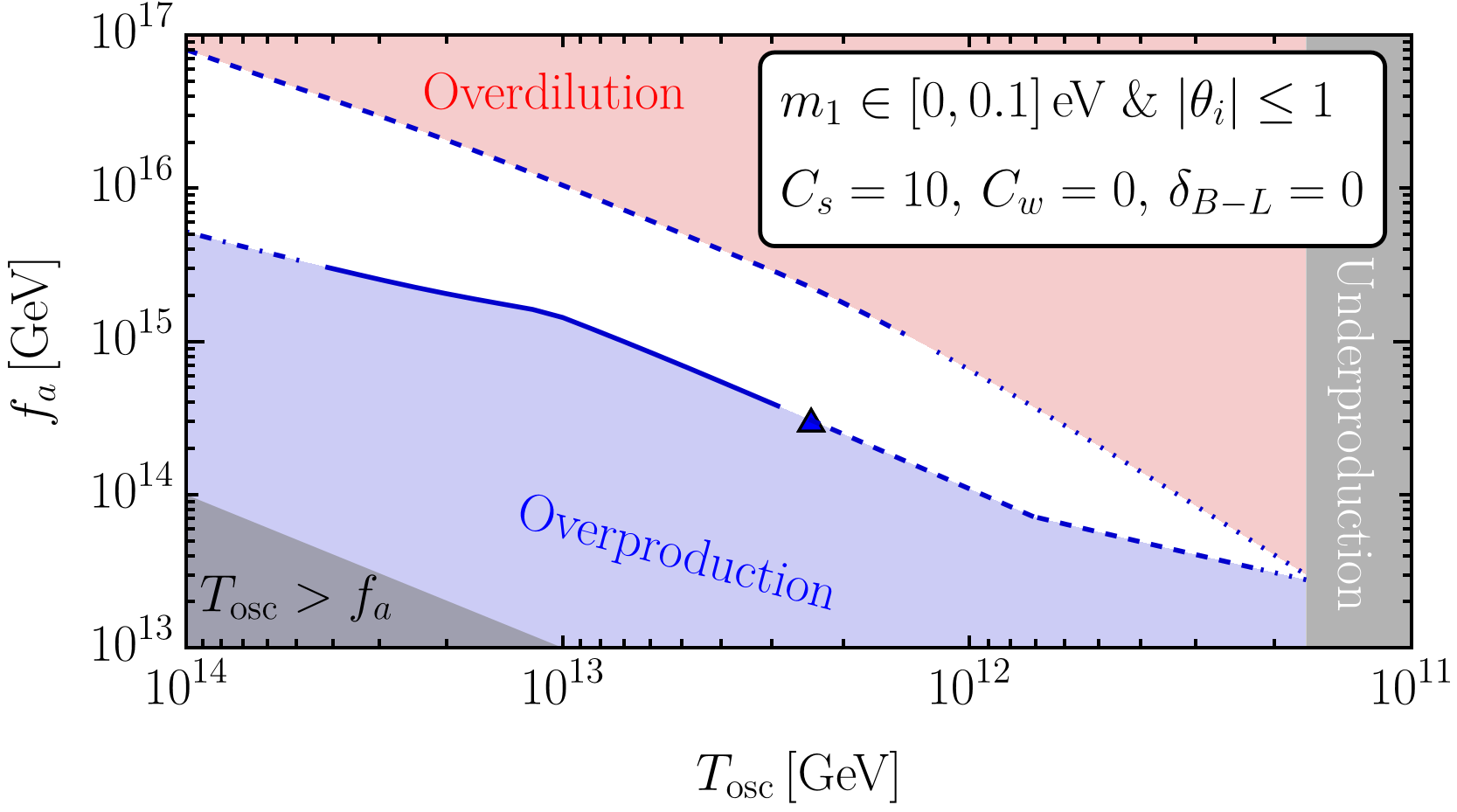}}\hfill
    {{\includegraphics[width=0.499\textwidth]{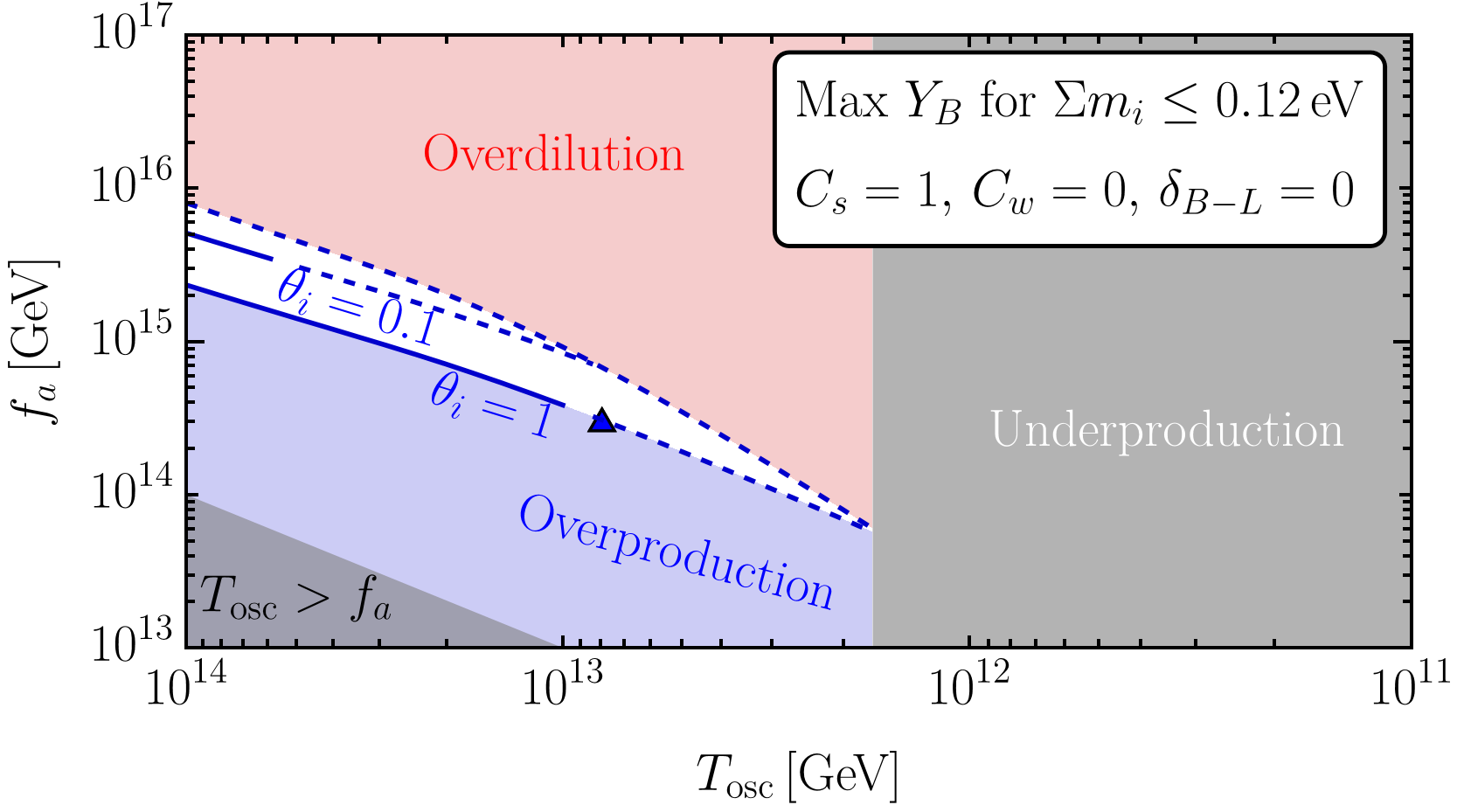}}}\hfill
    {{\includegraphics[width=0.499\textwidth]{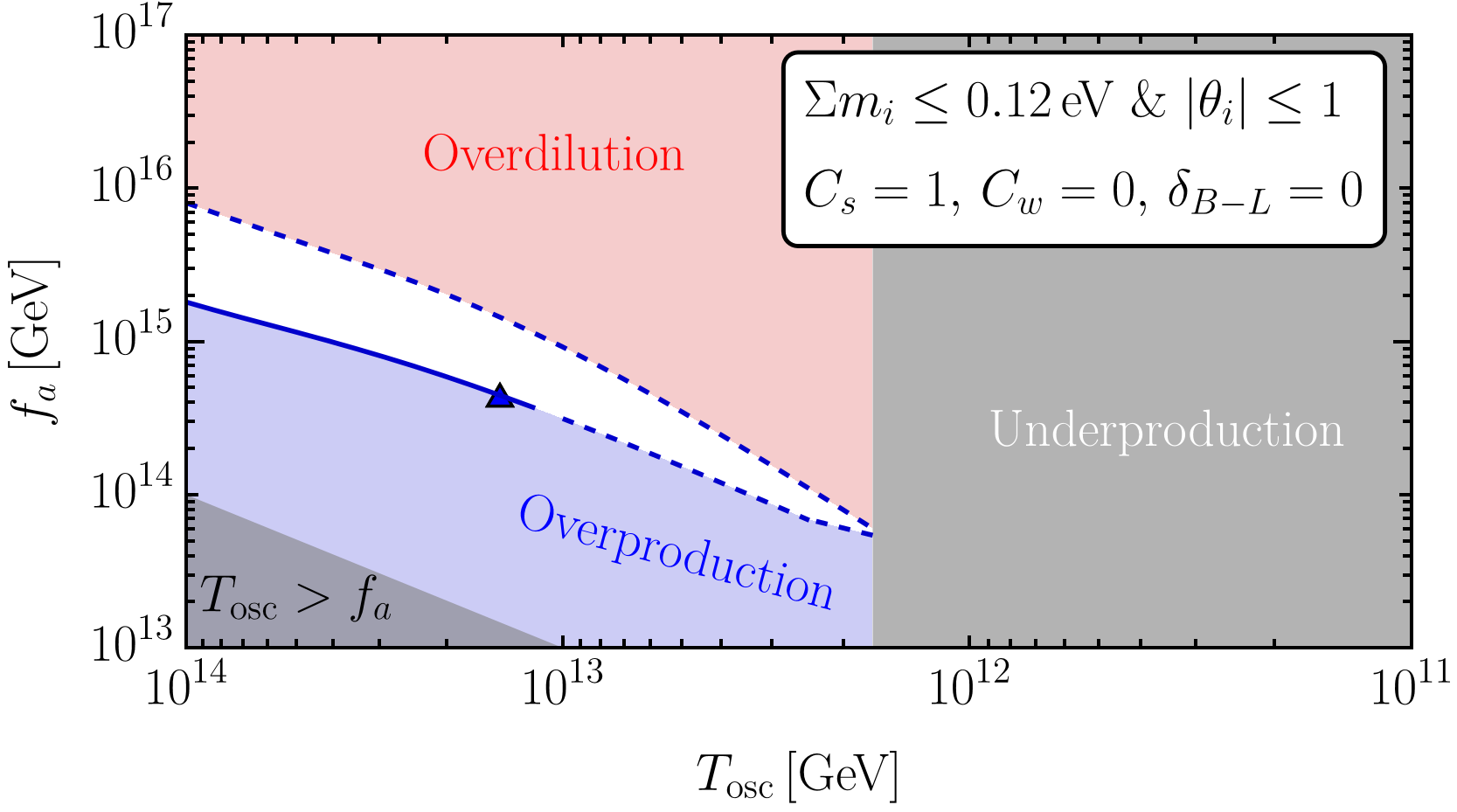}}}\hfill
    {{\includegraphics[width=0.499\textwidth]{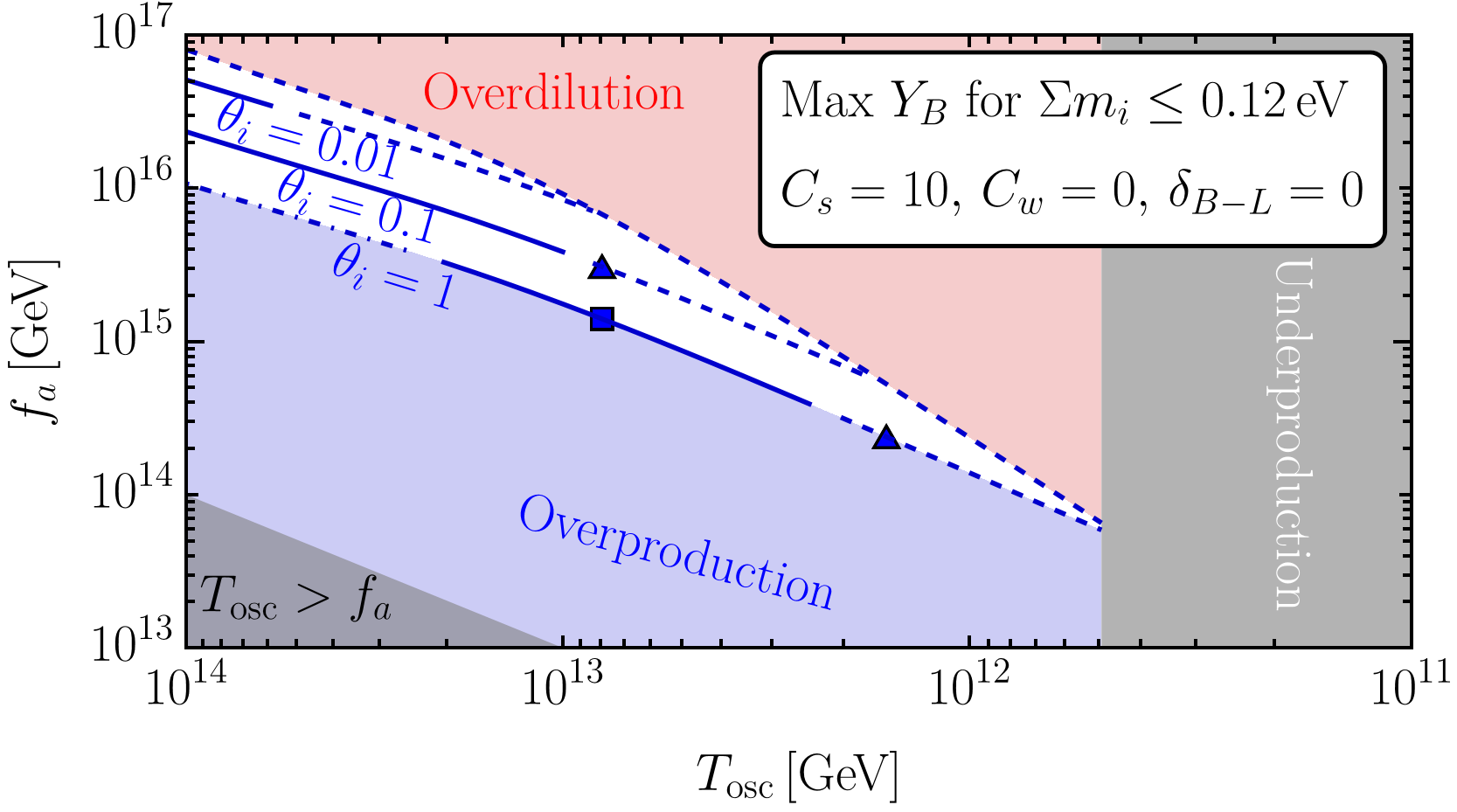}}}\hfill
    {{\includegraphics[width=0.499\textwidth]{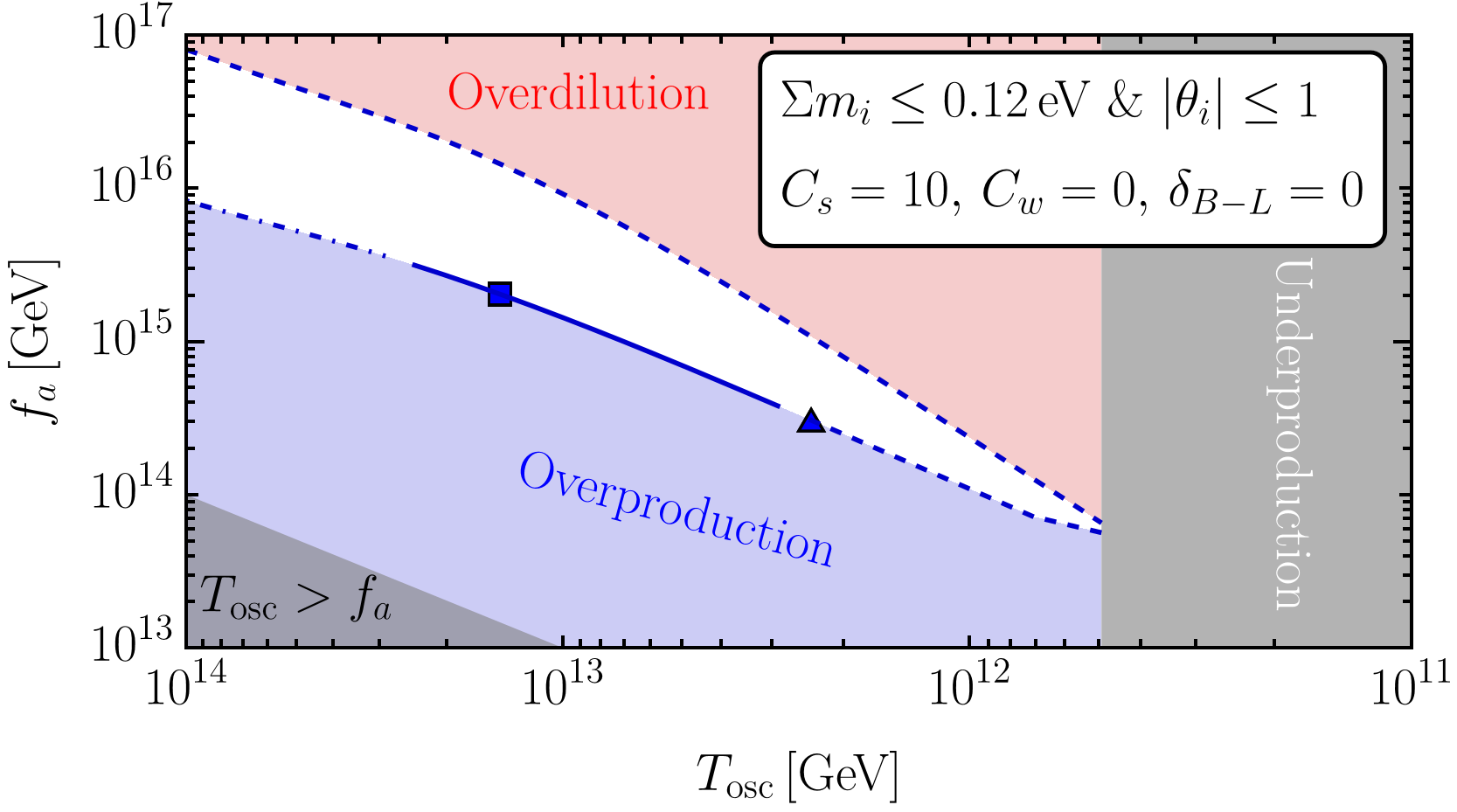}}}\hfill
    \caption{Panels illustrating the $T_{\text{osc}}$--$f_a$ plane for various (nonzero) values of $C_s$ and $m_1$, as indicated in the respective labels. Regions corresponding to BAU overproduction are shaded in blue. Areas of BAU underproduction are shaded in gray if the initial asymmetry production falls below the observed value, and in red if entropy dilution from axion decays drives the final BAU below the observed value. The white regions denote the parameter space where the observed BAU can be successfully reproduced, while the blue lines delineate solutions for select fixed values of $\theta_i$. Baryonic isocurvature constraints are represented by dotted, dashed, solid, and dot-dashed lines, which correspond to the intervals $3 \times 10^{-4} \, \theta_i \, (f_a / \text{GeV}) \in (10^9, 10^{10})$, $(10^{10}, 10^{11})$, $(10^{11}, 10^{12})$, and $(10^{12}, 10^{13})$, respectively (cf.\ the isocurvature bound in Eq.~\eqref{eq:isocurvbound}). Triangles and boxes indicate $\Delta_a=0.1$ and $\Delta_a=0.01$, respectively. See the main text for details.}
    \label{fig:SSParameterSpace}
\end{figure}

\paragraph{Axion coupled to \boldmath{$G\widetilde{G}$}:} Our final results for axions coupled to $G\widetilde{G}$ are summarized in Fig.~\ref{fig:SSParameterSpace}, which illustrates the allowed regions in the $T_{\text{osc}}$--$f_a$ plane. The anomaly coefficient is fixed to $C_s=1$ ($10$) in the odd (even) rows. In the first column, $m_1$ is chosen to maximize the produced asymmetry within the ranges $m_1 \in [0, 0.1]$~eV (first and second rows) and $\sum m_i \leq 0.12$~eV (third and fourth rows). Specifically, the values of $m_1$ as a function of $T_{\rm osc}$ in the left-most panels correspond to those that determine the upper edges of the blue and red bands in the lower-left panel in Fig.~\ref{fig:Band plot}.\footnote{Analogous results for $m_1$ chosen to minimize the produced asymmetry within the constraints $m_1 \in [0, 0.1]$~eV and $\sum m_i \leq 0.12$~eV, corresponding to the lower edges of the blue and red bands in the lower-left panel in Fig.~\ref{fig:Band plot}, are provided in App.~\ref{appendix:FinalParameterSpace}.} In the second column, we present the total available parameter space for a fixed $C_s$, treating the initial misalignment angle $|\theta_i| \in [0, 1]$ as a free parameter and allowing $m_1$ to vary between zero and the upper bounds specified in the panel labels. The white region denotes the parameter space where the BAU can be successfully reproduced for some $\theta_i \in [0, 1]$, whereas the blue lines illustrate solutions for select fixed values of $\theta_i$. The uppermost blue line is determined by fixing $\theta_i$ to reproduce the BAU via $Y_{B-L}^{\text{sLG}}(\theta_i)$ and subsequently identifying the maximal $f_a$ that yields a dilution factor $\Delta_a \simeq 1$. Furthermore, blue triangles and squares on contour lines indicate where $\Delta_a = 0.1$ and $\Delta_a = 0.01$, respectively. Isocurvature constraints are encoded in the line styles: dotted, dashed, solid, and dot-dashed lines correspond to the intervals $3 \times 10^{-4} \, \theta_i \, (f_a / \text{GeV}) \in (10^9, 10^{10})$, $(10^{10}, 10^{11})$, $(10^{11}, 10^{12})$, and $(10^{12}, 10^{13})$, respectively. The inflationary Hubble scale $H_I$ is then bounded from above by these values.

Because the dilution factor scales as $\Delta_a' \propto f_a^{-3}$, larger values of $f_a$ induce stronger entropy dilution; consequently, the red region delineates the parameter space where the baryon asymmetry is diluted below the observed BAU for any values of $(T_{\text{osc}},\,\theta_i)$ for which $Y_{B-L}^{\text{sLG}}(T_{\text{osc}},\,\theta_i)\geq Y_{B-L}^{\text{obs}}$. Conversely, decreasing $f_a$ results in baryon overproduction within the blue-shaded regions. These parameter regions are not necessarily ruled out by our analysis. If combined with late-time entropy injection in another sector, these regions might become phenomenologically viable, after all. In the gray region on the right, the initial asymmetry production is insufficient to account for the BAU, even in the limit of negligible entropy dilution from axion decays. Finally, the starting assumption $f_a \gg T_{\text{osc}}$ (see also Eq.~\eqref{eq:f>}) breaks down near the gray region in the lower-left corner of the panels.

The boundary of the gray region is determined by the numerical solution to the Boltzmann equations detailed in Sec.~\ref{sec:3} and is the result of the interplay among the axion velocity, SM interactions, and LNV processes. Meanwhile, to the left of this gray boundary, it is instructive to examine the parametric dependence of the final asymmetry, $Y_{B-L} = \min(1, \Delta'_a) Y_{B-L}^{\text{sLG}}$, on the axion parameters. For a fixed value of $T_{\text{osc}}$, or equivalently, a fixed value for the axion mass, the asymmetry scales as $Y_{B-L}^{\text{sLG}} \propto \theta_i C_s$, while the dilution factor scales as $\Delta_a' \propto C_s \theta_i^{-2} f_a^{-3}$. This scaling behavior facilitates the mapping of constant-$\theta_i$ contours derived for a specific value of $C_s$ onto a parameter plane with a different $C_s$, provided the neutrino-mass parameters are held constant. For instance, a contour $f_a(T_{\text{osc}}, \theta_i)$ obtained for $C_s=1$ translates to the contour for $0.1 \theta_i$ and $C_s=10$ via a straightforward rescaling to $10 f_a(T_{\text{osc}}, \theta_i)$, as can be verified by comparing the contours across the left-most panels.\footnote{This statement implicitly assumes the system is in the regime where $\Delta_a < 1$.}

For a fixed $C_s$, the constant-$\theta_i$ contours derived under the constraints $m_1 \in [0, 0.1]$~eV and $\sum m_i \leq 0.12$~eV coincide for $T_{\text{osc}} \gtrsim 2 \times 10^{13}$~GeV when $m_1$ is selected to maximize the final baryon asymmetry. An analogous behavior is observed for $T_{\text{osc}} \lesssim 10^{13}$~GeV when $m_1$ is instead chosen to minimize the baryon asymmetry (cf.\ Fig.~\ref{fig:SSParameterSpaceMin}). This correspondence stems directly from the solutions to the relevant Boltzmann equations, as illustrated in the lower-left panel in Fig.~\ref{fig:Band plot}, where the upper (lower) edges of the blue and red bands are seen to overlap within these respective temperature regimes.

The right column of Fig.~\ref{fig:SSParameterSpace} illustrates that for neutrino masses satisfying the PLANCK 2018 cosmological bounds and an anomaly coefficient $C_s \sim \mathcal{O}(1\cdots10)$, the mechanism under consideration can successfully account for the BAU within the ranges $10^{15}$~GeV $\lesssim f_a \lesssim 10^{17}$~GeV and $T_{\text{osc}} \gtrsim \text{a few} \times 10^{11}$~GeV. These viable regions can be further extended in models that accommodate larger values of $m_1$, as demonstrated in the first two rows. Moreover, broader regions of the parameter space become accessible in scenarios where the anomaly coefficient $C_s$ is permitted to take on larger values, as both the initial asymmetry $Y_{B-L}^{\text{sLG}}$ and the dilution factor $\Delta_a'$ scale linearly with $C_s$. 

Finally, by studying the pattern of the contours shown in the various panels, it can be seen that suppressing baryonic isocurvature perturbations below the current observational bound while retaining a hierarchy $T_{\text{rh}}\gg T_{\text{osc}}$ requires an efficient period of reheating after inflation. 
In particular, if we parametrize the efficiency of reheating after inflation with a number $c$ via $H_I=c\sqrt{\pi^2g_*/90}\,T_{\text{rh}}^2/M_P$, where $c=1$ corresponds to instantaneous reheating, we see that for the range of $C_s$ and $m_1$ considered here, successful spontaneous LG typically requires $c\lesssim 100$. 
In this context, warm inflation~\cite{Berera:1995ie,Kamali:2023lzq} with a high inflationary scale may provide an interesting framework that could provide ideal initial conditions for spontaneous LG. We leave a more detailed study of this direction for future work. 

\begin{figure}
    \centering
    {\includegraphics[width=0.5\textwidth]{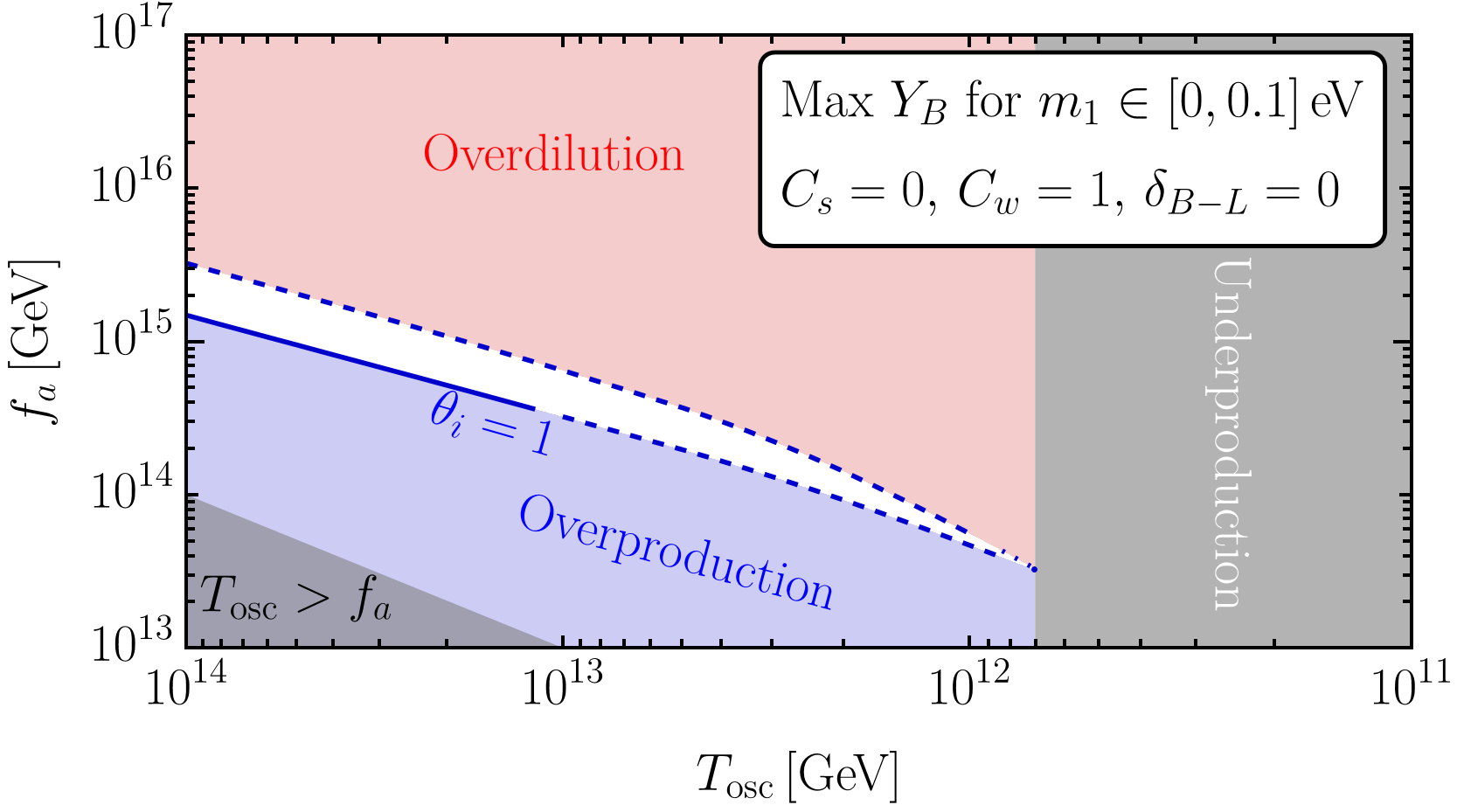}}\hfill
    {\includegraphics[width=0.5\textwidth]{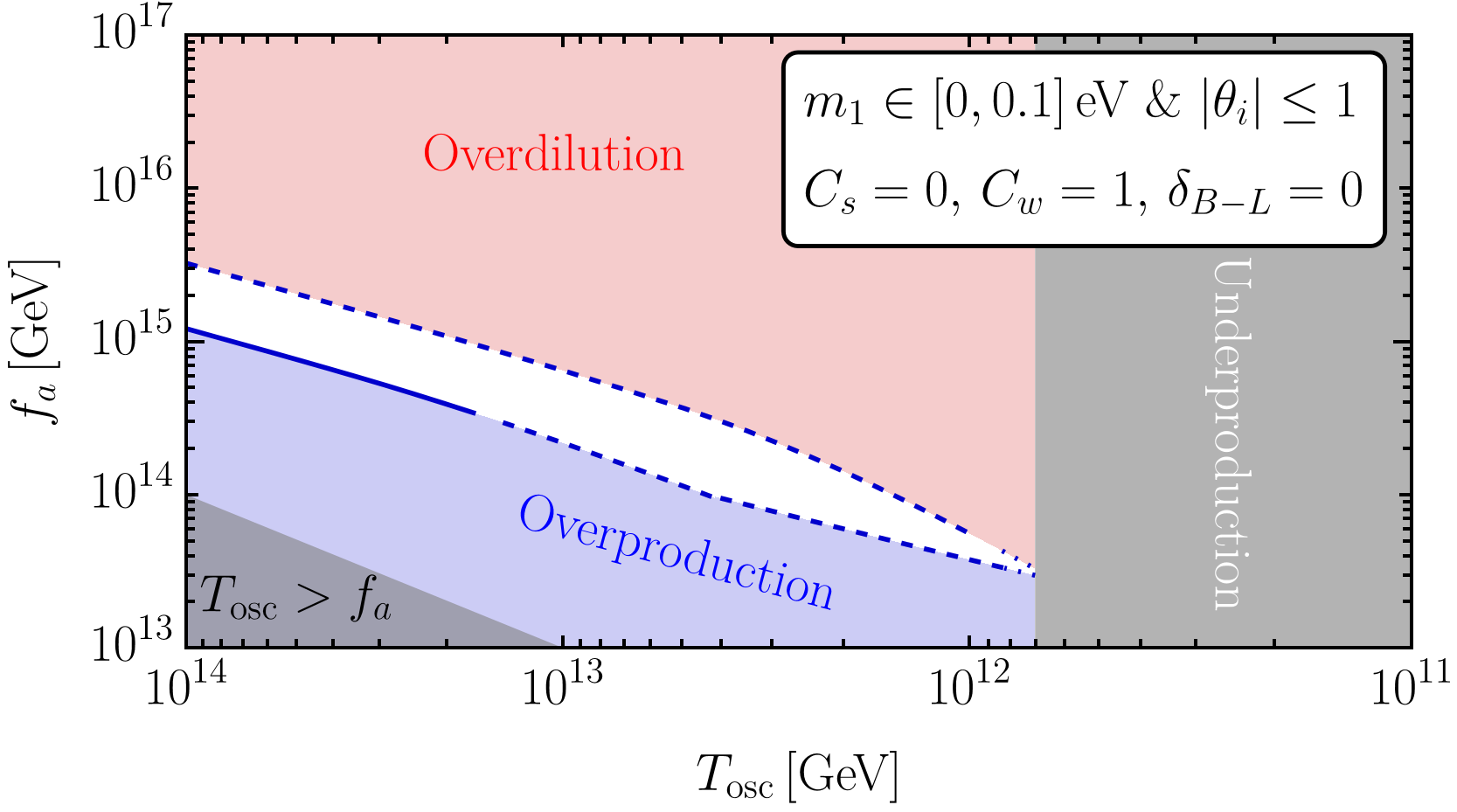}}\hfill
    {\includegraphics[width=0.5\textwidth]{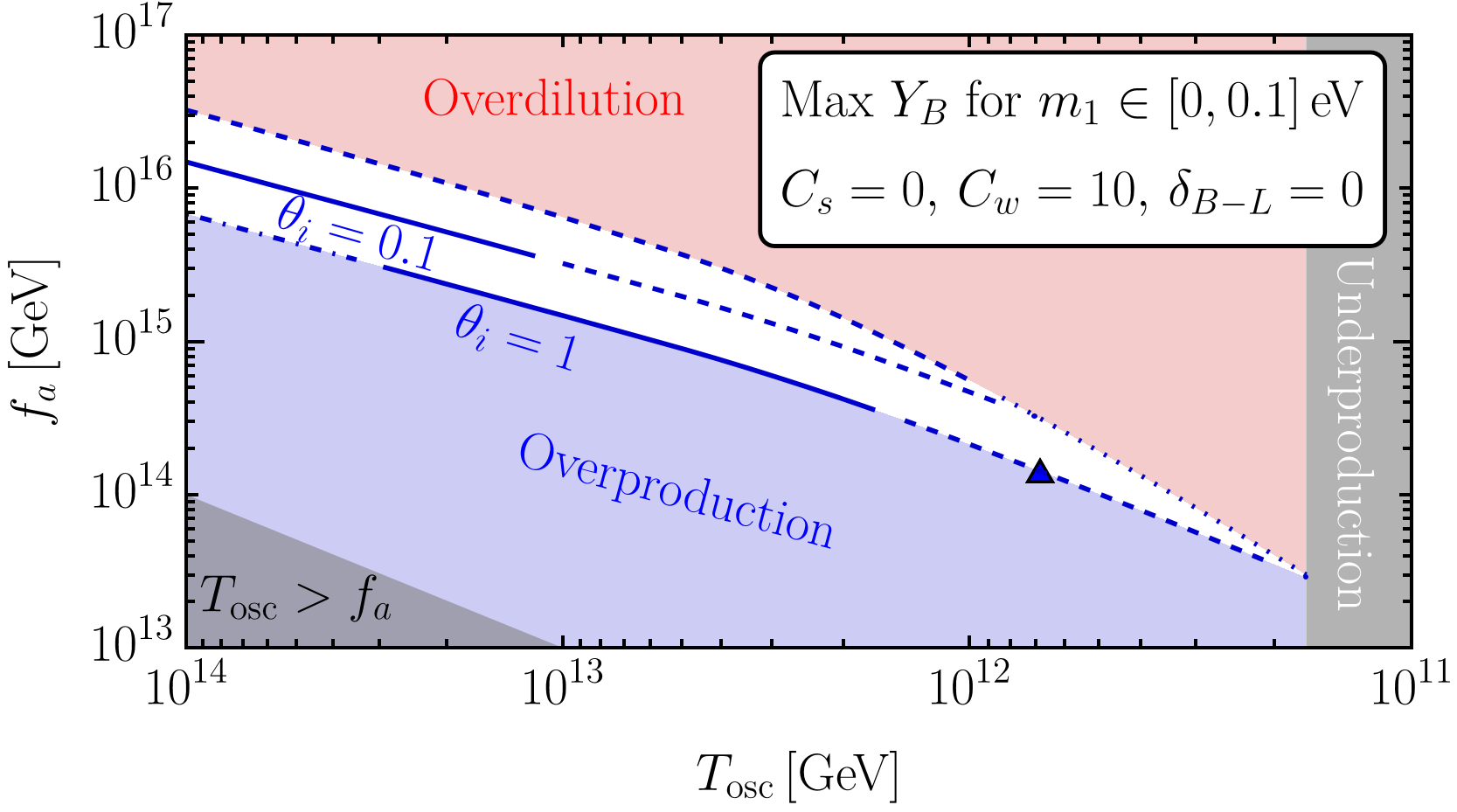}}\hfill
    {\includegraphics[width=0.5\textwidth]{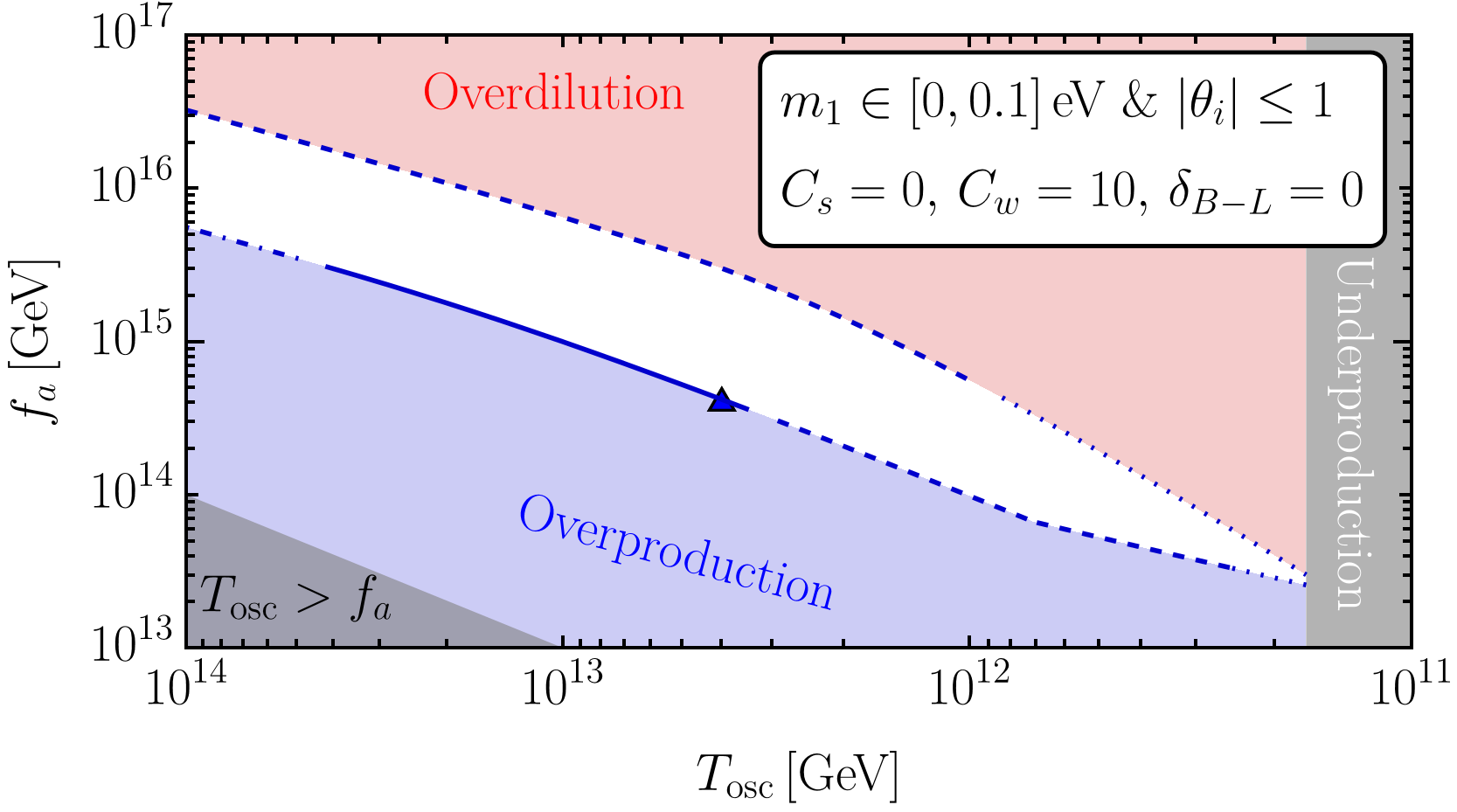}}\hfill
    {{\includegraphics[width=0.5\textwidth]{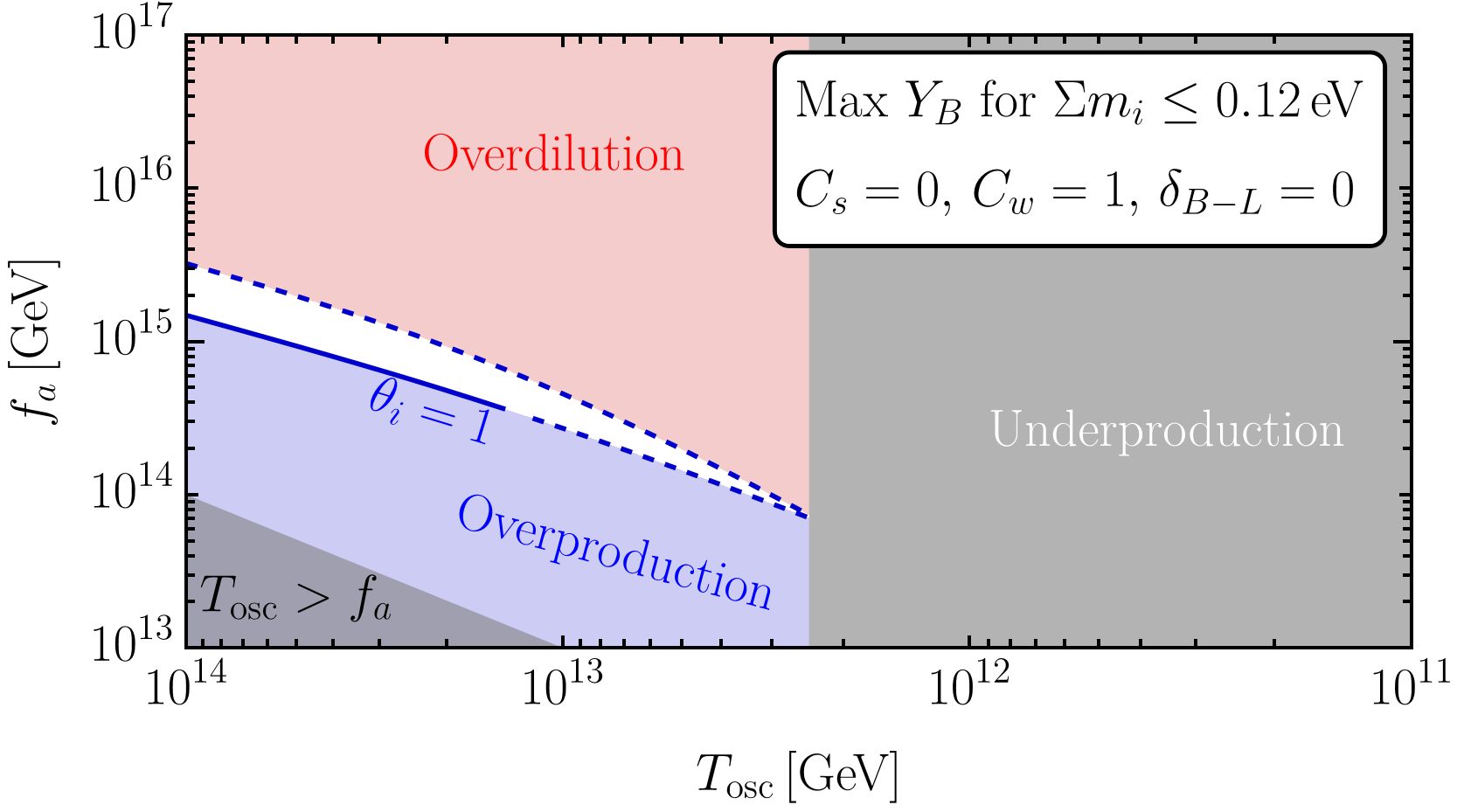}}}\hfill
    {{\includegraphics[width=0.5\textwidth]{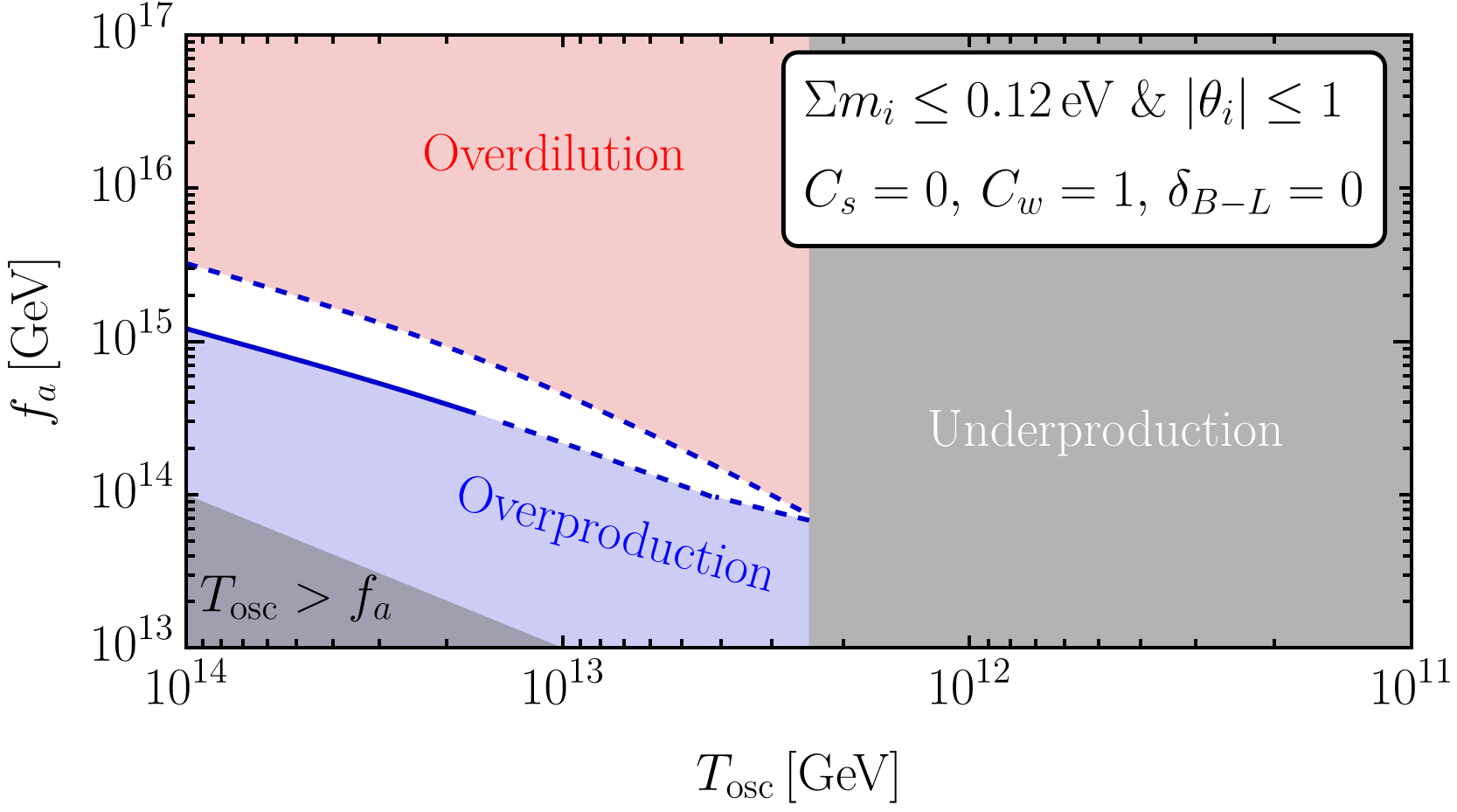}}}\hfill
    {{\includegraphics[width=0.5\textwidth]{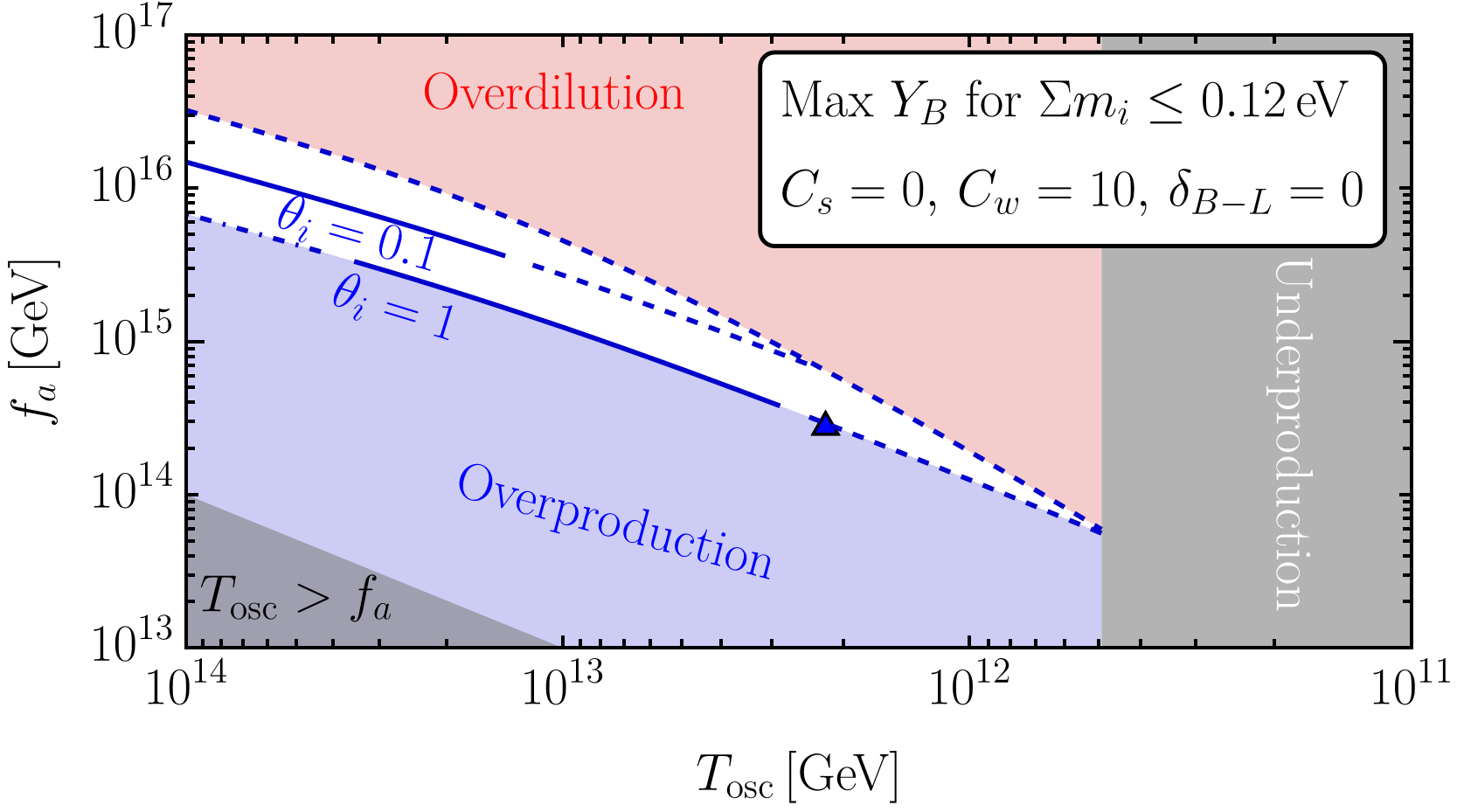}}}\hfill
    {{\includegraphics[width=0.5\textwidth]{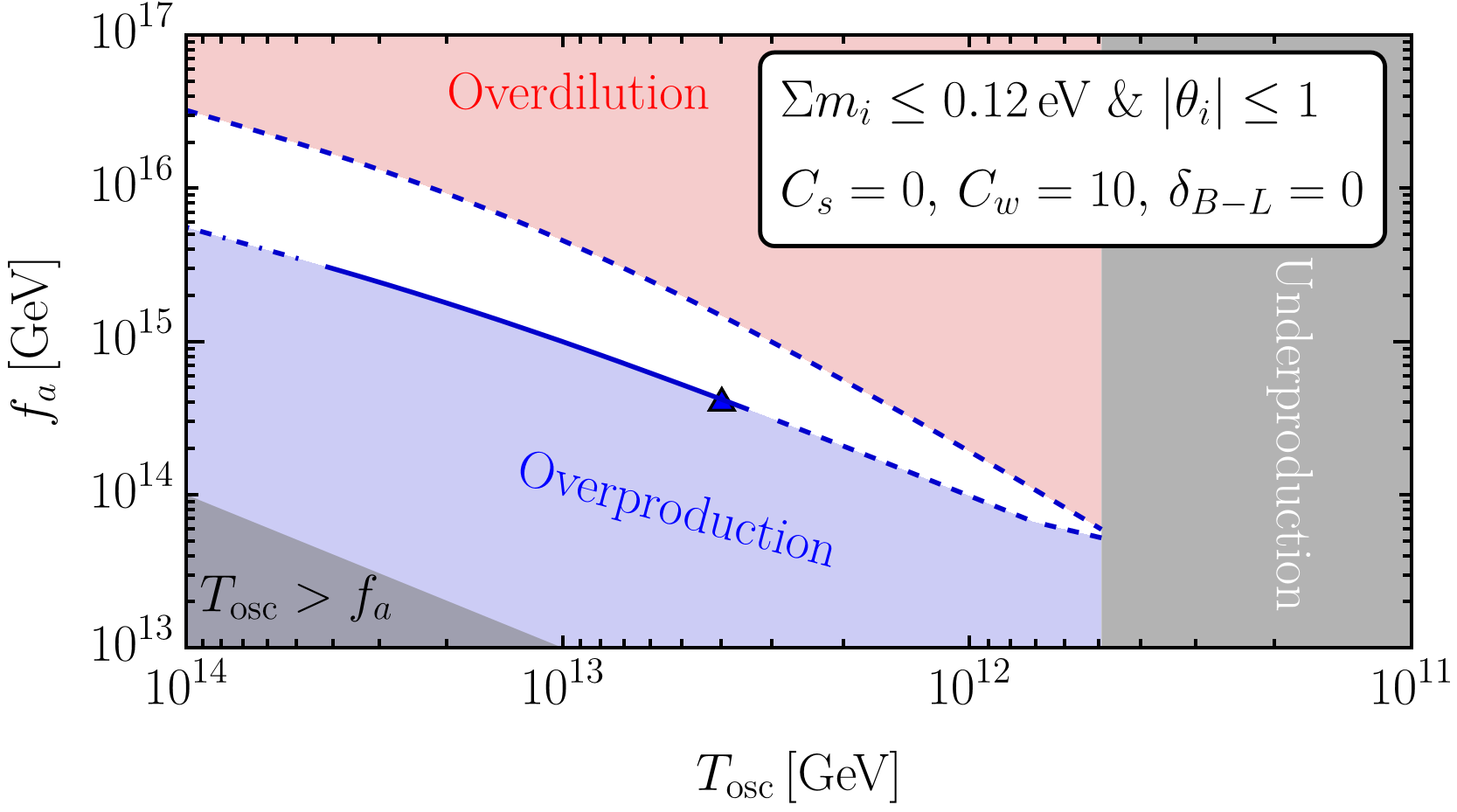}}}\hfill
    \caption{Similar to Fig.~\ref{fig:SSParameterSpace}, but with $C_s=0$ and $C_w \neq 0$ as shown in the respective labels. }
    \label{fig:WSParameterSpace}
\end{figure}

\paragraph{Axion coupled to \boldmath{$W\widetilde{W}$}:} Our final results for axions coupled to $W\widetilde{W}$ are summarized in Fig.~\ref{fig:WSParameterSpace}. This figure is structured similarly to Fig.~\ref{fig:SSParameterSpace}, but assumes nonzero $C_w$ rather than $C_s$. Because both $Y_{B-L}^{\text{sLG}}$ and $\Delta_a'$ scale with $\theta_i$, $f_a$, and the anomaly coefficient like in the strong-sphaleron case, the parameter spaces in the two figures share qualitative features. Moreover, the weak-sphaleron coupling accommodates a range of $T_{\rm osc}$ values comparable to that of an axion coupled to $G\widetilde{G}$. However, the maximum viable value of $f_a$ (associated with $T_{\text{osc}}\simeq 10^{14}$~GeV) is somewhat lower in this scenario. This reduction stems from a lower initial asymmetry production, which is a direct consequence of the lower equilibration temperature of weak sphalerons relative to their strong counterparts, see Sec.~\ref{sec:baryonasymmetry}.

Ultimately, Figures~\ref{fig:SSParameterSpace} and~\ref{fig:WSParameterSpace} establish the parameter space in which oscillating axions can successfully drive spontaneous LG, via the LNV Weinberg operator, and entirely account for the observed BAU for axions coupled to the strong and/or weak sectors with moderate anomaly coefficients of $\mathcal{O}(1\cdots10)$. The required asymmetry is naturally reproduced by axions featuring grand-unification (GUT)-scale decay constants and oscillation temperatures $T_{\text{osc}}\gtrsim \text{few}\cdot 10^{11}$~GeV, requiring only a standard initial misalignment angle of $\theta_i\gtrsim 10^{-2}$.

\section{Conclusions}
\label{sec:conclusion}

Axion-like particles provide a viable class of candidates to address several cosmological puzzles, including the nature of the inflaton, the origin and nature of dark matter, and the origin of the BAU. 
In particular, the motion of an axion in the early Universe provides a source of spontaneous $CPT$ symmetry violation. 
In combination with LNV processes, this is sufficient for a successful realization of baryogenesis via LG, referred to as spontaneous LG. In this paper, we provided a detailed study of a minimal realization of this mechanism, where the motion of the axion is accounted for through the standard misalignment mechanism, and LNV is provided by the five-dimensional Weinberg operator. An appealing feature of this scenario is that the BAU does not depend on the details of the UV physics that generates the Weinberg operator and, as found here, is insensitive to $CP$ violation in the light neutrino sector. For instance, if the UV sector responsible for generating the Weinberg operator is described by the type-I seesaw model, this LG scenario is independent of the heavy neutrino spectrum and the amount of $CP$ violation in that sector, and by construction, receives a negligible contribution from standard thermal LG. Finally, it remains entirely consistent with heavy neutrino masses near the GUT scale.

Throughout the paper, we focused on axions with couplings to either $G\widetilde{G}$, $W\widetilde{W}$, or $J^{\mu}_{B-L}$, where the latter allows us to incorporate the case where the axion is identified with a Majoron.  To this end, we numerically solved a full set of transport equations to determine the production of $B\!-\!L$ asymmetry through spontaneous LG for the different axion couplings. Our transport equations incorporate flavor-dependent interaction rates mediated by the Weinberg operator, expressed in terms of low-energy neutrino parameters. To the best of our knowledge, this is the first time expressions for fully flavor-dependent LNV rates mediated by the Weinberg operator appear in the literature. Our setup allowed us to perform a detailed study of the parameter space where the produced $B\!-\!L$ asymmetry is similar to or larger than the observed value for the various axion couplings. We find that the baryon asymmetry is controlled by the active-neutrino masses and their ordering, and is independent of the remaining parameters of the PMNS matrix. We quantified this dependence — on the absolute mass scale and on NO versus IO — in Sec.~\ref{sec:baryonasymmetry}.

On the analytical side, we derived an algebraic approximation of the generated $B\!-\!L$ asymmetry based on an instantaneous-equilibration limit, in which the network of transport equations collapses into a single (matrix) integral equation for the $B\!-\!L$ asymmetry. Explicit expressions for all terms entering the master equation were presented for completely generic shift-symmetric axion--SM couplings, arbitrary axion-velocities, and temperatures from the EWPT to $10^{15}$ GeV. We then compared our master formula with full numerical solutions of the transport equations, finding excellent agreement for the Majoron. For axions coupled to sphalerons, the agreement is typically within a relative factor of $\sim 2$ for oscillation temperatures below the respective sphaleron-equilibration temperatures, with higher oscillation temperatures exceeding the range of validity of the master formula.

Embedding a successful implementation of spontaneous LG into a consistent cosmological history requires considerations beyond simply quantifying the baryon asymmetry produced at high temperatures.  In particular, the induced baryonic isocurvature must respect observational bounds, and the entropy injection associated with axion decay must not dilute the baryon asymmetry below the observed value or spoil standard BBN predictions. 
To this end, we found that a Majoron without non-standard couplings cannot drive spontaneous LG if the Majoron velocity is dictated by standard misalignment. The reason is that, despite the large initial $B\!-\!L$ asymmetry produced by the Majoron, its very long lifetime causes it to dominate the energy budget of the Universe for an extended period. Its eventual decay then dilutes the generated asymmetry below the observed value. A potential avenue for saving the Majoron-driven spontaneous LG solution to the BAU is to reduce its lifetime by considering couplings to a dark sector.

The situation is different for axions with couplings to strong or weak sphalerons, which can decay on a shorter timescale and thereby avoid excessive entropy injection. 
For strong and weak anomaly coefficients in the range $\mathcal{O}(1\cdots10)$, we find that successful spontaneous LG becomes possible for $T_{\text{osc}}\gtrsim \text{few}\cdot 10^{11}$ GeV and $f_a$ around the GUT scale, as shown in Figs.~\ref{fig:SSParameterSpace} and~\ref{fig:WSParameterSpace} and discussed further in the main text. As shown there, a generic requirement for spontaneous LG to generate sufficient asymmetry without violating bounds on baryonic isocurvature is a high inflationary scale, followed by a period of efficient reheating after inflation, yielding a high reheating temperature. An interesting avenue for future research is embedding our framework into a complete model of inflation and reheating to track the asymmetry production from the end of inflation through the reheating phase, all the way down to $T\ll T_{\text{osc}}$. We leave such investigations for future work.

Finally, let us comment on the observational implications of the LG scenario studied in this paper. First of all, the need for a high energy scale of inflation and a high reheating temperature nourishes the hope of a large tensor-to-scalar ratio in the cosmic microwave background within the reach of future experiments. Second, an important outcome of the numerical analysis carried out in this work is a precise parametric understanding of the entropy dilution factor $\Delta_a$ that is required for successful baryogenesis. This factor is also crucial for other relics from the early Universe, notably, gravitational waves and dark matter. On the one hand, the stage in the cosmic expansion history driven by axion oscillations represents an era of early matter domination. As such, it leaves a direct imprint in the transfer function for primordial gravitational waves (e.g., from inflation)~\cite{DEramo:2019tit,Allahverdi:2020bys}, which may again be within the reach of future experiments. On the other hand, the entropy injection by late-time axion decays may also dilute the relic density of dark matter (depending on the time of dark-matter production in the early Universe), which may shift the viable parameter regions in specific dark-matter models. Finally, our scenario predicts (arguably weak) correlations between low-energy observables in the neutrino sector and the generation of the BAU at high energies as well as certain properties of the particle spectrum in GUTs (a heavy axion alongside heavy RHNs). The latter can provide guidance in GUT model building, which may then lead to concrete predictions, e.g., for proton-decay experiments. It will be important to make these statements more precise in future work; the results obtained in the present paper provide the basis for a rich research program fleshing out in more detail the phenomenological predictions of axion-driven spontaneous~LG.\\

\noindent\textbf{Acknowledgments:} We thank Kyohei Mukaida for collaboration at the early stages of this project. M.~A.~M.\ thanks Alfredo Stanzione and Sascha Weber for useful discussions. K.~S.\ is an affiliate member of the Kavli Institute for the Physics and Mathematics of the Universe (Kavli IPMU) at the University of Tokyo and supported by the World Premier International Research Center Initiative (WPI), MEXT, Japan (Kavli IPMU). K.~S.\ thanks Kavli IPMU for hospitality during the final period when this project was completed. H.W. was supported by JST SPRING, Japan Grant Number JPMJSP2104.

\begin{appendix}

    \section{Interaction rates}
    \label{appendix:BEq}
In this appendix, we specify the parameter values used in the numerical evaluation of the Boltzmann equations in Sec.~\ref{sec:baryonasymmetry}, based on the equations presented in Sec.~\ref{sec:BEqs}. Following Ref.~\cite{Domcke:2020kcp}, we parametrize the Yukawa rates as follows, 
\begin{align}
    \gamma_\tau(T)=6\kappa_{Y_e}y_\tau^2T, \quad \gamma_{t(c)}(T)=6\kappa_{Y_u}y_{t(c)}^2T,\quad \gamma_b(T)=6\kappa_{Y_d}y_b^2T,
\end{align}
where we use the following numerical values
\begin{align}
    &\kappa_{Y_e}=1.7\cdot10^{-3}, \quad \kappa_{Y_u}=10^{-2}, \quad \kappa_{Y_d}=10^{-2},\nonumber \\
    &y_\tau=10^{-2}, \quad y_t=0.49, \quad y_c=1.8\cdot 10^{-3}, \quad y_b=6.8\cdot10^{-3}. 
\end{align}
Both of the sphaleron rates are described by the following formulae~\cite{Moore:2010jd}, 
\begin{align}
    \gamma_{\text{sphal}_i}(T)&=3\,\kappa_{\text{sphal}_i} \alpha_i^5 T,\\
    \kappa_{\text{sphal}_i}&=0.21\left(\frac{N_{c_i}g_i^2T^2}{m_{D_i}^2}\right)\left(\ln\frac{m_{D_i}}{\gamma_i}+3.041\right)\left(\frac{N_{c_i}^2-1}{N_{c_i}^2}\right)N_{c_i}^5 , \label{eq:kappa}\\
    m_{D_2}^2&=\frac{11}{6}g_2^2T^2,\qquad  m_{D_3}^2=2g_3^2T^2, \\
    \gamma_i&=N_{c_i}\alpha_iT\left(\ln\frac{m_{D_i}}{\gamma_i}+3.041\right),
\end{align}
where $i=2$ corresponds to weak sphalerons and $i=3$ to strong sphalerons. Here, $N_{c_2}=2$, $N_{c_3}=3$, and $\alpha_{i}=g_{i}^2/(4\pi)$ with $g_{2}$ and $g_3$ the $SU(2)$ and $SU(3)$ gauge couplings, which we set to $g_2=0.55$ and $g_3=0.60$, respectively. 

\section{Algebraic results}
\label{appendix:algebraic}
In this appendix, we provide additional details about the algebraic approach outlined in Sec.~\ref{subsec:algebraic}. In particular, we provide explicit expressions for the matrices $\bm{C},\,\bm{S},\,$ and $\bm{S}'$, defined in Eqs.~\eqref{eq:L+H} and~\eqref{eq:mu0}. Our results can be readily applied to temperature regimes ranging from $T\sim 10^{15}$ GeV all the way down to the EWPT, extending well beyond the regime considered in the main text. Importantly, this analysis remains fully general: it applies to spontaneous LG in any temperature regime and accommodates completely arbitrary axion–SM couplings. The treatment builds on the linear algebra formalism presented in Ref.~\cite{Domcke:2020quw} (see also Ref.~\cite{Domcke:2020kcp}), and below we briefly summarize some necessary definitions and refer the reader to Ref.~\cite{Domcke:2020quw} for further details. 

Let us consider a $16$-dimensional vector space spanned by ${\bm{\mu}}_i$, with $i$ running over the $16$ individual fields that make up the scalar and fermion content of the SM  
\begin{align}
    \label{eq:i}
    i=e,\, \mu,\, \tau,\, L_e ,\, L_\mu,\, L_\tau,\, u,\, c,\, t,\, d,\, s,\, b,\, Q_1,\, Q_2,\, Q_3,\, H.
\end{align}
The configuration of $\bm{\mu}$ at any temperature is dictated by $M+N=16$ linear conditions, where $M$ represents the rank of the SM equilibrium interaction matrix $I$ and $N$ represents the number of independent conserved charges $C$,
\begin{equation}
\label{eq:constraints}
\sum_i n_i^I\,\mu_i = \bm{n}^I\cdot\bm{\mu} = \bm{n}_S^I\,\dot{\theta} \,,\qquad \sum_i n_i^C\,g_i\,\mu_i = \left(\bm{n}^C \circ \bm{g}\right) \cdot \bm{\mu} = \bar{\mu}_C \,.
\end{equation}
Here, the $\circ$ symbol denotes the entry-wise Hadamard product, and the source vector $\bm{n}_S^I$ is defined below Eq.~\eqref{eq:mu0}. The vector $\bm{g}$ encodes the electroweak and color multiplicities of the individual SM fields, in the basis of Eq.~\eqref{eq:i},
\begin{equation}
\bm{g} = \qty( 1, 1, 1,  2, 2, 2,  3, 3, 3,  3, 3, 3,  6, 6, 6,  4 ) \,.
\end{equation}
The $M$ vectors $\bm{n^I}$ characterize the particle content participating in each of the equilibrated interactions $I$. Following Refs.~\cite{Domcke:2020kcp,Domcke:2020quw}, we adopt a basis of $12$ linearly independent charge vectors, ordered by the temperature (from high to low) at which the corresponding interactions reach chemical equilibrium, 
\begin{align}
\label{eq:nyt}
\bm{n}^{y_t}    = & \left(0,  0,  0, 0, 0, 0,  0,  0, -1,  0,  0,  0, 0, 0, 1,  1\right) \,, \\
\label{eq:nySS}
\bm{n}^{\rm SS} = & \left(0,  0,  0, 0, 0, 0, -1, -1, -1, -1, -1, -1, 2, 2, 2,  0\right) \,, \\
\bm{n}^{\rm WS} = & \left(0,  0,  0, 1, 1, 1,  0,  0,  0,  0,  0,  0, 3, 3, 3,  0\right) \,, \\
\bm{n}^{y_b}    = & \left(0,  0,  0, 0, 0, 0,  0,  0,  0,  0,  0, -1, 0, 0, 1, -1\right) \,, \\
\bm{n}^{y_\tau} = & \left(0,  0, -1, 0, 0, 1,  0,  0,  0,  0,  0,  0, 0, 0, 0, -1\right) \,, \\
\bm{n}^{y_c}    = & \left(0,  0,  0, 0, 0, 0,  0, -1,  0,  0,  0,  0, 0, 1, 0,  1\right) \,, \\
\bm{n}^{y_\mu}  = & \left(0, -1,  0, 0, 1, 0,  0,  0,  0,  0,  0,  0, 0, 0, 0, -1\right) \,, \\  
\bm{n}^{y_{sb}} = & \left(0,  0,  0, 0, 0, 0,  0,  0,  0,  0, -1,  0, 0, 0, 1, -1\right) \,, \\
\bm{n}^{y_s}    = & \left(0,  0,  0, 0, 0, 0,  0,  0,  0,  0, -1,  0, 0, 1, 0, -1\right) \,, \\
\bm{n}^{y_d}    = & \left(0,  0,  0, 0, 0, 0,  0,  0,  0, -1,  0,  0, 1, 0, 0, -1\right) \,, \\
\bm{n}^{y_{ds}} = & \left(0,  0,  0, 0, 0, 0,  0,  0,  0, -1,  0,  0, 0, 1, 0, -1\right) \,, \\
\label{eq:nye}
\bm{n}^{y_e}    = & \left(-1, 0,  0, 1, 0, 0,  0,  0,  0,  0,  0,  0, 0, 0, 0, -1\right) \,.
\end{align}
Meanwhile, the $N$ charge vectors $\bm{n}^C$ represent the particle charges associated with the linearly independent global $U(1)_C$ symmetries that remain preserved by the equilibrated SM interactions $I$. Each interaction $\bm{n}^I$ that reaches chemical equilibrium effectively violates a corresponding conserved charge $\bm{n}^C$. In particular, for the $12$ interaction vectors defined in Eqs.~\eqref{eq:nyt}--\eqref{eq:nye}, there are $12$ associated charge vectors,
\begin{align}
\label{eq:nie}
\bm{n}^{y_e}    ~\leftrightarrow~ & \bm{n}^e              = \left(1, 0, 0, 0, 0, 0, 0, 0, 0, 0, 0, 0, 0, 0, 0, 0\right)                                                                      \,, \\
\bm{n}^{y_{ds}} ~\leftrightarrow~ & \bm{n}^{2B_1-B_2-B_3} = \left(0, 0, 0, 0, 0, 0, \sfrac23, -\sfrac13,\right.\nonumber \\
&\left.-\sfrac13, \sfrac23, -\sfrac13, -\sfrac13, \sfrac23, -\sfrac13, -\sfrac13, 0\right) \,, \\
\bm{n}^{y_d}    ~\leftrightarrow~ & \bm{n}^{u-d}          = \left(0, 0, 0, 0, 0, 0, 1, 0, 0, -1, 0, 0, 0, 0, 0, 0\right)                                                                     \,, \\
\bm{n}^{y_s}    ~\leftrightarrow~ & \bm{n}^{d-s}          = \left(0, 0, 0, 0, 0, 0, 0, 0, 0, 1, -1, 0, 0, 0, 0, 0\right)                                                                     \,, \\
\bm{n}^{y_{sb}} ~\leftrightarrow~ & \bm{n}^{B_1-B_2}      = \left(0, 0, 0, 0, 0, 0, \sfrac13, -\sfrac13, 0, \sfrac13, -\sfrac13, 0, \sfrac13, -\sfrac13, 0, 0\right)                         \,, \\
\bm{n}^{y_\mu}  ~\leftrightarrow~ & \bm{n}^\mu            = \left(0, 1, 0, 0, 0, 0, 0, 0, 0, 0, 0, 0, 0, 0, 0, 0\right)                                                                      \,, \\
\bm{n}^{y_c}    ~\leftrightarrow~ & \bm{n}^{u-c}          = \left(0, 0, 0, 0, 0, 0, 1, -1, 0, 0, 0, 0, 0, 0, 0, 0\right)                                                                     \,, \\
\bm{n}^{y_\tau} ~\leftrightarrow~ & \bm{n}^\tau           = \left(0, 0, 1, 0, 0, 0, 0, 0, 0, 0, 0, 0, 0, 0, 0, 0\right)                                                                      \,, \\
\bm{n}^{y_b}    ~\leftrightarrow~ & \bm{n}^{d-b}          = \left(0, 0, 0, 0, 0, 0, 0, 0, 0, 1, 0, -1, 0, 0, 0, 0\right)                                                                     \,, \\
\bm{n}^{\rm WS} ~\leftrightarrow~ & \bm{n}^B              = \left(0, 0, 0, 0, 0, 0, \sfrac13, \sfrac13, \sfrac13, \sfrac13, \sfrac13, \sfrac13, \sfrac13, \sfrac13, \sfrac13, 0\right)       \,, \\
\bm{n}^{\rm SS} ~\leftrightarrow~ & \bm{n}^u              = \left(0, 0, 0, 0, 0, 0, 1, 0, 0, 0, 0, 0, 0, 0, 0, 0\right)                                                                      \,,\\
\label{eq:niu}
\bm{n}^{\rm y_t} ~\leftrightarrow~ & \bm{n}^t              = \left(0, 0, 0, 0, 0, 0, 0, 0, 1, 0, 0, 0, 0, 0, 0, 0\right)                                                                      \,.
\end{align}
Additionally, all SM interactions above the EWPT conserve SM hypercharge, $Y$,\footnote{Here, we are referring to the global symmetry $U(1)_Y$ associated with the SM gauge symmetry.} and the three lepton-flavor asymmetries $\Delta_\alpha=B/3-L_\alpha$. Their associated charge vectors read, 
\begin{align}
\label{eq:Y}\bm{n}^Y                    & = \left(-1, -1, -1, -\sfrac12, -\sfrac12, -\sfrac12, \sfrac23, \sfrac23, \sfrac23, -\sfrac13, -\sfrac13, -\sfrac13, \sfrac16, \sfrac16, \sfrac16, \sfrac12 \right) \,, \\
\label{eq:Deltae}\bm{n}^{\Delta_e}      & = \left(-1,  0,  0,        -1,         0,         0, \sfrac19, \sfrac19, \sfrac19,  \sfrac19,  \sfrac19,  \sfrac19, \sfrac19, \sfrac19, \sfrac19, 0        \right) \,, \\
\label{eq:Deltamu}\bm{n}^{\Delta_\mu}   & = \left( 0, -1,  0,         0,        -1,         0, \sfrac19, \sfrac19, \sfrac19,  \sfrac19,  \sfrac19,  \sfrac19, \sfrac19, \sfrac19, \sfrac19, 0        \right) \,, \\
\label{eq:Deltatau}\bm{n}^{\Delta_\tau} & = \left( 0,  0, -1,         0,         0,        -1, \sfrac19, \sfrac19, \sfrac19,  \sfrac19,  \sfrac19,  \sfrac19, \sfrac19, \sfrac19, \sfrac19, 0        \right) \,.
\end{align}
At any given temperature, Eq.~\eqref{eq:constraints} along with the explicit basis vectors in Eqs.~\eqref{eq:nyt}--\eqref{eq:Deltatau}, define a linear set of equations of the following form, 
\begin{equation}
\label{eq:Mmu}
\bm{M}\,\bm{\mu} = \bm{m} \,,\qquad \bm{M} = \begin{pmatrix}\left(\bm{n}^I\right)^{\rm T} \\ \left(\bm{n}^C \circ\bm{g}\right)^{\rm T}\end{pmatrix} \,,\qquad \bm{m} = \begin{pmatrix} n_S^I\dot{\theta} \\ \bar{\mu}_C \end{pmatrix} \,.
\end{equation}
By solving this linear system of equations for the chemical potentials $\bm{\mu}$, one can readily identify the matrices $\bm{S}$, $\bm{S}'$, and $\bm{C}$ through Eqs.~\eqref{eq:L+H}-\eqref{eq:mu0}.

Finally, as discussed in Sec.~\ref{subsec:algebraic}, the presence of LNV interactions drives the thermal plasma towards an equilibrium attractor where, 
\begin{align}
    \label{qdeltaalphaeq2}
    \mu_{\Delta_\alpha}^{\rm{eq}}&=\mu_{\Delta_\alpha}^{{\rm{eq}},C}+\mu_{\Delta_\alpha}^{{\rm{eq}},\theta}, \\
    \label{qWILG2}
    \mu_{\Delta_\alpha}^{{\rm{eq}},C}(T)&=\sum_\beta\sum_{C\neq \Delta_\alpha}C_{\alpha\beta}^{-1}(T)S_{\beta C}\bar{\mu}_C, \\
    \label{qTheta2}\mu_{\Delta_\alpha}^{{\rm{eq}},\theta}(T)&=\dot{\theta}(T)\sum_\beta C_{\alpha\beta}^{-1}(T)\left[\delta_{B-L}+\sum_{I\neq \text{LNV}} S'_{\beta I}(T)n_S^I\right].
\end{align}
In the limit where LNV-interactions are rapid compared to the Hubble expansion, hereafter referred to as the \textit{strong wash-in limit}, one obtains $\mu_{B-L}\approx \Tr{\mu_{\Delta_\alpha}^{\rm{eq}}}$. 

Below, we present explicit expressions for the matrices $\bm{S}$, $\bm{S}'$, and $\bm{C}$. We also present the result for $\mu_{B-L}$ in the strong wash-in limit. For $T\gtrsim T_\tau$, we provide one set of results that apply to scenarios described by the $(1,\,2,\,3)$ and $(1,\,2,\,\tau)$ bases, which are identical in this temperature range upon identifying $\bar{\mu}_3\leftrightarrow \bar{\mu}_\tau$. We also provide one set of results for scenarios where the LNV interaction probes a single direction in flavor space, which can e.g. occur in a thermal history featuring a highly hierarchical RHN spectrum. For $T_\tau\gtrsim T\gtrsim T_\mu$, we provide one set of results for scenarios described by the $(1,\,2,\,\tau)$ basis and one set of results where the LNV interaction only probes a single direction in flavor space. Finally, for $T\lesssim T_\mu$ we work in the $(e,\,\mu,\,\tau)$ basis. Our results for scenarios where the LNV interaction only probes a single direction in flavor space represents an extension of the results in Ref.~\cite{Domcke:2020quw}, in the presence of axion-SM interactions. In particular, the reader is referred to the supplement material of Ref.~\cite{Domcke:2020quw} for definitions of the chemical potentials $\bar{\mu}_\perp,\,\bar{\mu}_\parallel, \, \bar{\mu}_{\Delta_\parallel},\, \bar{\mu}_{\Delta_\perp}$ entering some of expressions below.\\


\noindent{\boldmath{$T_{\rm{top}}\gtrsim T\gtrsim T_{\rm{SS}}$:}} In the high-temperature regime, where the top-quark Yukawa interaction is the only equilibrated charge-violating process, we have

\begin{align}
    \bm{C}&=\begin{pmatrix}
\frac{2}{3} & \frac{1}{6} & \frac{1}{6} \\
\frac{1}{6} & \frac{2}{3} & \frac{1}{6} \\
\frac{1}{6} & \frac{1}{6} & \frac{2}{3}
\end{pmatrix},\qquad {\bm{C}}=\left(\frac{2}{3}\right),
\end{align}

\begin{align}
\bm{S} &= \bordermatrix{ & \bar{\mu}_u & \bar{\mu}_B & \bar{\mu}_{d-b} & \bar{\mu}_{3,\tau} & \bar{\mu}_{u-c} & \bar{\mu}_2 & \bar{\mu}_{B_1-B_2} & \bar{\mu}_{d-s} & \bar{\mu}_{u-d} & \bar{\mu}_{2B_1-B_2-B_3} & \bar{\mu}_{1} \cr
                          & \frac29 & \frac19 & -\frac29 & \frac16 & \frac16 & \frac16 & -\frac16 & -\frac16 & -\frac59 & \frac19 & -\frac13 \cr
                          & \frac29 & \frac19 & -\frac29 & \frac16 & \frac16 & -\frac13 & -\frac16 & -\frac16 & -\frac59 & \frac19 & \frac16 \cr
                          & \frac29 & \frac19 & -\frac29 & -\frac13 & \frac16 & \frac16 & -\frac16 & -\frac16 & -\frac59 & \frac19 & \frac16 \cr } \,,\\
\bm{S}&=\bordermatrix{ & \bar{\mu}_u   & \bar{\mu}_B   & \bar{\mu}_{d-b} & \bar{\mu}_{u-c} & \bar{\mu}_{B_1-B_2} & \bar{\mu}_{d-s} & \bar{\mu}_{u-d} & \bar{\mu}_{2B_1-B_2-B_3} & \bar{\mu}_\parallel & \bar{\mu}_\perp & \bar{\mu}_{\Delta_\perp} \cr
                         & \frac29 & \frac19 & -\frac29  & \frac16   & -\frac16      & -\frac16  & -\frac59  & \frac19            & -\frac13      & \frac16   & -\frac16 \cr } \,,
\end{align}

\begin{align}
    \bm{S}' &= \bordermatrix{ & n_S^{y_t} \cr
                         & \frac13 \cr
                         & \frac13 \cr
                         & \frac13 \cr } \,,\qquad
    \bm{S}' =\bordermatrix{ & n_S^{y_t} \cr
                         & \frac13 } \,,
\end{align}
where the first [second] index in $\bar{\mu}_{(3,\tau)}$ applies to the $(1,\,2,\,3)$ [$(1,\,2,\,\tau)$] basis. The final $B\!-\!L$ asymmetry in the strong wash-in limit is given by
\begin{align}
    \mu_{B-L}&=\left(\frac{2\bar{\mu}_u}{3}+\frac{\bar{\mu}_B}{3}-\frac{2\bar{\mu}_{db}}{3}+\frac{\bar{\mu}_{uc}}{2}-\frac{\bar{\mu}_{B_1-B_2}}{2}-\frac{\bar{\mu}_{d-s}}{2}-\frac{5\bar{\mu}_{u-d}}{3}+\frac{\bar{\mu}_{2B_1-B_2-B_3}}{3}\right)\nonumber \\
    &+(3\delta_{B-L}+n_S^{y_t})\dot{\theta}, \\
    \mu_{B-L}&=\left(\frac13\,\bar{\mu}_u + \frac16\,\bar{\mu}_B - \frac13\,\bar{\mu}_{d-b} + \frac14\,\bar{\mu}_{u-c} - \frac14\,\bar{\mu}_{B_1-B_2} - \frac14\,\bar{\mu}_{d-s} - \frac56\,\bar{\mu}_{u-d} \right. \nonumber \\
    &\left. + \frac16\,\bar{\mu}_{2B_1-B_2-B_3} + \frac{1-3P}{4}\left(\bar{\mu}_e+\bar{\mu}_\mu+\bar{\mu}_\tau\right) + \frac34\,\bar{\mu}_{\Delta_\perp}\right) \nonumber \\
    &+\left(\frac{3\delta_{B-L}}{2}+\frac{n_S^{y_t}}{3}\right)\dot{\theta}.
\end{align}


\noindent{\boldmath{$T_{\rm SS}\gtrsim T\gtrsim T_{\rm WS}$:}}
At slightly lower temperatures, where also strong-sphaleron interactions are entering equilibrium, one has
\begin{align}
    \bm{C}=\begin{pmatrix}
\frac{15}{23} & \frac{7}{46} & \frac{7}{46} \\
\frac{7}{46} & \frac{15}{23} & \frac{7}{46} \\
\frac{7}{46} & \frac{7}{46} & \frac{15}{23}
\end{pmatrix}, \qquad \bm{C}=\left(\frac{15}{23}\right),
\end{align}

\begin{align}
    \bm{S} &= \bordermatrix{ 
    & \bar{\mu}_B & \bar{\mu}_{d-b} & \bar{\mu}_{(3,\tau)} & \bar{\mu}_{u-c} & \bar{\mu}_2 & \bar{\mu}_{B_1-B_2} & \bar{\mu}_{d-s} & \bar{\mu}_{u-d} & \bar{\mu}_{2B_1-B_2-B_3} & \bar{\mu}_{1} \cr
    & \frac{1}{6} & -\frac{4}{23} & \frac{7}{46} & \frac{9}{46} & \frac{7}{46} & -\frac{9}{46} & -\frac{5}{46} & -\frac{9}{23} & \frac{3}{23} & -\frac{8}{23} \cr
    & \frac{1}{6} & -\frac{4}{23} & \frac{7}{46} & \frac{9}{46} & -\frac{8}{23} & -\frac{9}{46} & -\frac{5}{46} & -\frac{9}{23} & \frac{3}{23} & \frac{7}{46} \cr
    & \frac{1}{6} & -\frac{4}{23} & -\frac{8}{23} & \frac{9}{46} & \frac{7}{46} & -\frac{9}{46} & -\frac{5}{46} & -\frac{9}{23} & \frac{3}{23} & \frac{7}{46} \cr
}, \\
\bm{S} &= \bordermatrix{ & \bar{\mu}_B   & \bar{\mu}_{d-b}     & \bar{\mu}_{u-c}    & \bar{\mu}_{B_1-B_2} & \bar{\mu}_{d-s}     & \bar{\mu}_{u-d}     & \bar{\mu}_{2B_1-B_2-B_3} & \bar{\mu}_\parallel & \bar{\mu}_\perp    & \bar{\mu}_{\Delta_\perp}  \cr
                         & \frac16 & -\frac{4}{23} & \frac{9}{46} & -\frac{9}{46} & -\frac{5}{46} & -\frac{9}{23} & \frac{3}{23}       & -\frac{8}{23} & \frac{7}{46} & -\frac{7}{46} \cr } \,,
\end{align}

\begin{align}
    \bm{S}' = \bordermatrix{ & n_S^{y_t} & n_S^{\rm SS} \cr
                          & \frac{9}{23} & -\frac{3}{46} \cr
                          & \frac{9}{23} & -\frac{3}{46} \cr
                          & \frac{9}{23} & -\frac{3}{46} \cr }, \qquad \bm{S}' =\bordermatrix{ & n_S^{y_t} & n_S^{\rm SS} \cr
                          & \frac{9}{23} & -\frac{3}{46} },
\end{align}

\begin{align}
    \mu_{B-L}&=\left(\frac{23\bar{\mu}_B}{44} -\frac{6\bar{\mu}_{d-b}}{11} - \frac{\bar{\mu}_{(3,\tau)}}{22} + \frac{27\bar{\mu}_{u-c}}{44} -\frac{\bar{\mu}_2}{22} - \frac{27\bar{\mu}_{B_1-B_2}}{44} - \frac{15\bar{\mu}_{d-s}}{44} -\frac{27\bar{\mu}_{u-d}}{22}\right.\nonumber \\
    &\left.+ \frac{9\bar{\mu}_{2B_1-B_2-B_3}}{22} -\frac{\bar{\mu}_{1}}{22}\right) +\left(\frac{69}{22}\delta_{B-L}+\frac{27}{22}n_S^{y_t}-\frac{9}{44}n_S^{\rm SS}\right)\dot{\theta},  \\
    \mu_{B-L}&=\left(\frac{23}{90} \,\bar{\mu}_{B} - \frac{4}{15} \,\bar{\mu}_{d-b} + \frac{3}{10} \,\bar{\mu}_{u-c} - \frac{3}{10}\,\bar{\mu}_{B_1-B_2} - \frac{1}{6}\,\bar{\mu}_{d-s} - \frac{3}{5}\,\bar{\mu}_{u-d}+ \frac{1}{5} \,\bar{\mu}_{2B_1-B_2-B_3} \right.\nonumber\\
    &\left.+ \frac{7-23P}{30}\left(\bar{\mu}_e+\bar{\mu}_\mu+\bar{\mu}_\tau\right) + \frac{23}{30}\,\bar{\mu}_{\Delta_\perp}\right)+\left(\frac{23\delta_{B-L}}{15}+\frac{3n_S^{y_t}}{5}-\frac{n_S^{\rm SS}}{10}\right)\dot{\theta}.
\end{align}


\noindent{\boldmath{$T_{\rm{WS}}\gtrsim T\gtrsim T_{b\tau}$:}} Next, we consider the temperature regime where also weak sphalerons are approximately equilibrated, while Yukawa interactions associated with the bottom and tau still remain out of equilibrium. Here, we obtain
\begin{align}
    \bm{C}&=\begin{pmatrix}
\frac{202}{345} & \frac{59}{690} & \frac{59}{690} \\
\frac{59}{690} & \frac{202}{345} & \frac{59}{690} \\
\frac{59}{690} & \frac{59}{690} & \frac{202}{345}
\end{pmatrix}, \qquad \bm{C}=\left(\frac{202}{345}\right),
\end{align}

\begin{align}
    \bm{S} &= \bordermatrix{ 
    & \bar{\mu}_{d-b} & \bar{\mu}_{(3,\tau)} & \bar{\mu}_{u-c} & \bar{\mu}_2 & \bar{\mu}_{B_1-B_2} & \bar{\mu}_{d-s} & \bar{\mu}_{u-d} & \bar{\mu}_{2B_1-B_2-B_3} & \bar{\mu}_{1} \cr
    & -\frac{4}{23} & \frac{151}{690} & \frac{9}{46} & \frac{151}{690} & -\frac{9}{46} & -\frac{5}{46} & -\frac{9}{23} & \frac{3}{23} & -\frac{97}{345} \cr
    & -\frac{4}{23} & \frac{151}{690} & \frac{9}{46} & -\frac{97}{345} & -\frac{9}{46} & -\frac{5}{46} & -\frac{9}{23} & \frac{3}{23} & \frac{151}{690} \cr
    & -\frac{4}{23} & -\frac{97}{345} & \frac{9}{46} & \frac{151}{690} & -\frac{9}{46} & -\frac{5}{46} & -\frac{9}{23} & \frac{3}{23} & \frac{151}{690} \cr
}, \\
\bm{S} &= \bordermatrix{ & \bar{\mu}_{d-b}     & \bar{\mu}_{u-c}    & \bar{\mu}_{B_1-B_2} & \bar{\mu}_{d-s}     & \bar{\mu}_{u-d}     & \bar{\mu}_{2B_1-B_2-B_3} & \bar{\mu}_\parallel & \bar{\mu}_\perp    & \bar{\mu}_{\Delta_\perp}  \cr
                         & -\frac{4}{23} & \frac{9}{46} & -\frac{9}{46} & -\frac{5}{46} & -\frac{9}{23} & \frac{3}{23}       & -\frac{97}{345} & \frac{151}{690} & -\frac{59}{690} \cr } \,,
\end{align}

\begin{align}
    \bm{S}' = \bordermatrix{ 
    & n_S^{y_t} & n_S^{\rm SS} & n_S^{\rm WS} \cr
    & \frac{9}{23} & -\frac{19}{115} & \frac{2}{15} \cr
    & \frac{9}{23} & -\frac{19}{115} & \frac{2}{15} \cr
    & \frac{9}{23} & -\frac{19}{115} & \frac{2}{15} \cr 
}, \qquad \bm{S}' =\bordermatrix{ 
    & n_S^{y_t} & n_S^{\rm SS} & n_S^{\rm WS} \cr
    & \frac{9}{23} & -\frac{19}{115} & \frac{2}{15}
},
\end{align}

\begin{align}
    \mu_{B-L}&=\left(-\frac{20\bar{\mu}_{d-b}}{29} + \frac{6\bar{\mu}_{(3,\tau)}}{29} + \frac{45\bar{\mu}_{u-c}}{58} +\frac{6\bar{\mu}_2}{29} - \frac{45\bar{\mu}_{B_1-B_2}}{58} - \frac{25\bar{\mu}_{d-s}}{58} -\frac{45\bar{\mu}_{u-d}}{29}\right.\nonumber \\
    &\left.+ \frac{15\bar{\mu}_{2B_1-B_2-B_3}}{29} +\frac{6\bar{\mu}_{1}}{29}\right) +\left(\frac{115}{29}\delta_{B-L}+\frac{45}{29}n_S^{y_t}-\frac{19}{29}n_S^{\rm SS}+\frac{46n_S^{\rm WS}}{87}\right)\dot{\theta},  \\
    \mu_{B-L}&=\left(- \frac{30}{101} \,\bar{\mu}_{d-b} + \frac{135}{404} \,\bar{\mu}_{u-c} - \frac{135}{404}\,\bar{\mu}_{B_1-B_2} - \frac{75}{404}\,\bar{\mu}_{d-s} - \frac{135}{202}\,\bar{\mu}_{u-d}+ \frac{45}{202} \,\bar{\mu}_{2B_1-B_2-B_3} \right.\nonumber\\
    &\left.+ \frac{(151-345P)}{404}\left(\bar{\mu}_e+\bar{\mu}_\mu+\bar{\mu}_\tau\right) + \frac{345}{404}\,\bar{\mu}_{\Delta_\perp}\right)\nonumber \\
    &+\left(\frac{345\delta_{B-L}}{202}+\frac{135n_S^{y_t}}{202}-\frac{57n_S^{\rm SS}}{202}+\frac{23n_S^{\rm WS}}{101}\right)\dot{\theta}.
\end{align}


\noindent{\boldmath{$T_{b\tau}\gtrsim T\gtrsim T_c$:}} In the next temperature regime, also tau and bottom Yukawa interactions are equilibrated,
\begin{align}
    \bm{C}=\begin{pmatrix}
\frac{237}{460} & \frac{7}{460} & \frac{8}{115} \\
\frac{7}{460} & \frac{237}{460} & \frac{8}{115} \\
\frac{8}{115} & \frac{8}{115} & \frac{53}{115}
\end{pmatrix}, \qquad \bm{C}=\begin{pmatrix}
\frac{237}{460} & \frac{8}{115}  \\
\frac{8}{115} & \frac{53}{115} 
\end{pmatrix},
\end{align}

\begin{align}
    \bm{S}&=\bordermatrix{ 
    & \bar{\mu}_{u-c} & \bar{\mu}_2 & \bar{\mu}_{B_1-B_2} & \bar{\mu}_{d-s} & \bar{\mu}_{u-d} & \bar{\mu}_{2B_1-B_2-B_3} & \bar{\mu}_{1} \cr
    & \frac{41}{460} & \frac{15}{92} & 0 & -\frac{41}{460} & -\frac{41}{230} & 0 & -\frac{31}{92} \cr
    & \frac{41}{460} & -\frac{31}{92} & 0 & -\frac{41}{460} & -\frac{41}{230} & 0 & \frac{15}{92} \cr
    & \frac{14}{115} & \frac{4}{23} & 0 & -\frac{14}{115} & -\frac{28}{115} & 0 & \frac{4}{23} \cr
}, \\
    \bm{S} &= \bordermatrix{ 
    & \bar{\mu}_{u-c} & \bar{\mu}_{B_1-B_2}  & \bar{\mu}_{d-s} & \bar{\mu}_{u-d} & \bar{\mu}_{2B_1-B_2-B_3} & \bar{\mu}_{\parallel_\tau}  & \bar{\mu}_\perp &  \bar{\mu}_{\Delta_\perp} \cr
    & \frac{41}{460} & 0 & -\frac{41}{460} & -\frac{41}{230} & 0 & -\frac{31}{92}  & \frac{15}{92}  & -\frac{7}{460} \cr
    & \frac{14}{115} & 0 & -\frac{14}{115} & -\frac{28}{115} & 0 & \frac{4}{23}  & \frac{4}{23} &  -\frac{8}{115} \cr
},
\end{align}

\begin{align}
    \bm{S}' &= \bordermatrix{ 
    & n_S^{y_t} & n_S^{\rm SS} & n_S^{\rm WS} & n_S^{y_b} & n_S^{y_\tau} \cr
    & \frac{123}{460} & -\frac{51}{460} & \frac{17}{115} & -\frac{123}{460} & -\frac{5}{46} \cr
    & \frac{123}{460} & -\frac{51}{460} & \frac{17}{115} & -\frac{123}{460} & -\frac{5}{46} \cr
    & \frac{42}{115} & -\frac{9}{115} & \frac{12}{115} & -\frac{42}{115} & \frac{5}{23} \cr
}, \\
    \bm{S}'&=\bordermatrix{ 
    & n_S^{y_t} & n_S^{\rm SS} & n_S^{\rm WS} & n_S^{y_b} & n_S^{y_\tau} \cr
    & \frac{123}{460} & -\frac{51}{460} & \frac{17}{115} & -\frac{123}{460} & -\frac{5}{46} \cr
    & \frac{42}{115} & -\frac{9}{115} & \frac{12}{115} & -\frac{42}{115} & \frac{5}{23} \cr
},
\end{align}

\begin{align}
    \mu_{B-L}&=\left(\frac{\bar{\mu}_{u-c}}{2} -\frac{\bar{\mu}_{d-s}}{2} -\bar{\mu}_{u-d}\right)\nonumber \\
    &+\left(5\delta_{B-L}+\frac{3}{2}n_S^{y_t}-\frac{n_S^{\rm SS}}{2}+\frac{2n_S^{\rm WS}}{3}-\frac{3n_S^{y_b}}{2}\right)\dot{\theta},  \\
    \mu_{B-L}&=\left(\frac{41\bar{\mu}_{u-c}}{107} -\frac{41\bar{\mu}_{d-s}}{107} -\frac{82\bar{\mu}_{u-d}}{107}+\frac{90\bar{\mu}_{\Delta_\perp}}{107}+\frac{(65-90P_\tau)}{107}(\bar{\mu}_e+\bar{\mu}_\mu)\right)\nonumber \\
    &+\left(\frac{385}{107}\delta_{B-L}+\frac{123}{107}n_S^{y_t}-\frac{36n_S^{\rm SS}}{107}+\frac{48n_S^{\rm WS}}{107}-\frac{123n_S^{y_b}}{107}+\frac{25n_S^{y_\tau}}{107}\right)\dot{\theta}.
\end{align}


\noindent{\boldmath{$T_c\gtrsim T\gtrsim T_\mu$:}} Here, the difference compared to the previous regime is that also the charm Yukawa interaction is equilibrated. This represents the lowest temperature regime considered in the main text, 

\begin{align}
    \bm{C}&=\begin{pmatrix}
\frac{585}{1178} & -\frac{2}{589} & \frac{26}{589} \\
-\frac{2}{589} & \frac{585}{1178} & \frac{26}{589} \\
\frac{26}{589} & \frac{26}{589} & \frac{251}{589}
\end{pmatrix}, \qquad \bm{C}=\begin{pmatrix}
\frac{585}{1178} & \frac{26}{589} \\
\frac{26}{589} & \frac{251}{589}
\end{pmatrix}.
\end{align}

\begin{align}
    \bm{S}&=\bordermatrix{ 
    & \bar{\mu}_2 & \bar{\mu}_{B_1-B_2} & \bar{\mu}_{d-s} & \bar{\mu}_{u-d} & \bar{\mu}_{2B_1-B_2-B_3} & \bar{\mu}_{1} \cr
    & \frac{84}{589} & \frac{123}{2356} & -\frac{41}{589} & -\frac{123}{1178} & 0 & -\frac{421}{1178} \cr
    & -\frac{421}{1178} & \frac{123}{2356} & -\frac{41}{589} & -\frac{123}{1178} & 0 & \frac{84}{589} \cr
    & \frac{86}{589} & \frac{42}{589} & -\frac{56}{589} & -\frac{84}{589} & 0 & \frac{86}{589} \cr
},\\
    \bm{S} &= \bordermatrix{ 
    & \bar{\mu}_{B_1-B_2} & \bar{\mu}_{d-s} & \bar{\mu}_{u-d} & \bar{\mu}_{2B_1-B_2-B_3} & \bar{\mu}_{\parallel_\tau} & \bar{\mu}_\perp & \bar{\mu}_{\Delta_\perp} \cr
    & \frac{123}{2356} & -\frac{41}{589} & -\frac{123}{1178} & 0 & -\frac{421}{1178} & \frac{84}{589} & \frac{2}{589} \cr
    & \frac{42}{589} & -\frac{56}{589} & -\frac{84}{589} & 0 & \frac{86}{589} & \frac{86}{589} & -\frac{26}{589} \cr
},
\end{align}

\begin{align}
    \bm{S}'&=\bordermatrix{ 
    & n_S^{y_t} & n_S^{\rm SS} & n_S^{\rm WS} & n_S^{y_b} & n_S^{y_\tau} & n_S^{y_c} \cr
    & \frac{615}{2356} & -\frac{381}{2356} & \frac{86}{589} & -\frac{369}{2356} & -\frac{56}{589} & \frac{123}{589} \cr
    & \frac{615}{2356} & -\frac{381}{2356} & \frac{86}{589} & -\frac{369}{2356} & -\frac{56}{589} & \frac{123}{589} \cr
    & \frac{210}{589} & -\frac{87}{589} & \frac{60}{589} & -\frac{126}{589} & \frac{139}{589} & \frac{168}{589} \cr
},\\
    \bm{S}' &= \bordermatrix{ 
    & n_S^{y_t} & n_S^{\rm SS} & n_S^{\rm WS} & n_S^{y_b} & n_S^{y_\tau} & n_S^{y_c} \cr
    & \frac{615}{2356} & -\frac{381}{2356} & \frac{86}{589} & -\frac{369}{2356} & -\frac{56}{589} & \frac{123}{589} \cr
    & \frac{210}{589} & -\frac{87}{589} & \frac{60}{589} & -\frac{126}{589} & \frac{139}{589} & \frac{168}{589} \cr
},
\end{align}

\begin{align}
    \mu_{B-L}&=\left(-\frac{\bar{\mu}_2}{9} + \frac{\bar{\mu}_{B_1-B_2}}{3} - \frac{4\bar{\mu}_{d-s}}{9} -\frac{2\bar{\mu}_{u-d}}{3} -\frac{\bar{\mu}_{1}}{9}\right),\nonumber\\
    &+\left(\frac{17}{3}\delta_{B-L}+\frac{5n_S^{y_t}}{3}-\frac{8n_S^{\rm SS}}{9}+\frac{20n_S^{\rm WS}}{27}-n_S^{y_b}+\frac{n_S^{y_\tau}}{9}+\frac{4n_S^{y_c}}{3}\right)\dot{\theta},\\
    \mu_{B-L}&=\left(\frac{123\bar{\mu}_{B_1-B_2}}{494} - \frac{82\bar{\mu}_{d-s}}{247} -\frac{123\bar{\mu}_{u-d}}{247} +\frac{225\bar{\mu}_{\Delta_\perp}}{247}+\frac{(142-225P_\tau)}{247}(\bar{\mu}_e+\bar{\mu}_\mu)\right),\nonumber\\
    &+\left(\frac{983}{247}\delta_{B-L}+\frac{615n_S^{y_t}}{494}-\frac{303n_S^{\rm SS}}{494}+\frac{120n_S^{\rm WS}}{247}-\frac{369n_S^{y_b}}{494}+\frac{83n_S^{y_\tau}}{247}+\frac{246n_S^{y_c}}{247}\right)\dot{\theta},
\end{align}


\noindent{\boldmath{$T_\mu\gtrsim T\gtrsim T_{ds}$:}}
In this regime, all SM interactions except for the Yukawa interactions involving first-generation SM fermions are equilibrated. As a result, the lepton-flavor basis $\alpha=(e,\mu,\tau)$ becomes independent of how many directions in lepton-flavor space the LNV interaction probes. 

\begin{align}
    \bm{C}=\begin{pmatrix}
\frac{339}{716} & \frac{3}{179} & \frac{3}{179} \\
\frac{3}{179} & \frac{211}{537} & \frac{32}{537} \\
\frac{3}{179} & \frac{32}{537} & \frac{211}{537}
\end{pmatrix},
\end{align}

\begin{align}
    \bm{S}= \bordermatrix{ 
    & \bar{\mu}_{u-d} & \bar{\mu}_{2B_1-B_2-B_3} & \bar{\mu}_{e} \cr
    & -\frac{37}{716} & 0 & -\frac{265}{716} \cr
    & -\frac{13}{179} & 0 & \frac{23}{179} \cr
    & -\frac{13}{179} & 0 & \frac{23}{179} \cr
},
\end{align}

\begin{align}
    \bm{S}' = \bordermatrix{ 
    & n_S^{y_t} & n_S^{\rm SS} & n_S^{\rm WS} & n_S^{y_b} & n_S^{y_\tau} & n_S^{y_c} & n_S^{y_\mu} & n_S^{y_{sb}} & n_S^{y_s} \cr
    & \frac{111}{716} & -\frac{21}{179} & \frac{28}{179} & -\frac{111}{716} & -\frac{31}{358} & \frac{111}{716} & -\frac{31}{358} & 0 & -\frac{111}{716} \cr
    & \frac{39}{179} & -\frac{15}{179} & \frac{20}{179} & -\frac{39}{179} & -\frac{46}{537} & \frac{39}{179} & -\frac{133}{537} & 0 & -\frac{39}{179} \cr
    & \frac{39}{179} & -\frac{15}{179} & \frac{20}{179} & -\frac{39}{179} & -\frac{133}{537} & \frac{39}{179} & -\frac{46}{537} & 0 & -\frac{39}{179} \cr
}
\end{align}

\begin{align}
    \mu_{B-L}&=\left(-\frac{7\bar{\mu}_{u-d}}{17} -\frac{3\bar{\mu}_{e}}{17}\right)+\left(\frac{106}{17}\delta_{B-L}+\frac{21n_S^{y_t}}{17}-\frac{10n_S^{\rm SS}}{17}+\frac{40n_S^{\rm WS}}{51}-\frac{21n_S^{y_b}}{17}+\frac{3n_S^{y_\tau}}{17}\right.\nonumber \\
    &\left.+\frac{21n_S^{y_c}}{17}+\frac{3n_S^{y_\mu}}{17}-\frac{21n_S^{y_s}}{17}\right)\dot{\theta}, 
\end{align}


\noindent{\boldmath{$T_{ds}\gtrsim T\gtrsim T_e$:}} In this regime, the electron Yukawa interaction is the only SM interaction that remains out of chemical equilibrium.
\begin{align}
\bm{C} = \begin{pmatrix}
\frac{6}{13}       & 0 & 0          \\
0 & \frac{41}{111} & \frac{4}{111}  \\
0 & \frac{4}{111}  & \frac{41}{111}
\end{pmatrix} \,,\qquad \bm{S} = \bordermatrix{ & \bar{\mu}_e \cr
                         & -\frac{5}{13} \cr
                         & \frac{4}{37}  \cr
                         & \frac{4}{37}  \cr } \,,
\end{align}

\begin{align}
    \bm{S}' = \bordermatrix{ 
    & n_S^{y_t} & n_S^{\rm SS} & n_S^{\rm WS} & n_S^{y_b} & n_S^{y_\tau} & n_S^{y_c} & n_S^{y_\mu} & n_S^{y_{sb}} & n_S^{y_s} & n_S^{y_d} & n_S^{y_{ds}} \cr
    & 0 & 0 & \frac{2}{13} & -\frac{3}{13} & -\frac{1}{13} & 0 & -\frac{1}{13} & 0 & -\frac{3}{13} & -\frac{3}{13} & 0 \cr
    & 0 & \frac{3}{37} & \frac{4}{37} & -\frac{12}{37} & -\frac{8}{111} & 0 & \frac{29}{111} & 0 & -\frac{12}{37} & -\frac{12}{37} & 0 \cr
    & 0 & \frac{3}{37} & \frac{4}{37} & -\frac{12}{37} & \frac{29}{111} & 0 & -\frac{8}{111} & 0 & -\frac{12}{37} & -\frac{12}{37} & 0 \cr
}
\end{align}

\begin{align}
    \mu_{B-L}&=- \frac{3}{10}\,\bar{\mu}_{e}+\left(\frac{71\delta_{B-L}}{10}+\frac{2n_S^{\rm SS}}{5}+\frac{13n_S^{\rm WS}}{15}-\frac{21n_S^{y_b}}{10}+\frac{3n_S^{y_\tau}}{10}\right.\nonumber \\
    &\left.+\frac{3n_S^{y_\mu}}{10}-\frac{21n_S^{y_s}}{10}-\frac{21n_S^{y_d}}{10}\right)\dot{\theta}
\end{align}


\noindent{\boldmath{$T_e\gtrsim T\gtrsim T_{\bf{EWPT}}$:}}
Finally, when all SM interactions are in chemical equilibrium, we obtain
\begin{align}
    \bm{C}=\begin{pmatrix}
\frac{257}{711} & \frac{20}{711} & \frac{20}{711} \\
\frac{20}{711} & \frac{257}{711} & \frac{20}{711} \\
\frac{20}{711} & \frac{20}{711} & \frac{257}{711}
\end{pmatrix},
\end{align}

\begin{align}
    \bm{S}' = \bordermatrix{ 
    & n_S^{y_t} & n_S^{\rm SS} & n_S^{\rm WS} & n_S^{y_b} & n_S^{y_\tau} & n_S^{y_c} & n_S^{y_\mu} & n_S^{y_{sb}} & n_S^{y_s} & n_S^{y_d} & n_S^{y_{ds}} & n_S^{y_e} \cr
    & 0 & \frac{5}{79} & \frac{28}{237} & -\frac{24}{79} & -\frac{52}{711} & 0 & -\frac{52}{711} & 0 & -\frac{24}{79} & -\frac{24}{79} & 0 & \frac{185}{711} \cr
    & 0 & \frac{5}{79} & \frac{28}{237} & -\frac{24}{79} & -\frac{52}{711} & 0 & \frac{185}{711} & 0 & -\frac{24}{79} & -\frac{24}{79} & 0 & -\frac{52}{711} \cr
    & 0 & \frac{5}{79} & \frac{28}{237} & -\frac{24}{79} & \frac{185}{711} & 0 & -\frac{52}{711} & 0 & -\frac{24}{79} & -\frac{24}{79} & 0 & -\frac{52}{711} \cr
},
\end{align}

\begin{align}
   \mu_{B-L}&=\left(\frac{79\delta_{B-L}}{11}+\frac{5n_S^{\rm SS}}{11}+\frac{28n_S^{\rm WS}}{33}-\frac{24n_S^{y_b}}{11}+\frac{3n_S^{y_\tau}}{11}\right.\nonumber \\
    &\left.+\frac{3n_S^{y_\mu}}{11}-\frac{24n_S^{y_s}}{11}-\frac{24n_S^{y_d}}{11}+\frac{3n_S^{y_e}}{11}\right)\dot{\theta}.
\end{align}


\section{Additional numerical results}
    \label{appendix:NumericResults}
The purpose of this appendix is to collect additional relevant numerical results that complement those discussed in the main text. In Sec.~\ref{appendix:asymprod} and~\ref{appendix:FinalParameterSpace}, we present results that are relevant to the discussions in Sec.~\ref{sec:baryonasymmetry} and~\ref{sec:FinalParameter}, respectively.

\subsection{Asymmetry production}
\label{appendix:asymprod}
\begin{figure}[h]
    \centering
    \includegraphics[width=\textwidth]{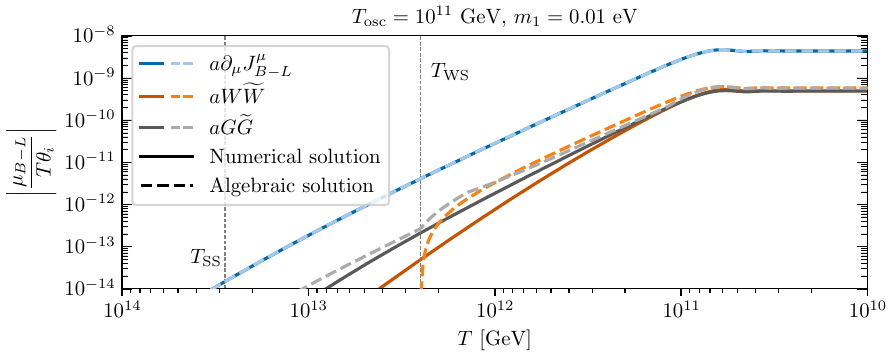}\\
    \includegraphics[width=\textwidth]{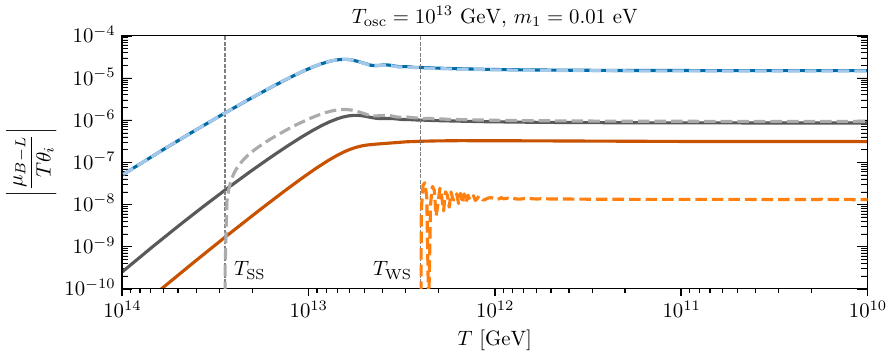}
    \caption{Same as Fig~\ref{fig:BPEvolution}, but for IO instead of NO. } 
    \label{fig:BPEvolutionIO}
\end{figure}

\begin{figure}
    \centering
    {{\includegraphics[width=0.99\textwidth]{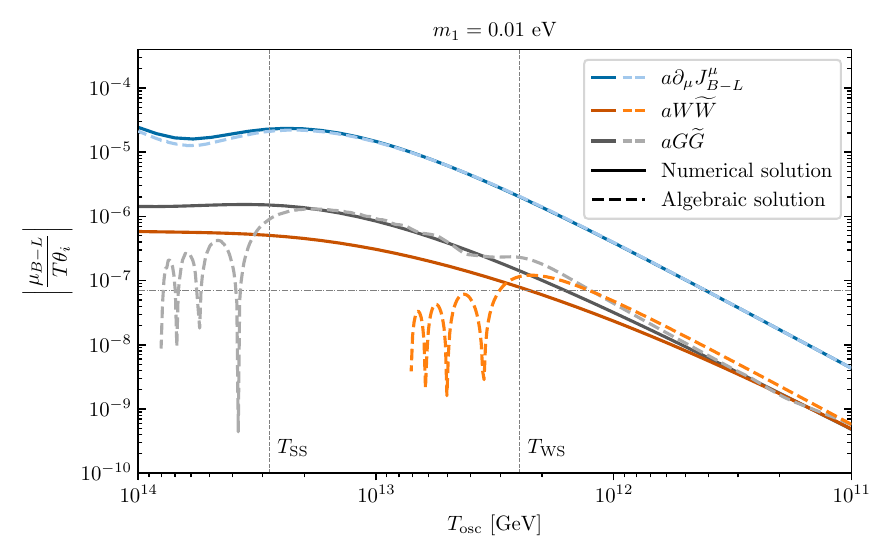}}}
    \caption{Same as Fig~\ref{fig:ParameterSpace1}, but for IO instead of NO. }
    \label{fig:ParameterSpace1_IO}
\end{figure}

\begin{figure}
    \centering
    {{\includegraphics[width=0.99\textwidth]{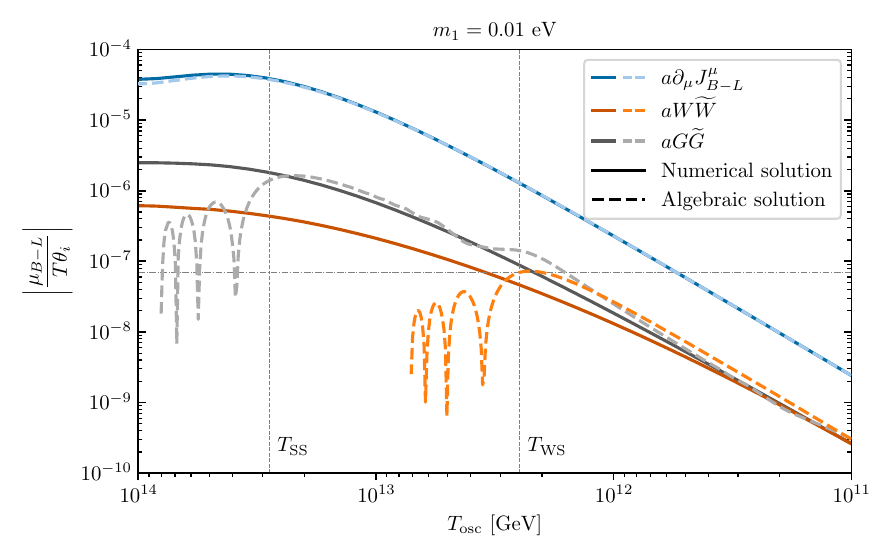}}}
    \caption{Same as Fig~\ref{fig:ParameterSpace1}, but in $(1,\ 2,\ \tau)$-basis instead of $(1,\ 2,\ 3)$-basis. }
    \label{fig:ParameterSpace1_Tau_Basis}
\end{figure}

Fig.~\ref{fig:BPEvolutionIO} shows example solutions of $\abs{\mu_{B-L}/(T\theta_i)}$ for IO. This result is qualitatively similar to the corresponding result for NO, which was shown in Fig.~\ref{fig:BPEvolution}. Fig.~\ref{fig:ParameterSpace1_IO} shows a parameter scan over $T_{\rm osc}$ for IO with the lightest neutrino mass fixed to $0.01$ eV, which should be compared with the corresponding NO result in Fig.~\ref{fig:ParameterSpace1}. The result in Fig.~\ref{fig:ParameterSpace1_IO} is similar to that in Fig.~\ref{fig:ParameterSpace1}, with the exception that for the Majorana-like case (blue curves) and large $T_{\rm osc}$, the algebraic solution deviates more strongly from the full numerical result than in the case of normal ordering. Fig.~\ref{fig:ParameterSpace1_Tau_Basis} is the same as Fig.~\ref{fig:ParameterSpace1} in the main text, except that the calculation has been performed in the $(1,\,2,\,\tau)$-basis instead of the neutrino-mass basis. 

\subsection{Final parameter space}
\label{appendix:FinalParameterSpace}
In this section, we present additional figures that complement the results presented in Sec.~\ref{sec:FinalParameter}. While the main text focuses exclusively on NO, Section~\ref{appendix:IOParameter} includes the corresponding results for IO. For completeness, additional NO results omitted from Sec.~\ref{sec:FinalParameter} are collected in Sec.~\ref{appendix:NOParameter}.

\subsubsection{Normal mass ordering}
\label{appendix:NOParameter}
\begin{figure}[]%
    \centering
    {\includegraphics[width=0.5\textwidth]{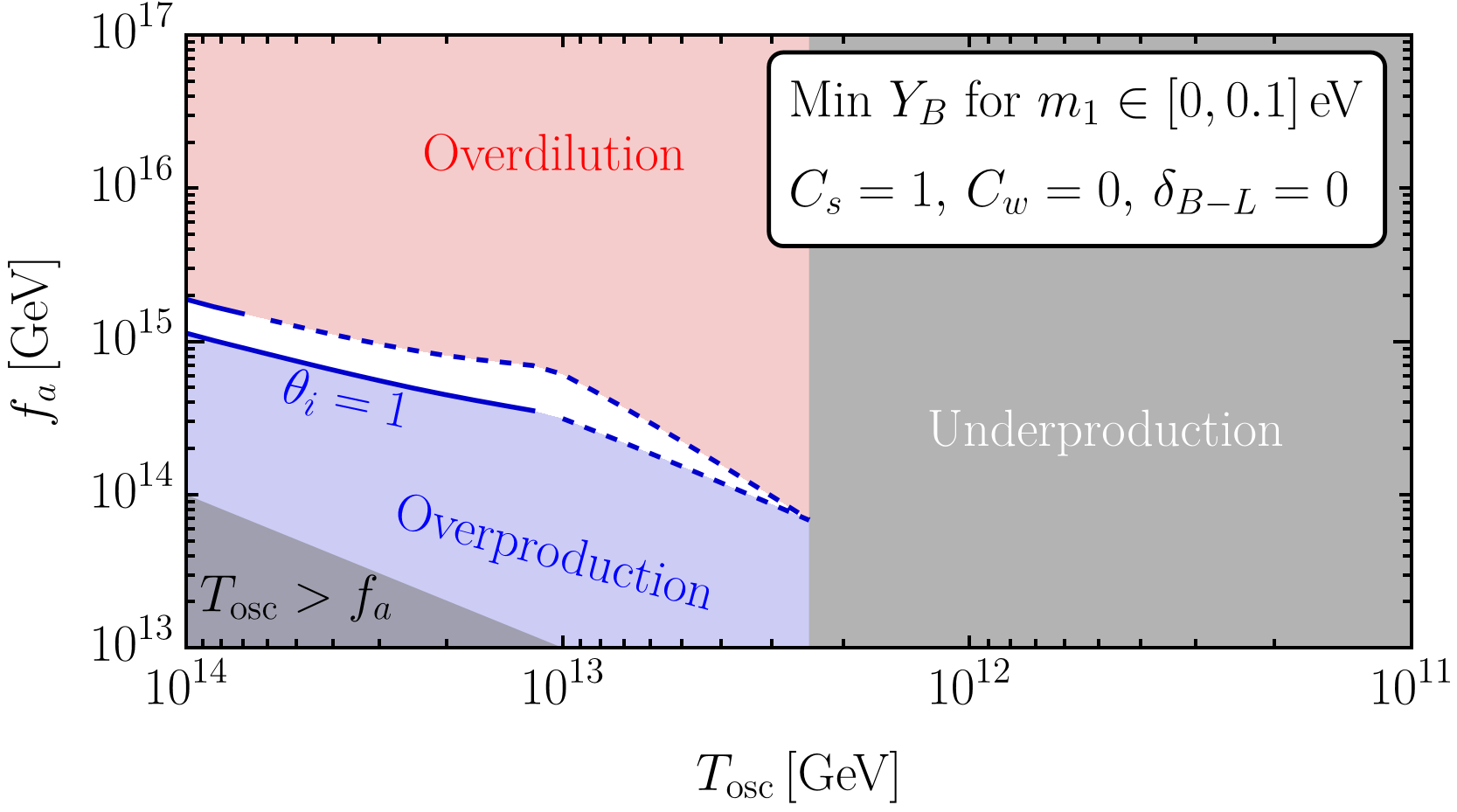}}\hfill
    {\includegraphics[width=0.5\textwidth]{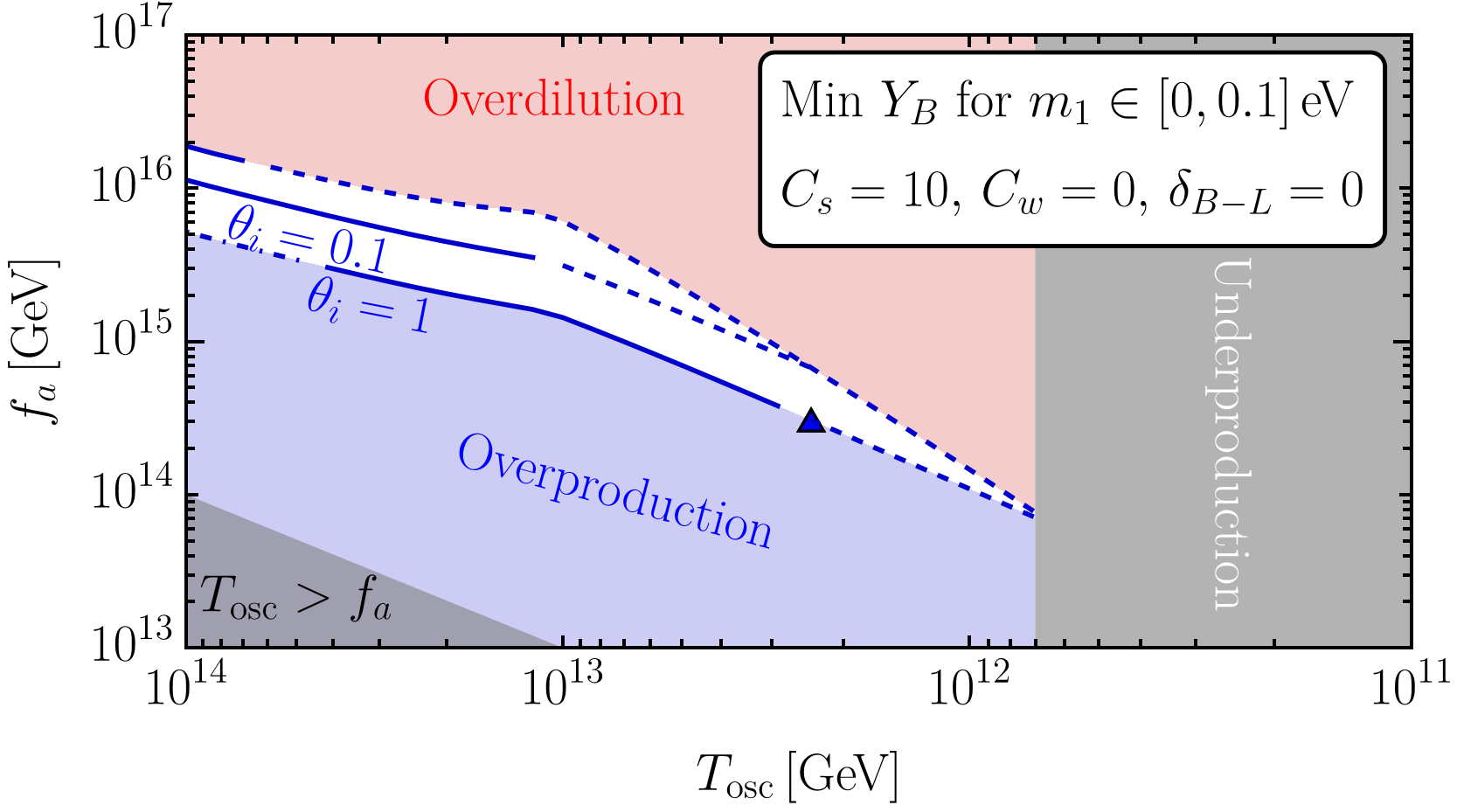}}\hfill
    {\includegraphics[width=0.5\textwidth]{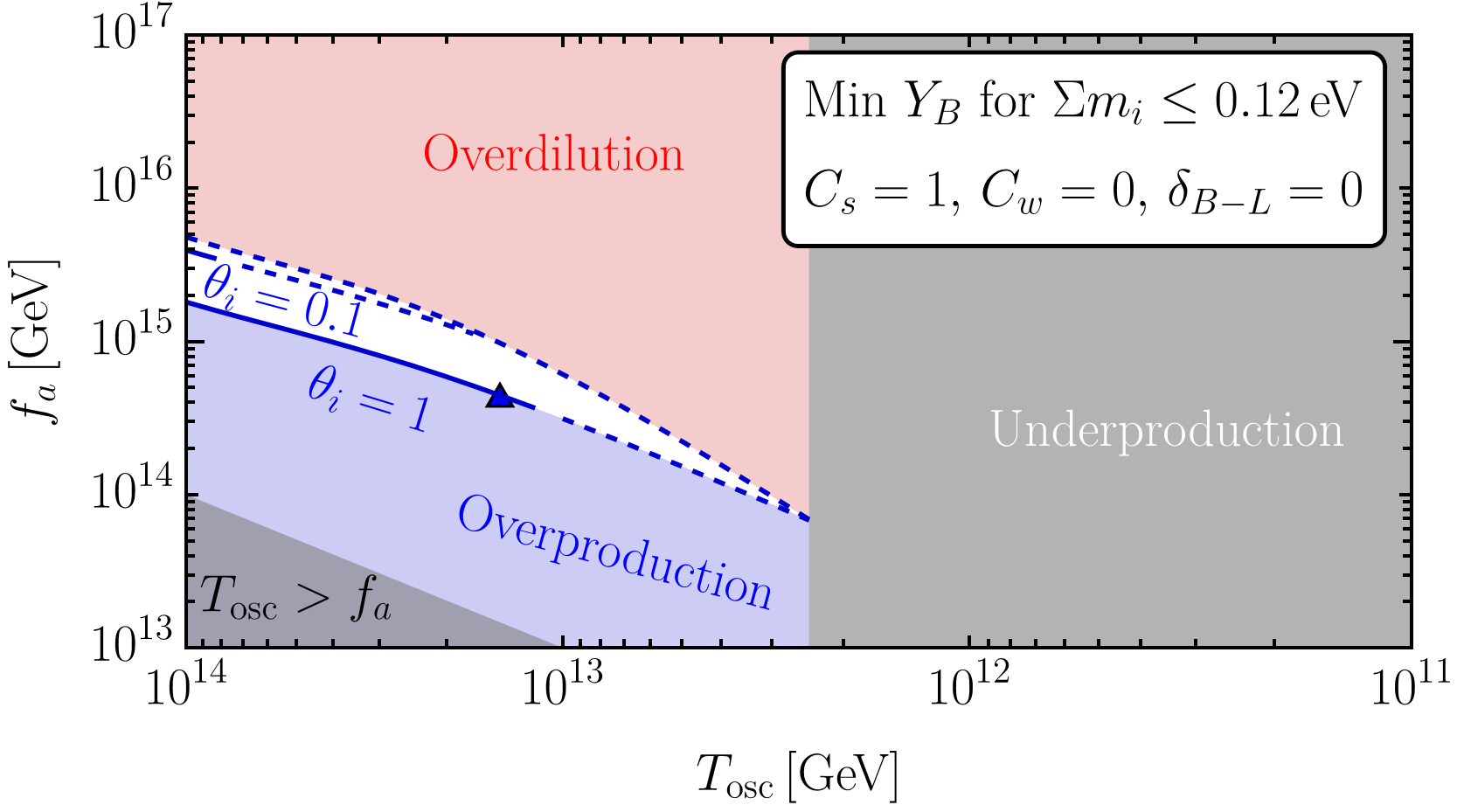}}\hfill
    {\includegraphics[width=0.5\textwidth]{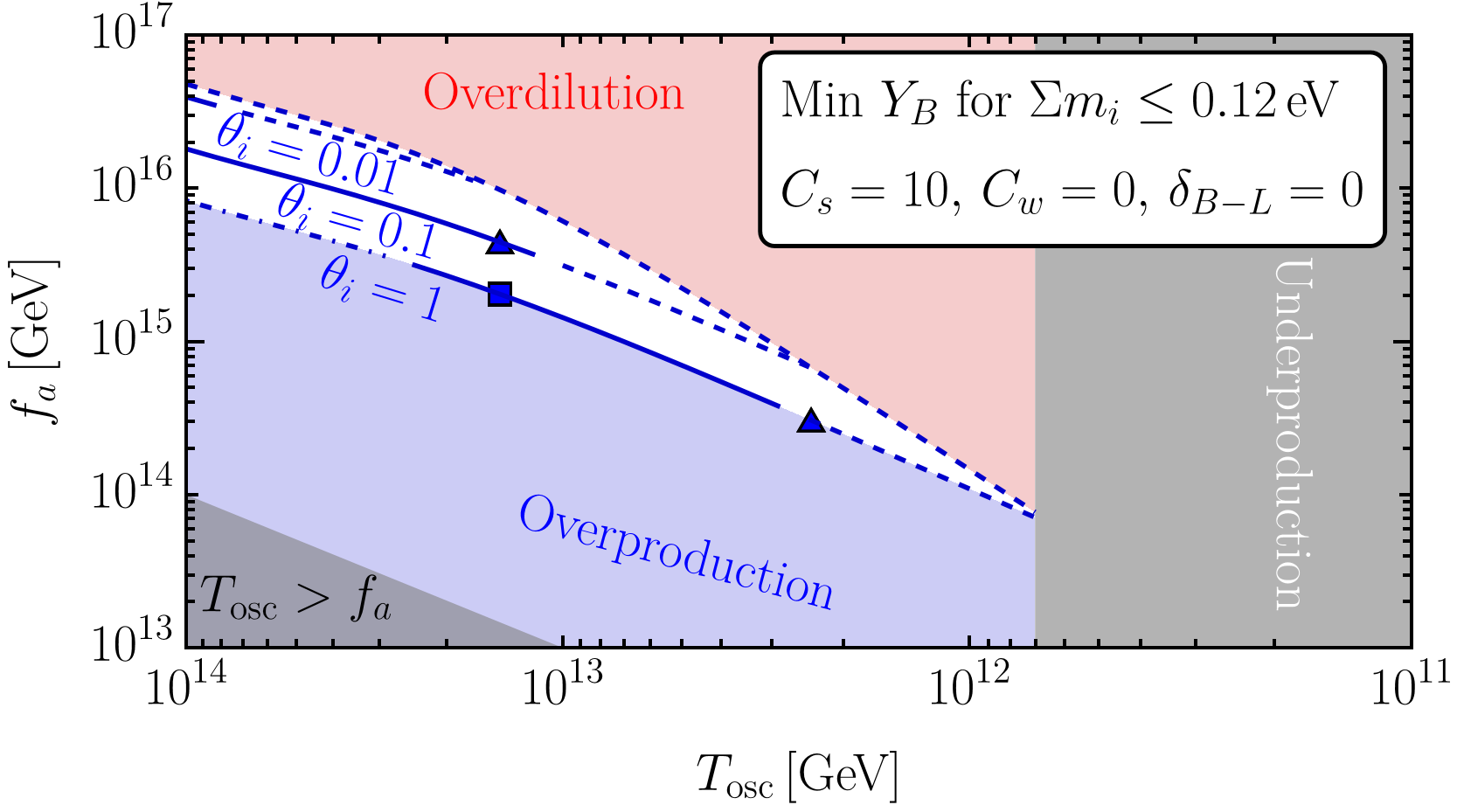}}\hfill
    \caption{Similar to left panels in Fig.~\ref{fig:SSParameterSpace}, but with $m_1$ chosen such that the produced BAU is minimized within $m_1\in [0,\,0.1]$ eV (first row) and $\sum m_i\leq 0.12$ eV (second row). }
    \label{fig:SSParameterSpaceMin}
\end{figure}

\begin{figure}[]%
    \centering
    {\includegraphics[width=0.5\textwidth]{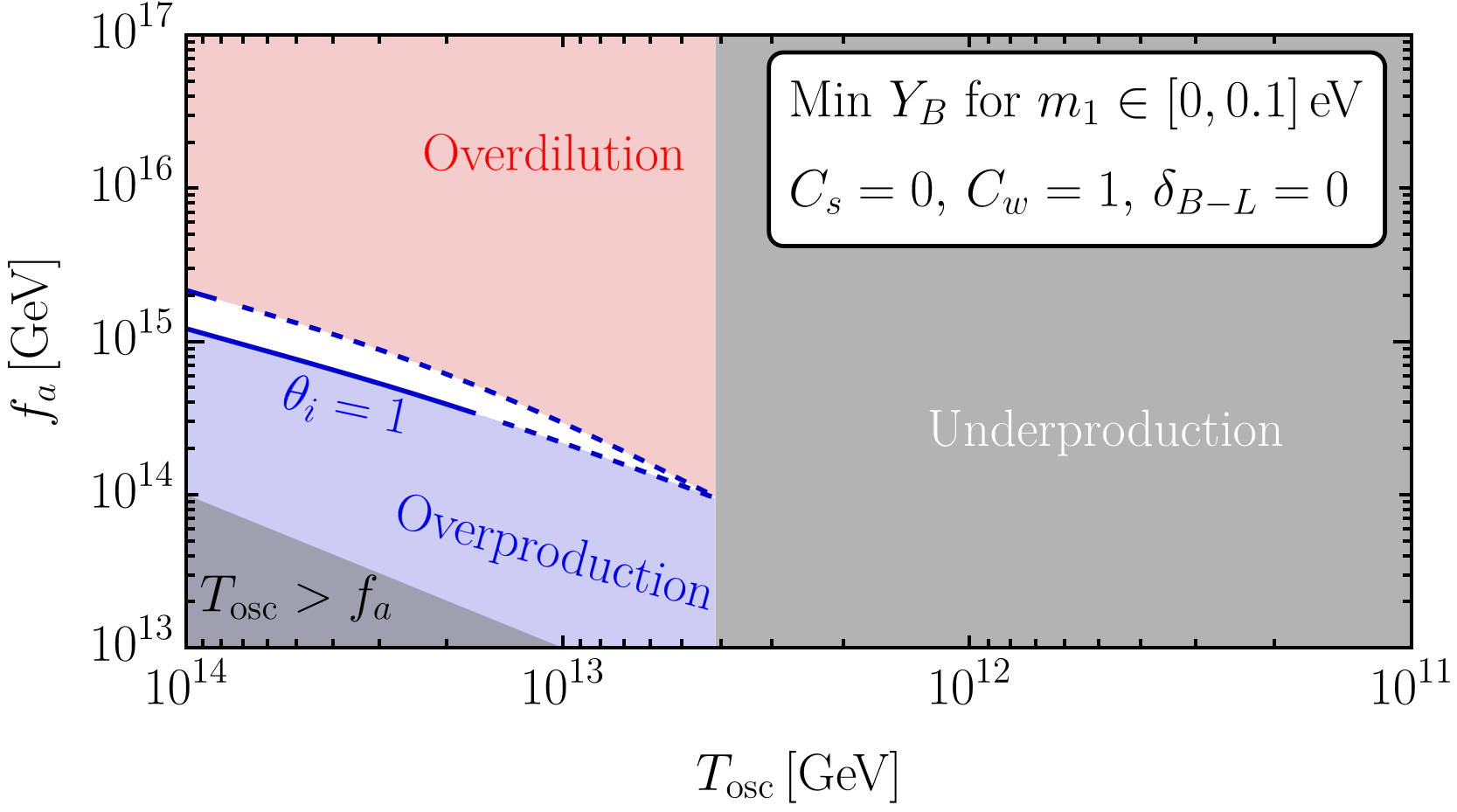}}\hfill
    {\includegraphics[width=0.5\textwidth]{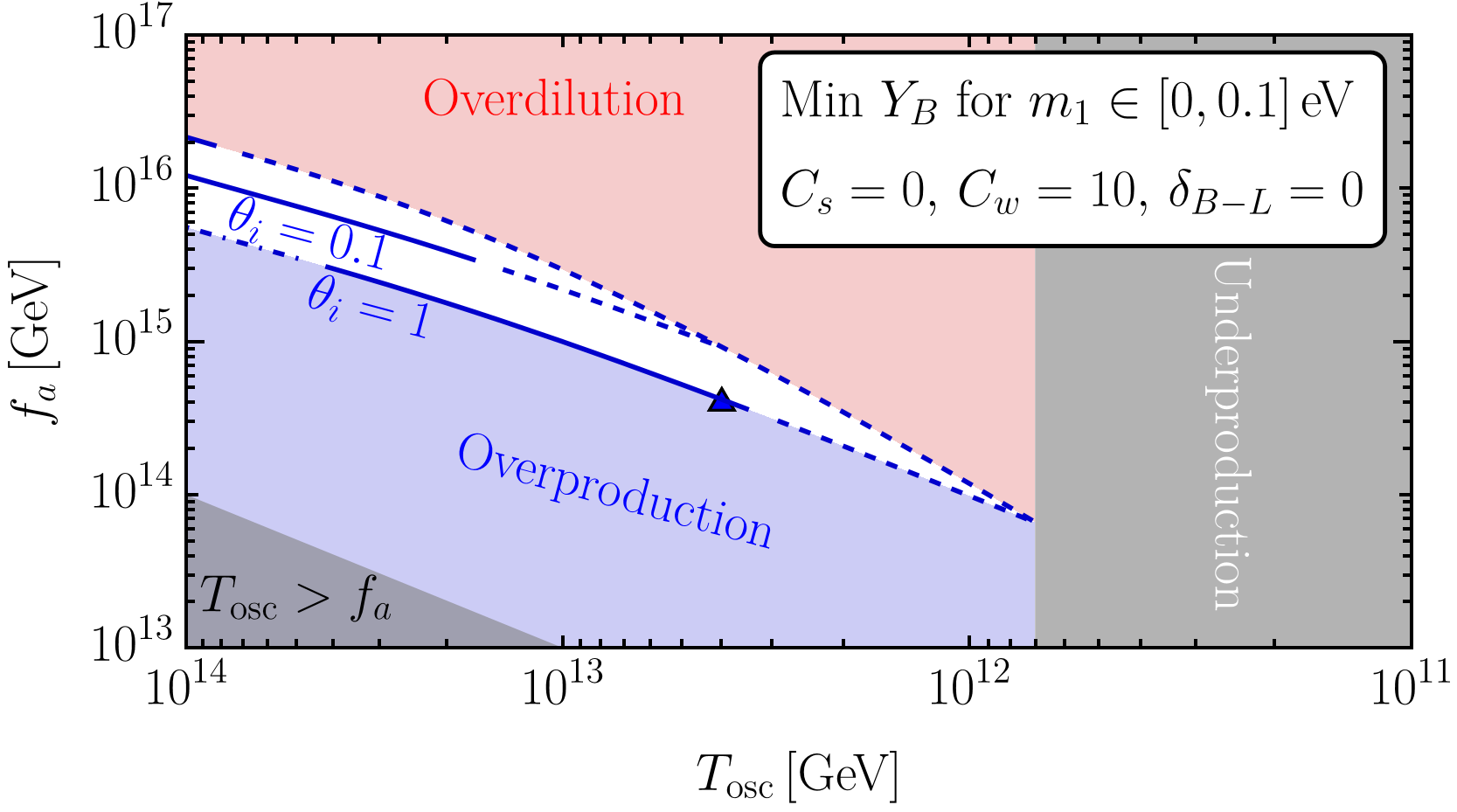}}\hfill
    {\includegraphics[width=0.5\textwidth]{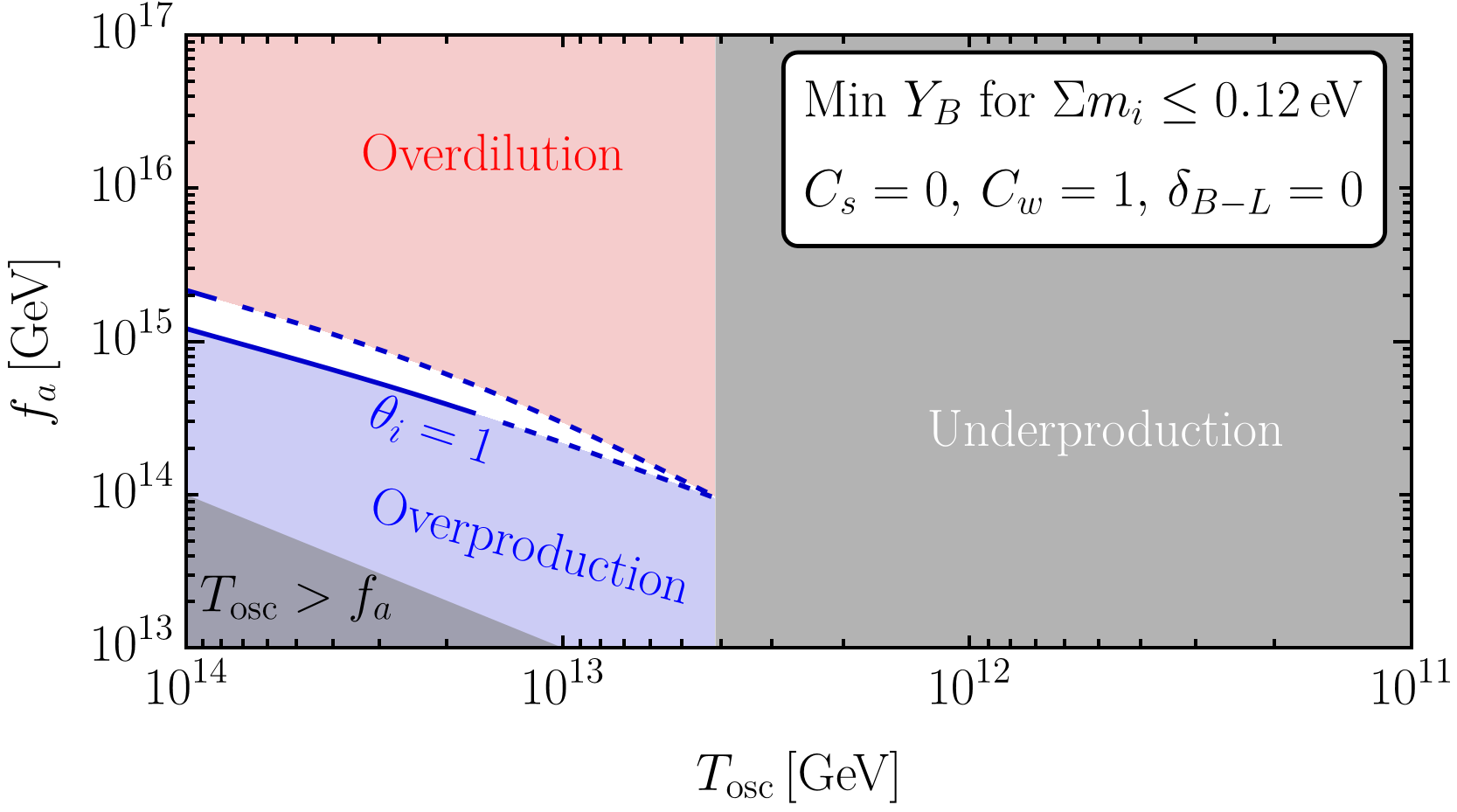}}\hfill
    {\includegraphics[width=0.5\textwidth]{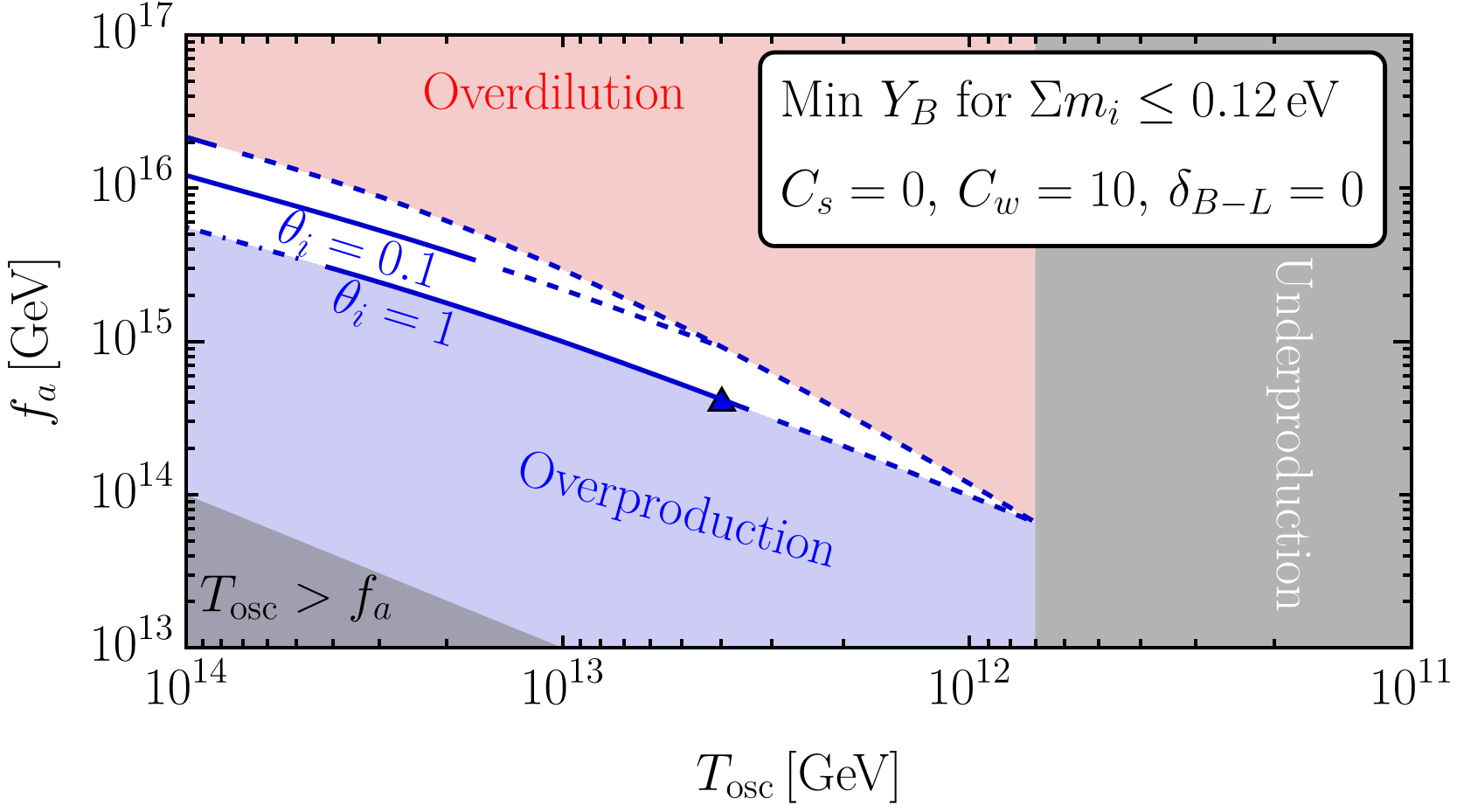}}\hfill
    \caption{Similar to left panels in Fig.~\ref{fig:WSParameterSpace}, but with $m_1$ chosen such that the produced BAU is minimized within $m_1\in [0,\,0.1]$ eV (first row) and $\sum m_i\leq 0.12$ eV (second row). }
    \label{fig:WSParameterSpaceMin}
\end{figure}

In the left panels of Figs.~\ref{fig:SSParameterSpace} and~\ref{fig:WSParameterSpace}, the value of $m_1$ was chosen to maximize the produced asymmetry within the ranges $m_1 \in [0, 0.1]$~eV (first and second rows) and $\sum m_i \leq 0.12$~eV (third and fourth rows) for nonzero $C_s$ and $C_w$, respectively. Fig.~\ref{fig:SSParameterSpaceMin} and~\ref{fig:WSParameterSpaceMin} display the opposite limit, in which the value of $m_1$ is chosen to minimize the produced asymmetry within the same ranges for $m_1$, as indicated in the different panels. 

\subsubsection{Inverted mass ordering}
\label{appendix:IOParameter}
\begin{figure}[]%
    \centering
    {\includegraphics[width=0.5\textwidth]{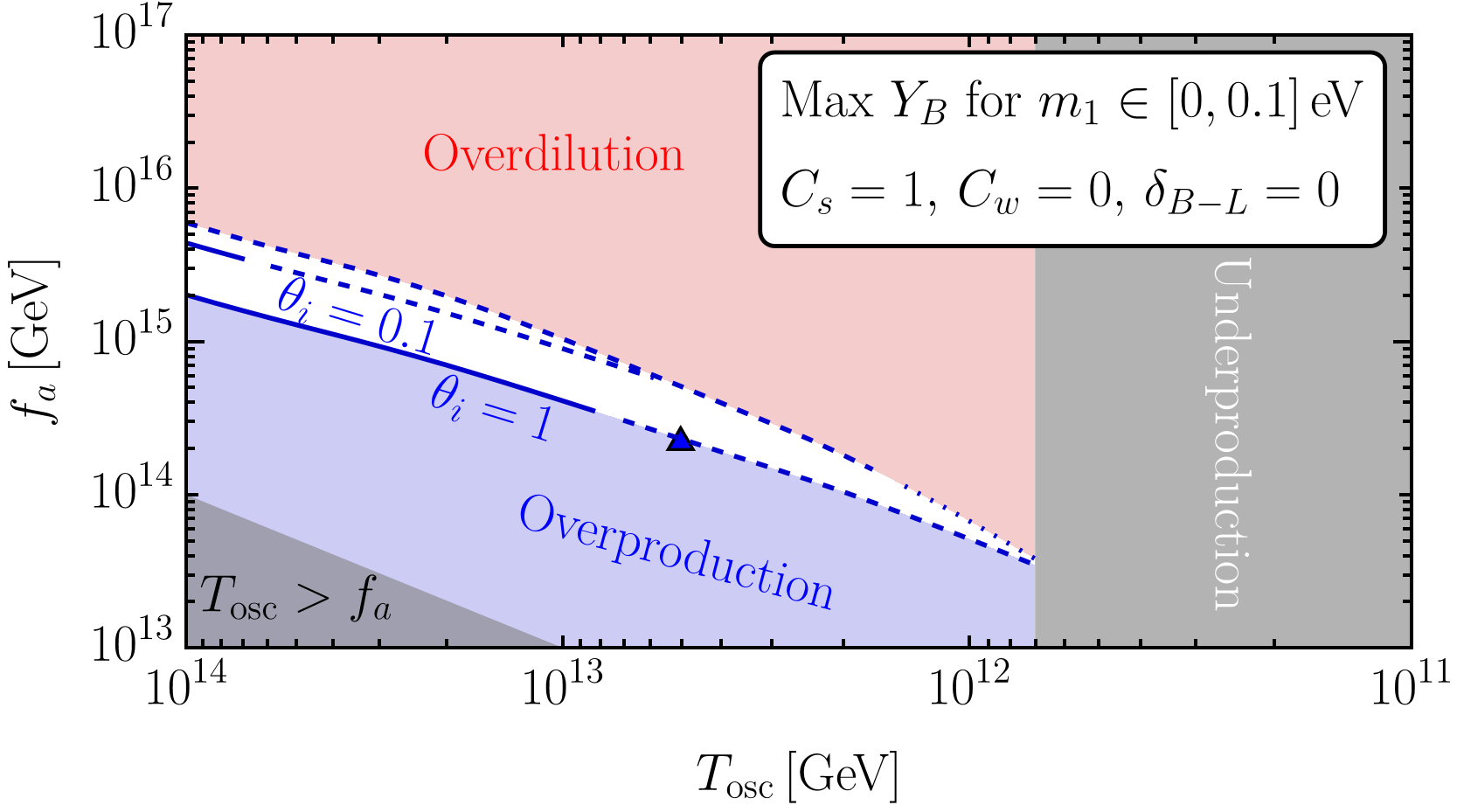}}\hfill
    {\includegraphics[width=0.5\textwidth]{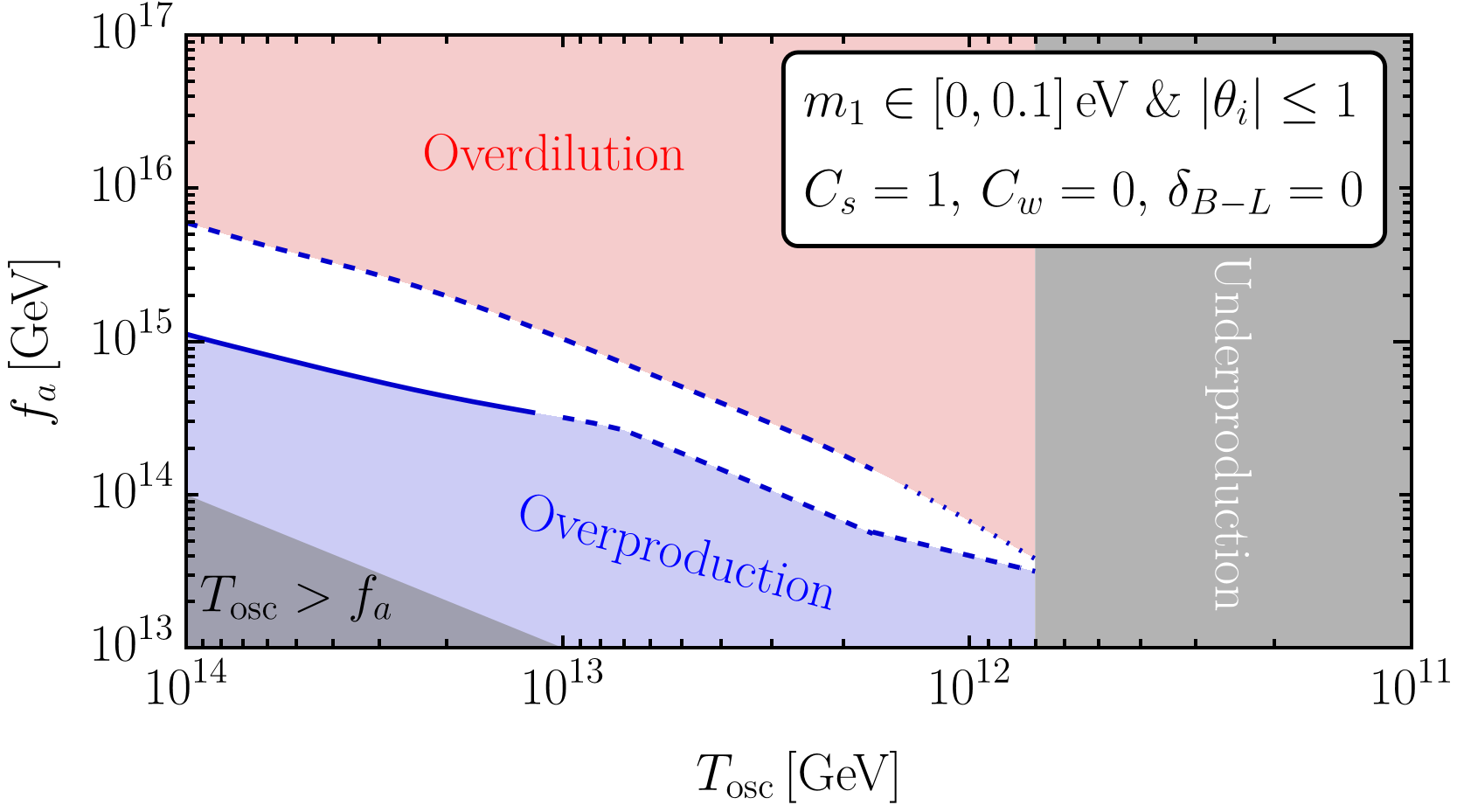}}\hfill
    {\includegraphics[width=0.5\textwidth]{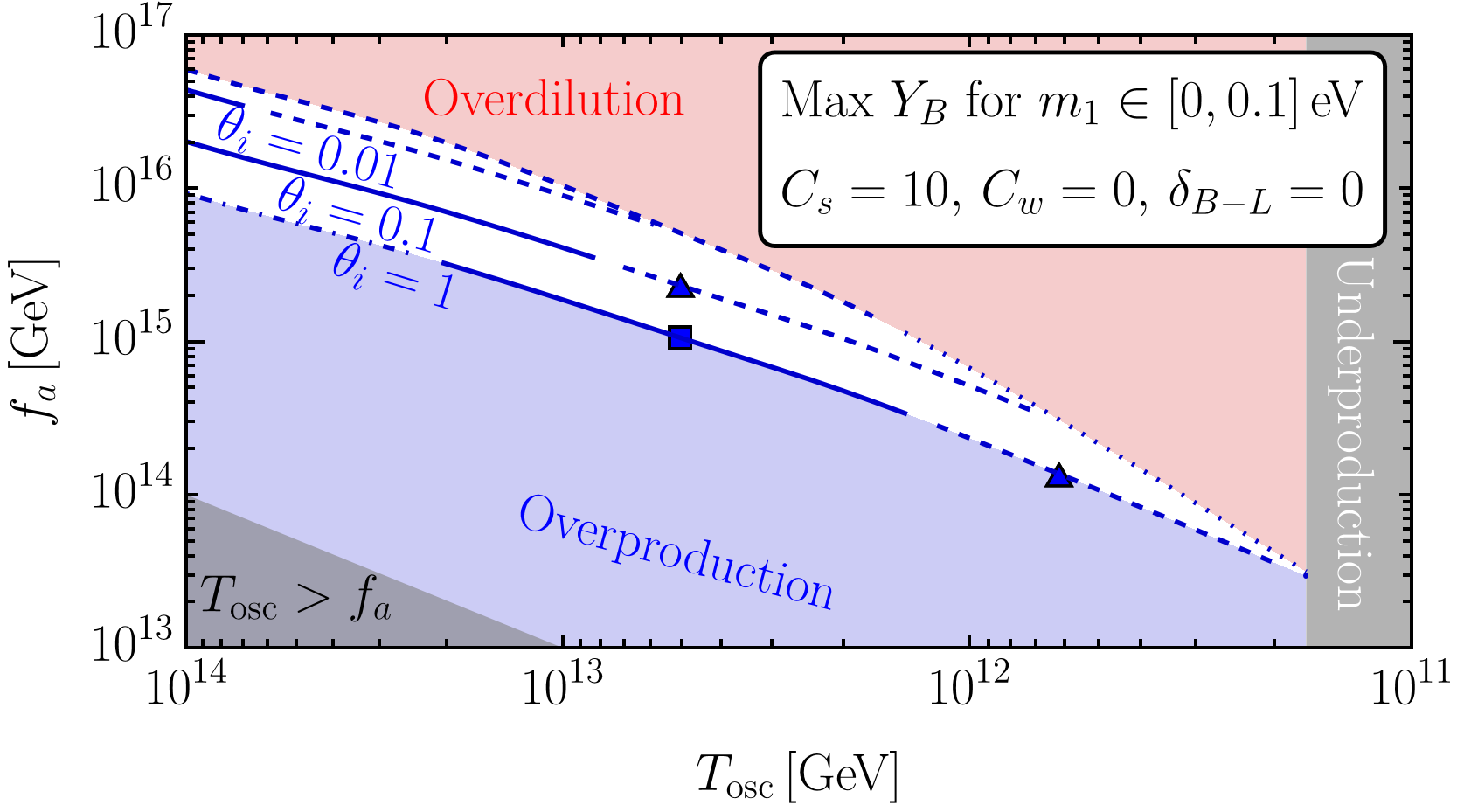}}\hfill
    {\includegraphics[width=0.5\textwidth]{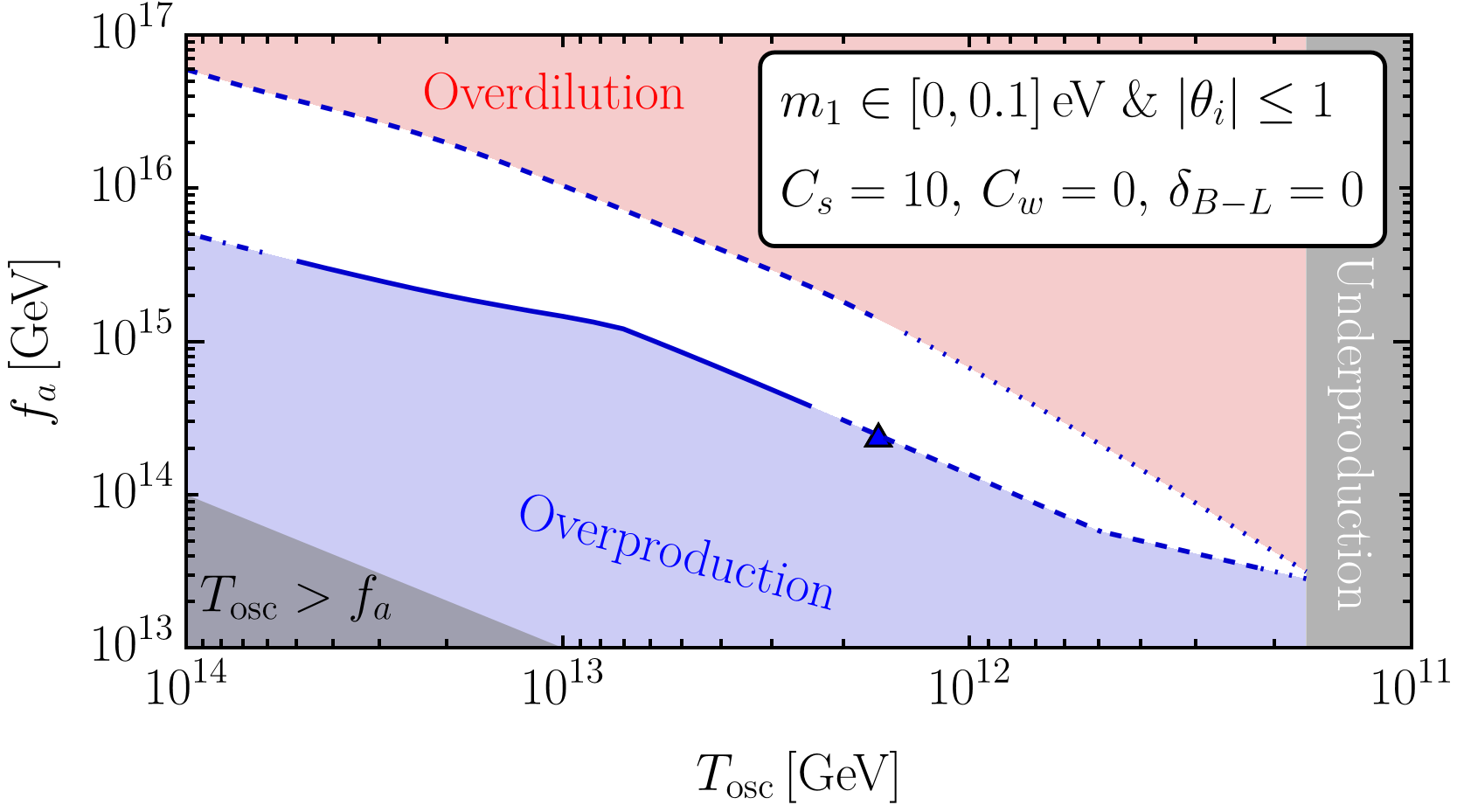}}\hfill
    {{\includegraphics[width=0.5\textwidth]{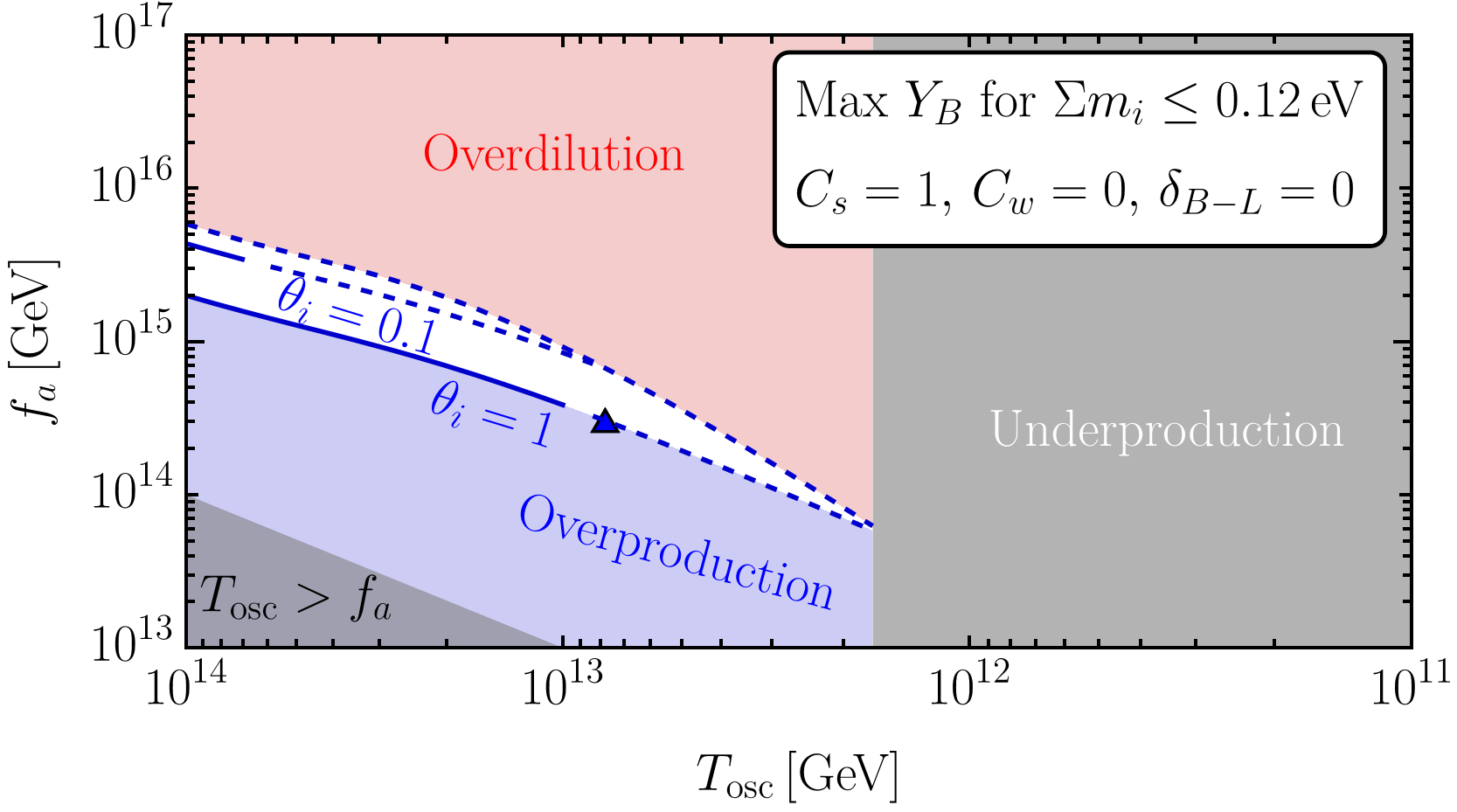}}}\hfill
    {{\includegraphics[width=0.5\textwidth]{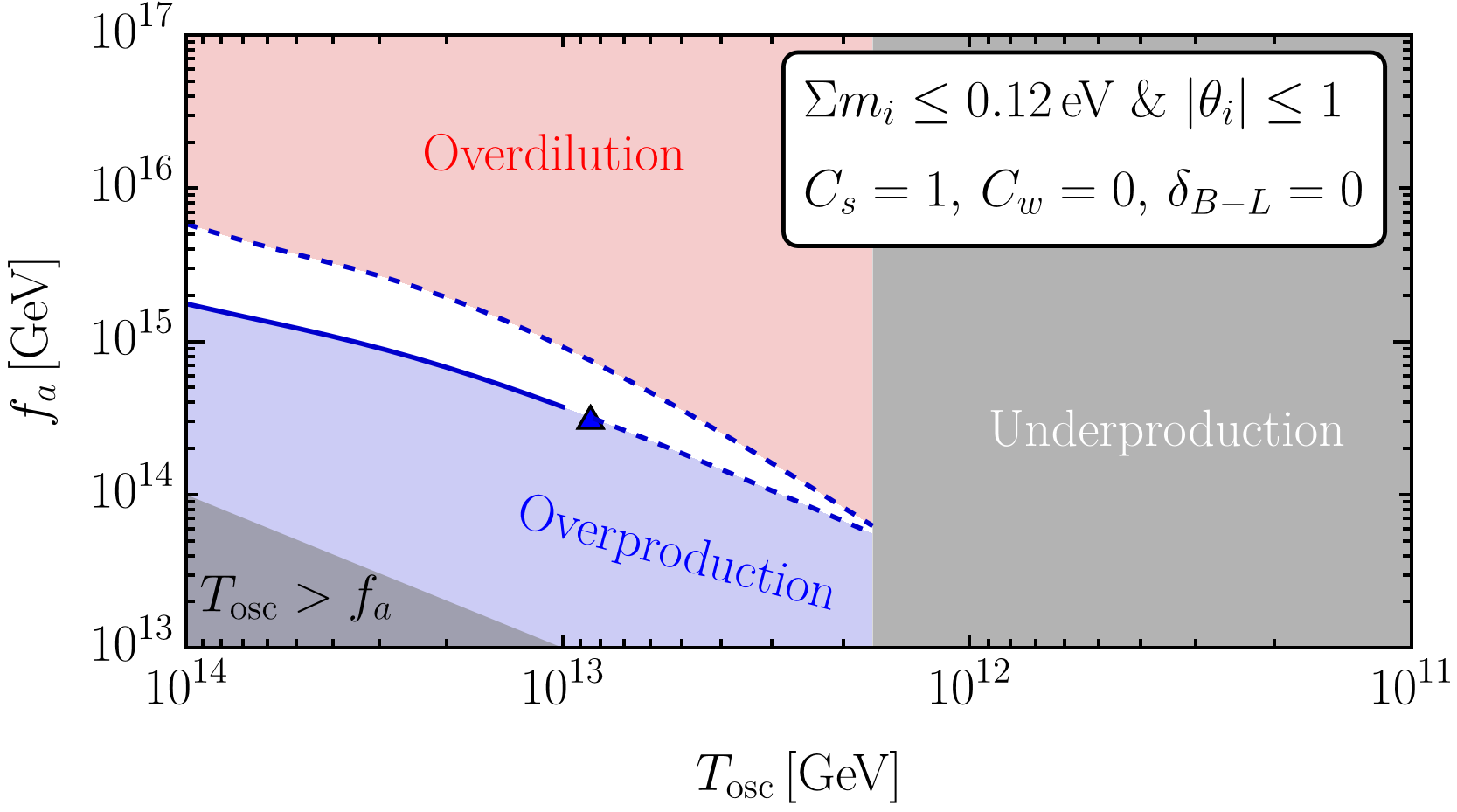}}}\hfill
    {{\includegraphics[width=0.5\textwidth]{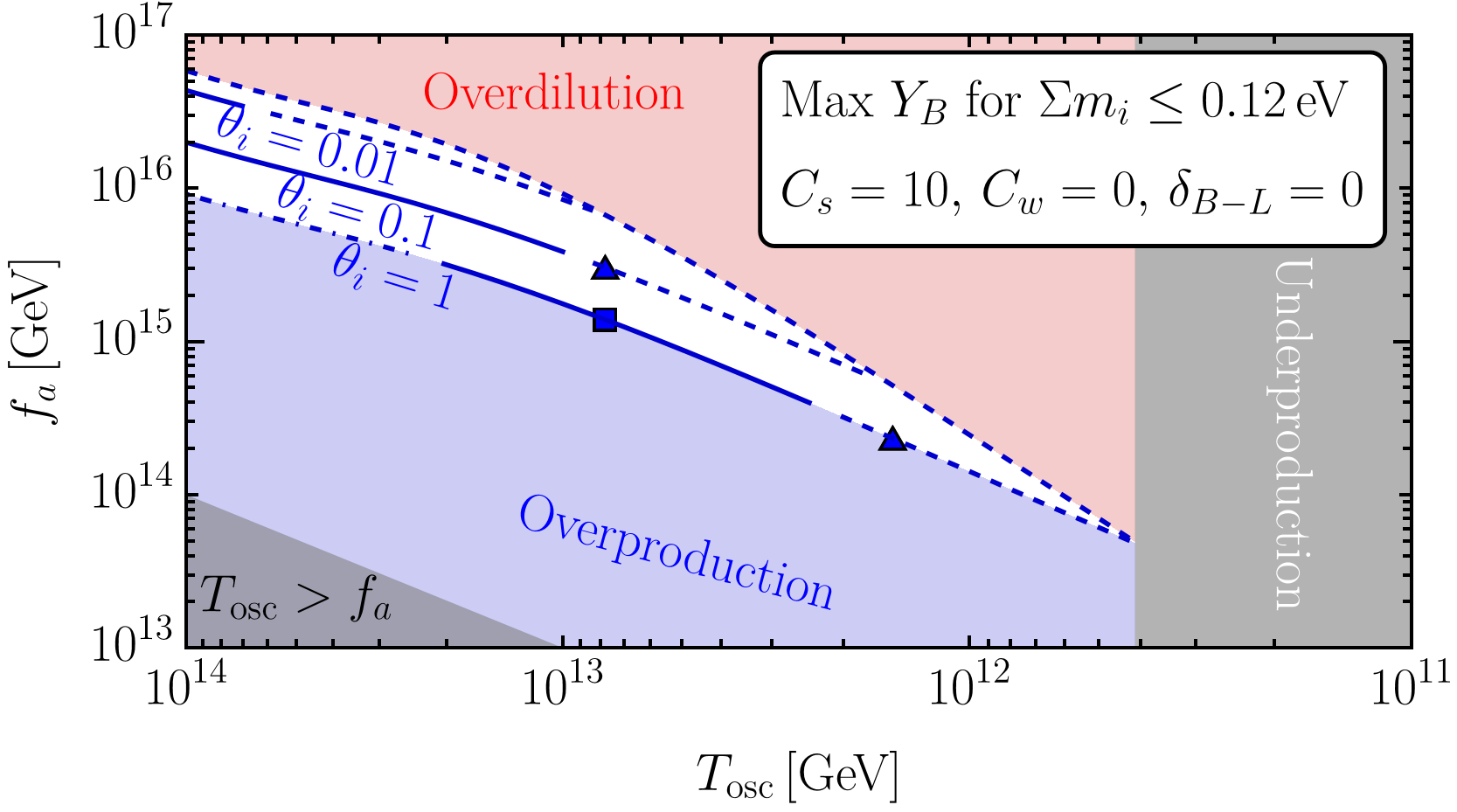}}}\hfill
    {{\includegraphics[width=0.5\textwidth]{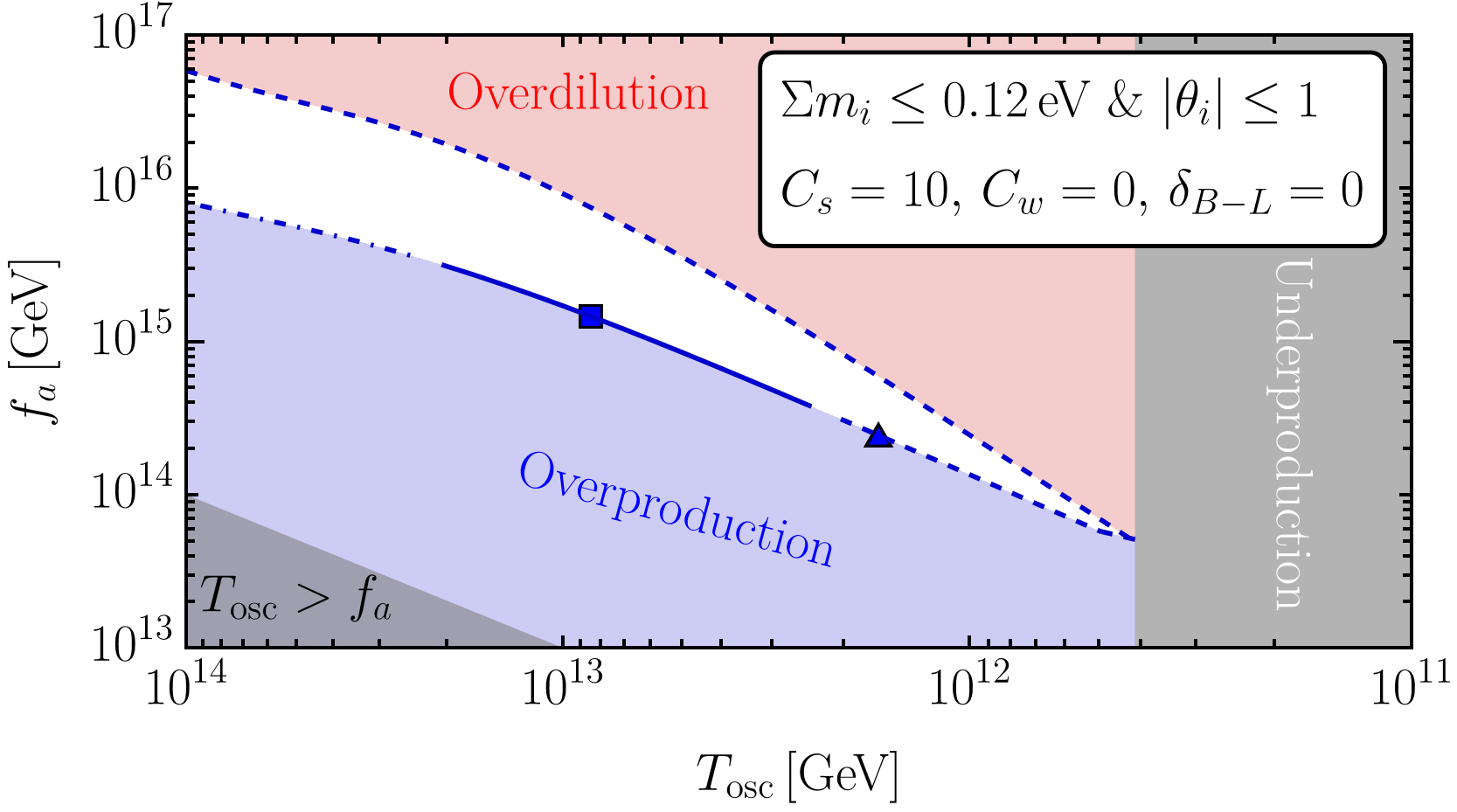}}}\hfill
    \caption{Same as Fig.~\ref{fig:SSParameterSpace}, but for IO instead of NO. }
    \label{fig:SSParameterSpaceIO}
\end{figure}

\begin{figure}[]%
    \centering
    {\includegraphics[width=0.5\textwidth]{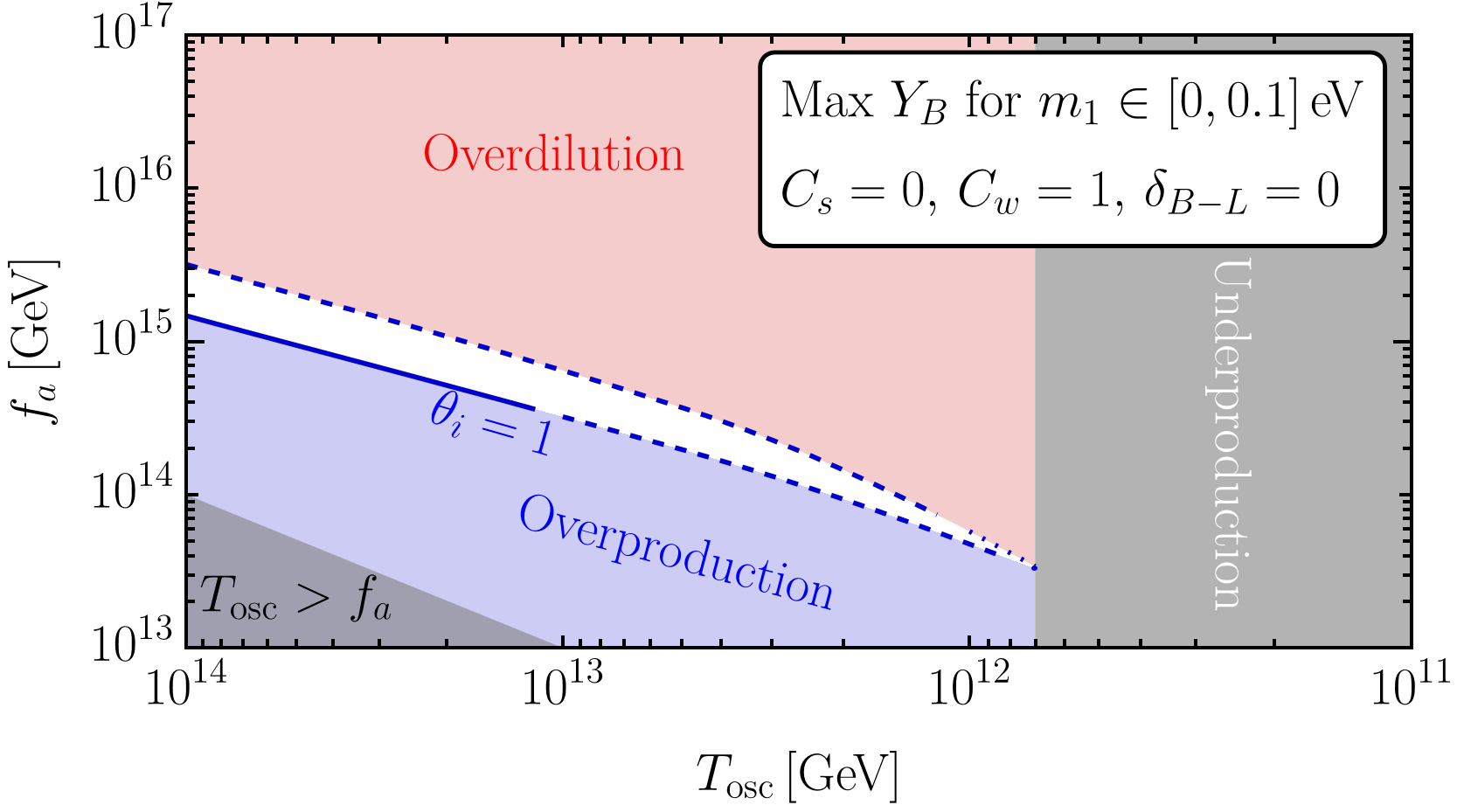}}\hfill
    {\includegraphics[width=0.5\textwidth]{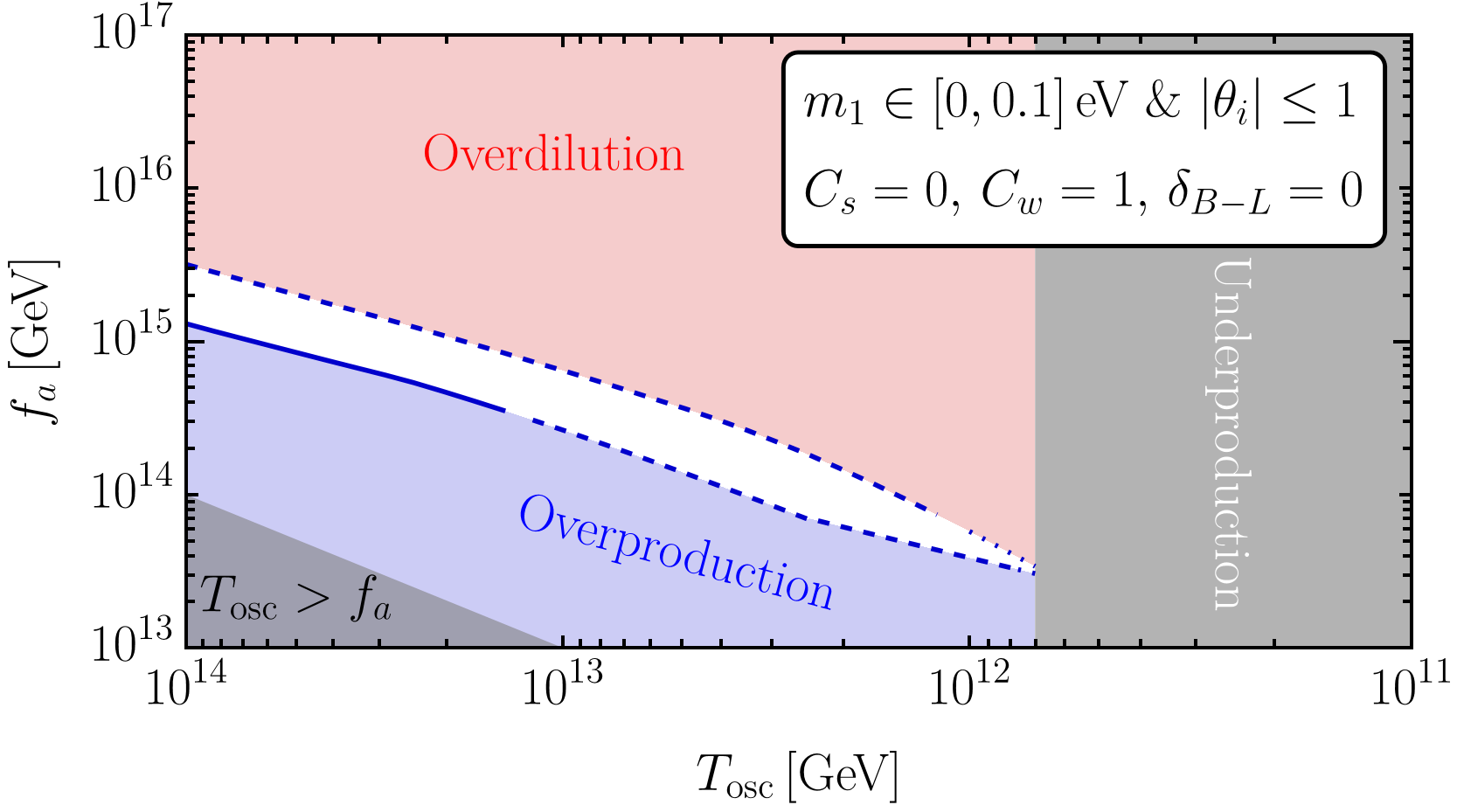}}\hfill
    {\includegraphics[width=0.5\textwidth]{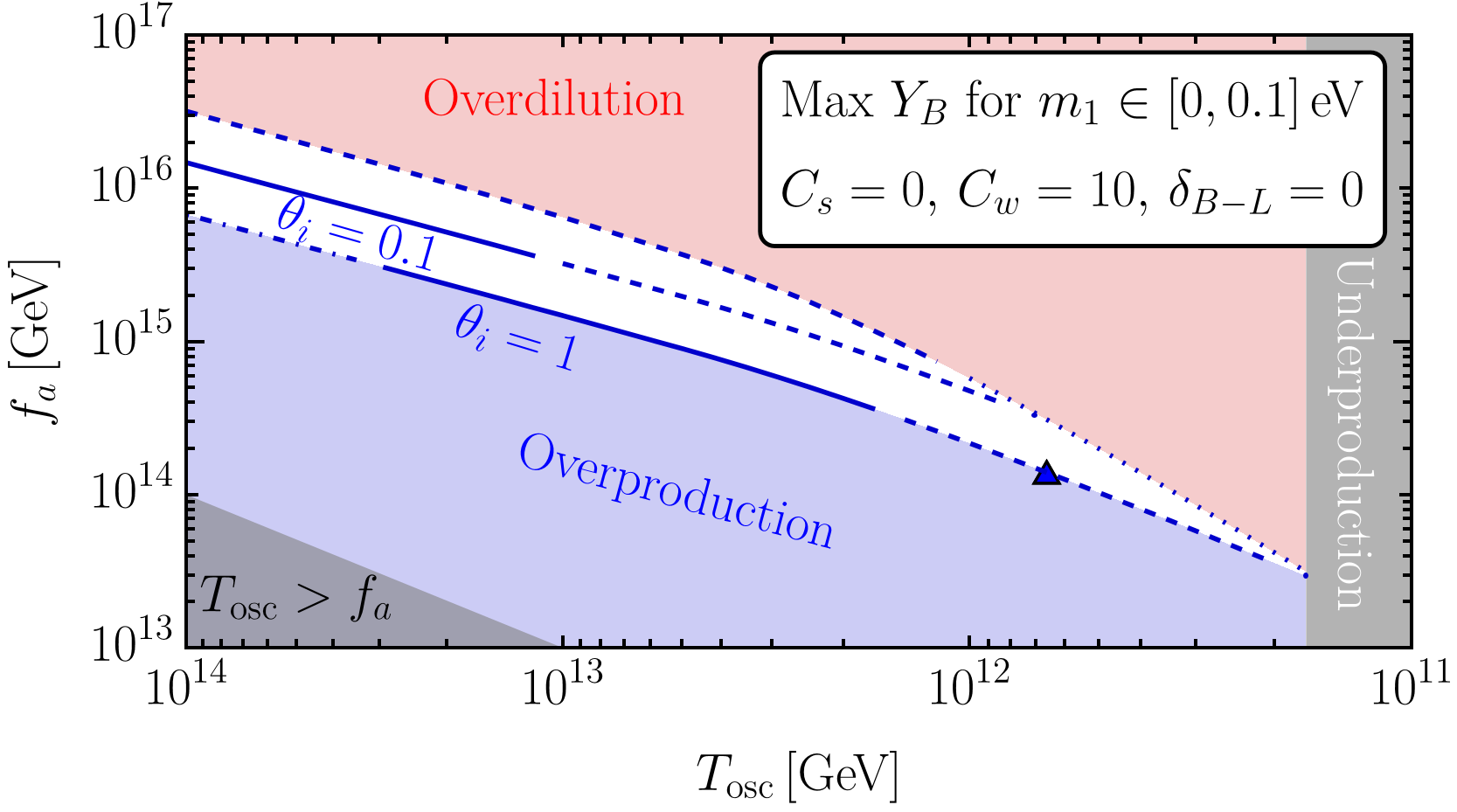}}\hfill
    {\includegraphics[width=0.5\textwidth]{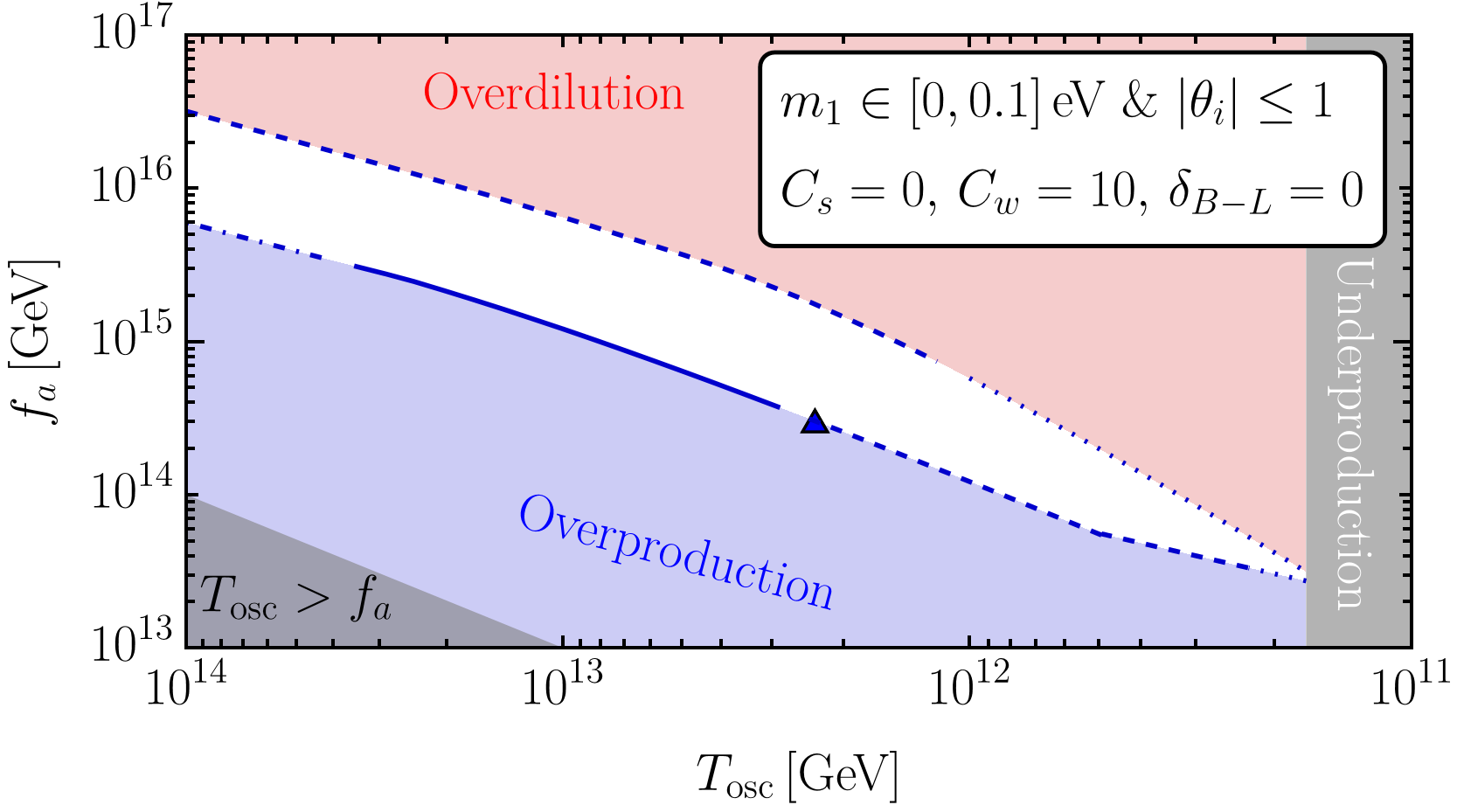}}\hfill
    {{\includegraphics[width=0.5\textwidth]{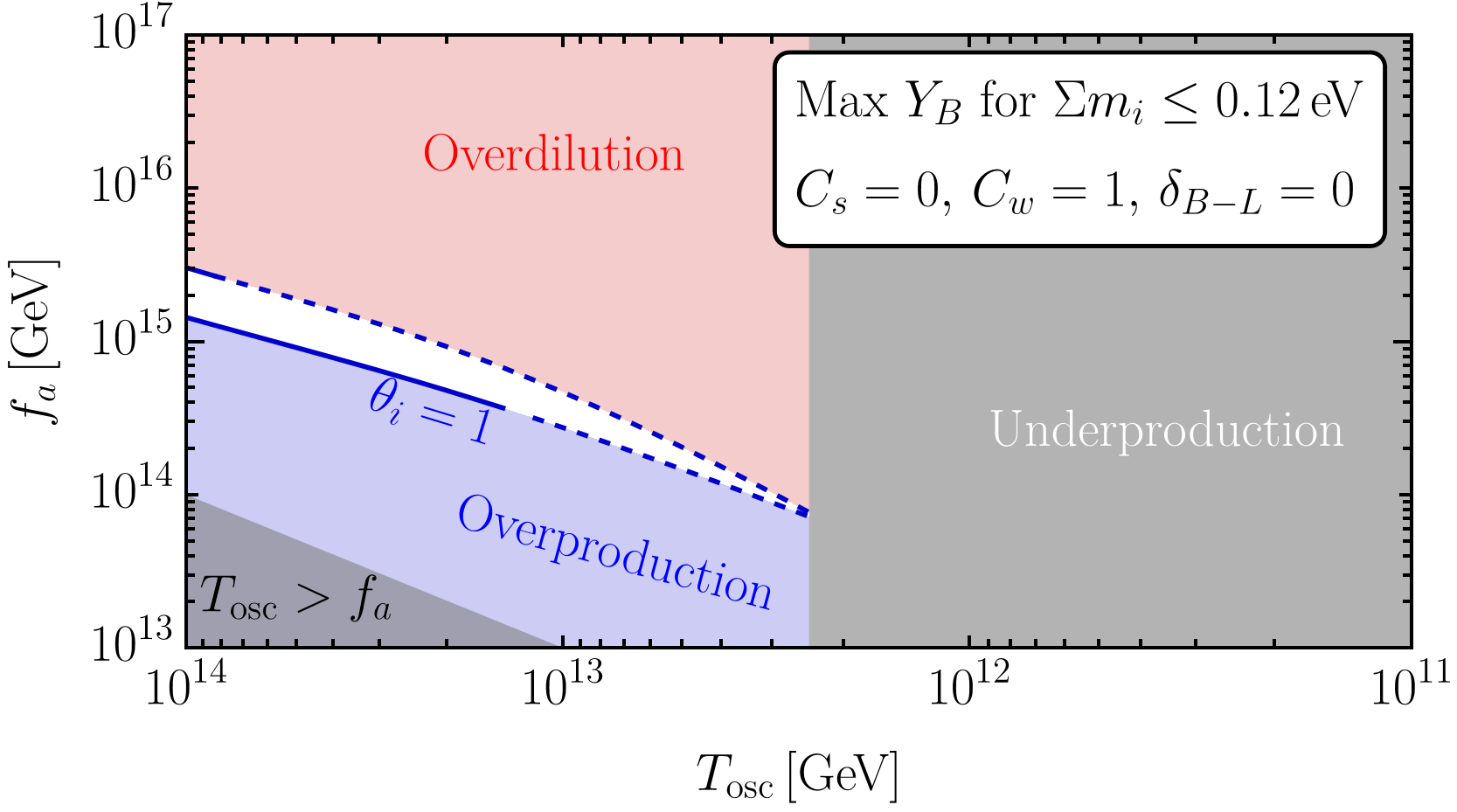}}}\hfill
    {{\includegraphics[width=0.5\textwidth]{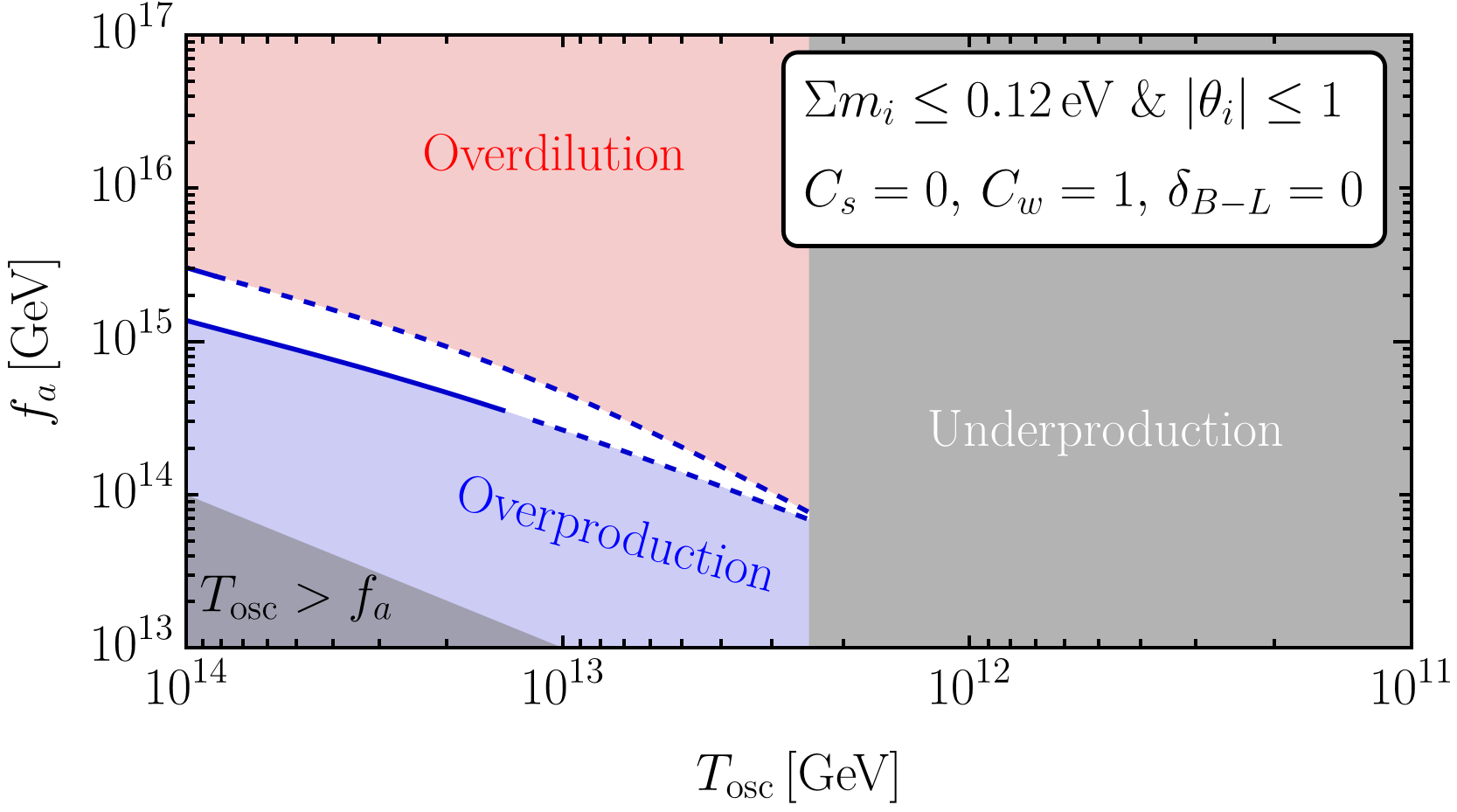}}}\hfill
    {{\includegraphics[width=0.5\textwidth]{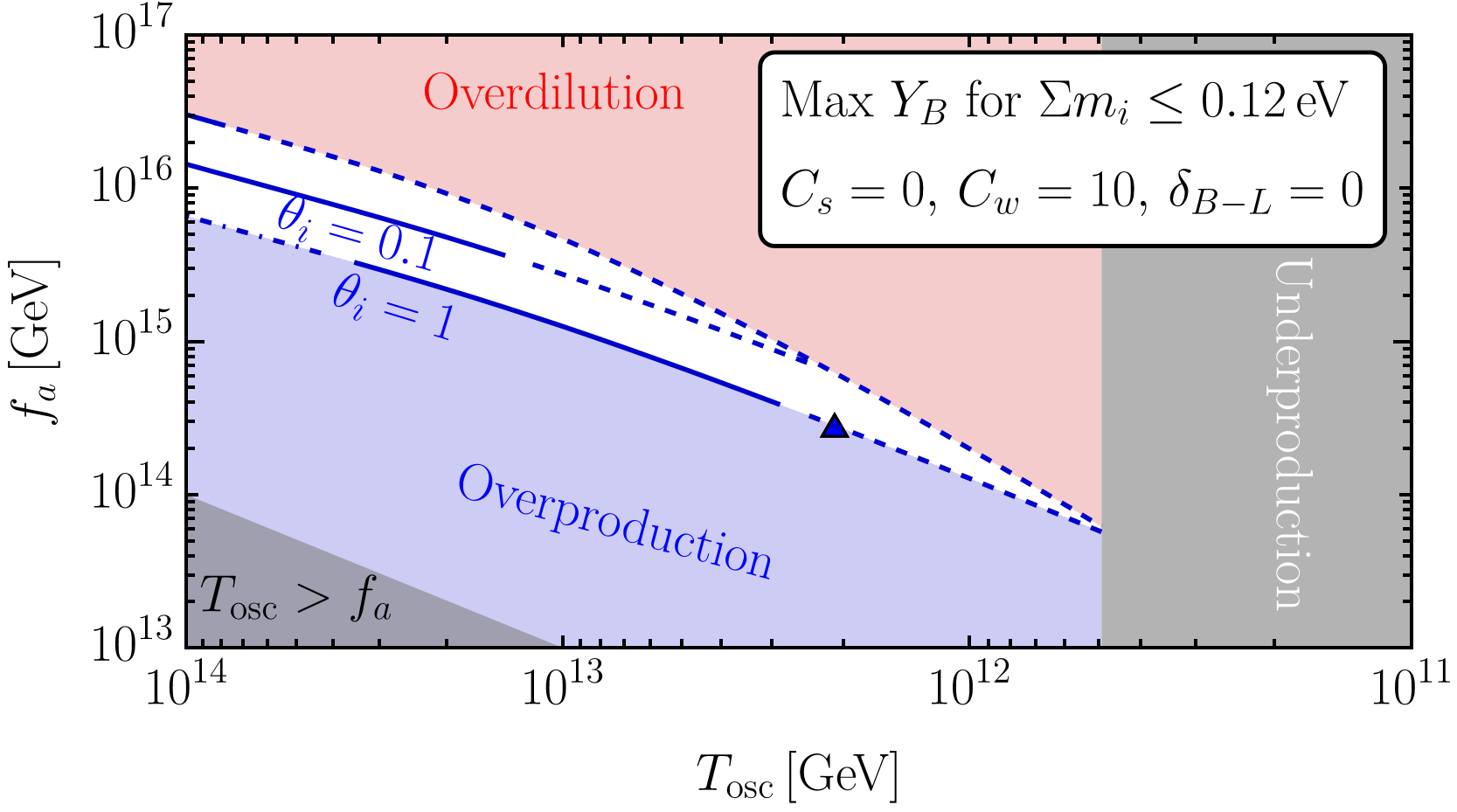}}}\hfill
    {{\includegraphics[width=0.5\textwidth]{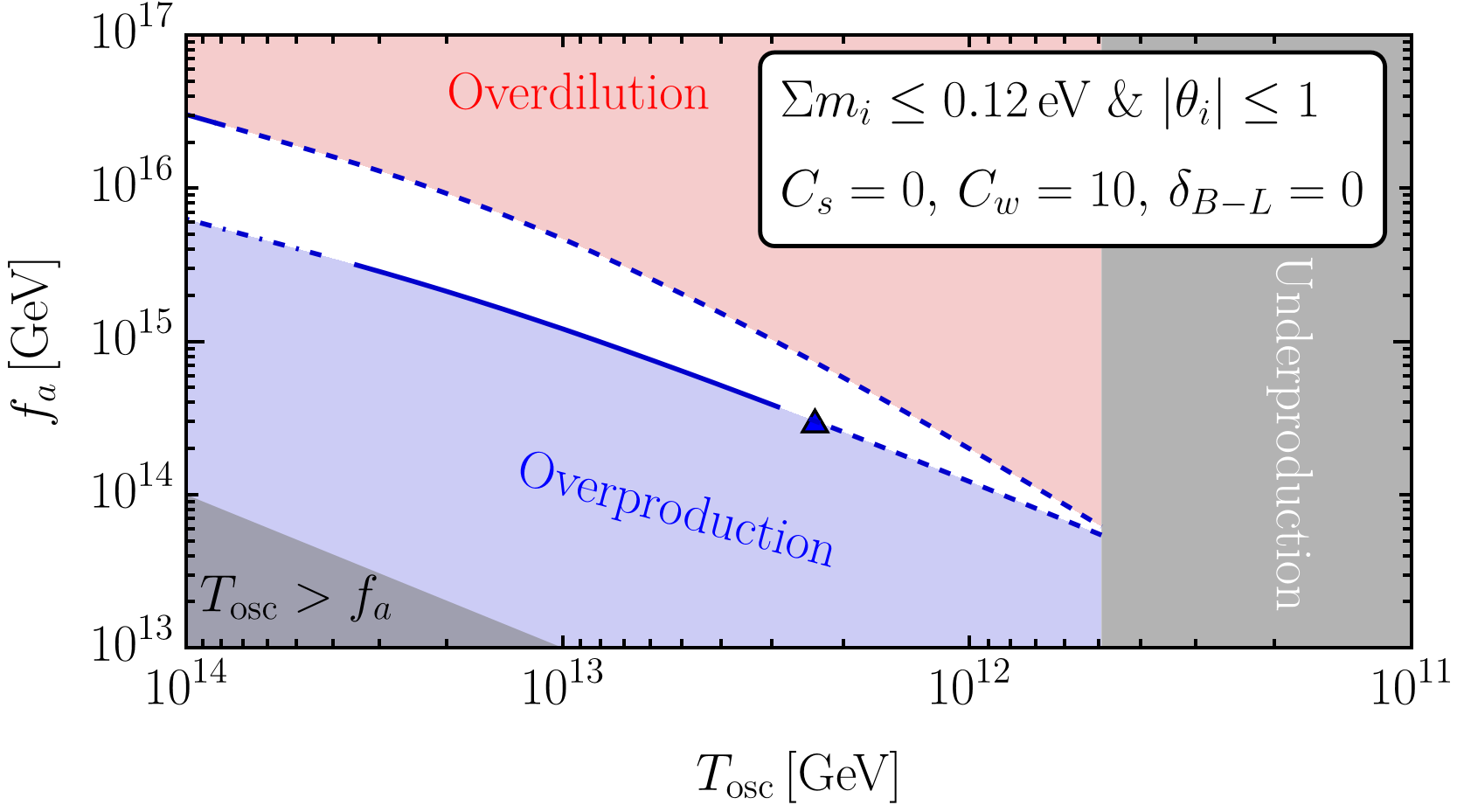}}}\hfill
    \caption{Same as Fig.~\ref{fig:WSParameterSpace}, but for IO instead of NO. }
    \label{fig:WSParameterSpaceIO}
\end{figure}

Thus far, our analysis of the viable parameter space for spontaneous LG has focused exclusively on NO. As noted in Sec.~\ref{sec:baryonasymmetry}, shifting from NO to IO may modify the asymmetry production by $\mathcal{O}(10\%\cdots 100\%)$ for axions coupled to $G\widetilde{G}$ or $W\widetilde{W}$. Consequently, the qualitative behavior of the results in Sec.~\ref{sec:FinalParameter} is independent of the neutrino mass ordering. Nevertheless, for completeness, we present the corresponding IO results below, where Fig.~\ref{fig:SSParameterSpaceIO} and~\ref{fig:WSParameterSpaceIO} serve as the counterparts to Fig.~\ref{fig:SSParameterSpace} and \ref{fig:WSParameterSpace} in the main text.

\end{appendix}

\clearpage


\addcontentsline{toc}{section}{References}

\small

\bibliographystyle{JHEP}
\bibliography{arxiv_2}

\end{document}